\documentclass[twocolumn]{aastex701}

\newcommand{\teff}{$T_{\rm eff}$}

\newcommand{\logg}{$\log g$}

\newcommand{\feh}{$\rm{[Fe/H]}$}

\newcommand{\vmic}{$\rm{v_{t}}$}

\newcommand{\sm}{$M_{\odot}$}
\newcommand{\sr}{$\rm{R_{\odot}}$}

\newcommand{\be}{\begin{equation}}
\newcommand{\ee}{\end{equation}}
\newcommand{\ben}{\begin{eqnarray}}
\newcommand{\een}{\end{eqnarray}}
\newcommand{\bfg}{\begin{figure}}
\newcommand{\efg}{\end{figure}}

\newcommand{\g}{$G$}

\usepackage{amsmath}

\usepackage[dvipsnames]{xcolor}

\shorttitle{{Giant Planets and Eccentric Orbits Are Common Around Galactic Thick Disc Stars}}
\shortauthors{T. Ferreira \& J. Yana Galarza et al.}

\begin{document}

\title{{Giant Planets and Eccentric Orbits Are Common Around Galactic Thick Disc Stars}\footnote{This paper includes data gathered with the 6.5-meter Magellan Telescopes located at Las Campanas Observatory, Chile}}

\correspondingauthor{Thiago Ferreira dos Santos}
\author[0000-0003-2059-470X]{Thiago Ferreira}
\affiliation{Department of Astronomy, Yale University, 219 Prospect St., New Haven, CT 06511, USA}
\email[show]{thiago.dossantos@yale.edu}

\author[orcid=0000-0001-9261-8366, gname=Jhon, sname=Yana Galarza]{Jhon Yana Galarza}
\altaffiliation{Carnegie Fellow}
\affiliation{The Observatories of the Carnegie Institution for Science, 813 Santa Barbara Street, Pasadena, CA 91101, USA}
\affiliation{Departamento de Astronomía, Facultad de Ciencias Físicas y Matemáticas Universidad de Concepción, Av. Esteban Iturra s/n Barrio Universitario, Casilla 160-C, Chile}
\email[]{jyanagalarza@carnegiescience.edu}

\author[0000-0001-6533-6179]{Henrique Reggiani}
\affiliation{Gemini South, Gemini Observatory, NSF's NOIRLab, Casilla 603, La Serena, Chile}
\email{henrique.reggiani@noirlab.edu}

\author[0000-0001-8153-639X]{Kiersten M. Boley}
\altaffiliation{NASA Sagan Fellow}
\affiliation{The Observatories of the Carnegie Institution for Science, 813 Santa Barbara Street, Pasadena, CA 91101, USA}
\email{kboley@carnegiescience.edu}

\author[0009-0004-8511-3721]{Isabelle Winnick}
\affiliation{Pomona College, 333 N College Way, Claremont, CA 91711, USA}
\email{iwaa2021@mymail.pomona.edu}

\author[0000-0002-4733-4994]{Joshua D. Simon}
\affiliation{The Observatories of the Carnegie Institution for Science, 813 Santa Barbara Street, Pasadena, CA 91101, USA}
\email{jsimon@carnegiescience.edu}

\author[0009-0008-2801-5040]{Johanna K. Teske}
\affiliation{Earth and Planets Laboratory, Carnegie Institution for Science, 5241 Broad Branch Road, NW, Washington, DC 20015, USA}
\affiliation{The Observatories of the Carnegie Institution for Science, 813 Santa Barbara Street, Pasadena, CA 91101, USA}
\email{jteske@carnegiescience.edu}

\author[0000-0002-5084-168X]{Eder Martioli}
\affiliation{Laborat\'{o}rio Nacional de Astrof\'{i}sica, Rua Estados Unidos 154, 37504-364, Itajub\'{a} - MG, Brazil}
\email{emartioli@lna.br}

\author[0000-0002-8177-7633]{Emiliano Jofr\'e}
\affiliation{Universidad Nacional de C\'ordoba - Observatorio Astron\'{o}mico de C\'{o}rdoba, Laprida 854, X5000BGR, C\'ordoba, Argentina}
\affiliation{Consejo Nacional de Investigaciones Cient\'{i}ficas y T\'{e}cnicas (CONICET), Godoy Cruz 2290, CABA, CPC 1425FQB, Argentina}
\email{emiliano.jofre@unc.edu.ar}

\author[orcid=0000-0003-0506-8269]{Verónica Loaiza-Tacuri}
\affiliation{Departamento de F\'isica, Universidade Federal de Sergipe, Av. Marcelo Deda Chagas, S/N Cep 49.107-230, S\~ao Crist\'ov\~ao, SE, Brazil}
\email{vloatac@gmail.com}

\author[0000-0002-2036-2311]{Yadira Gaibor}
\affiliation{Department of Physics, Massachusetts Institute of Technology, 77 Massachusetts Avenue, Cambridge, MA 02139, USA}
\affiliation{Kavli Institute for Astrophysics and Space Research, Massachusetts Institute of Technology, Cambridge, MA 02139, USA }
\email{ygaibor@mit.edu}

\author[0000-0002-8681-6136]{Stephen A. Shectman}
\affiliation{The Observatories of the Carnegie Institution for Science, 813 Santa Barbara Street, Pasadena, CA 91101, USA}
\email{shec@carnegiescience.edu}

\author[0000-0003-1305-3761]{R. Paul Butler}
\affiliation{Earth and Planets Laboratory, Carnegie Institution for Science, 5241 Broad Branch Road, NW, Washington, DC 20015, USA}
\email{bluaper@gmail.com}

\author[0000-0002-5226-787X]{Jeffrey D. Crane}
\affiliation{The Observatories of the Carnegie Institution for Science, 813 Santa Barbara Street, Pasadena, CA 91101, USA}
\email{jcrane@carnegiescience.edu}

\author[0000-0001-5130-6040]{Ian B. Thompson}
\affiliation{The Observatories of the Carnegie Institution for Science, 813 Santa Barbara Street, Pasadena, CA 91101, USA}
\email{ian@carnegiescience.edu}

\author[0000-0001-6637-5401]{Allyson Bieryla}
\affiliation{Center for Astrophysics \textbar \ Harvard \& Smithsonian, 60 Garden Street, Cambridge, MA 02138, USA}
\email{abieryla@cfa.harvard.edu}

\author[0000-0001-9911-7388]{David W. Latham}
\affiliation{Center for Astrophysics \textbar \ Harvard \& Smithsonian, 60 Garden Street, Cambridge, MA 02138, USA}
\email{dlatham@cfa.harvard.edu}

\author[0000-0002-0619-7639]{Carl Ziegler}
\affiliation{Department of Physics, Engineering and Astronomy, Stephen F. Austin State University, 1936 North St, Nacogdoches, TX 75962, USA}
\email{Carl.Ziegler@sfasu.edu}

\begin{abstract}

{Planet formation in the Galactic thick disc is expected to be inefficient---low solid reservoirs, short disc lifetimes, and harsh irradiation environments should conspire to inhibit the assembly of planetary bodies---yet, planets are there, and they are stranger than we expected. Here, we present a homogeneous characterisation of 32 exoplanetary systems orbiting chemically and kinematically confirmed thick disc stars, combining new detections with a systematic reassessment of archival systems, increasing the total number of exoplanets orbiting thick disc stars to 66. When planets form in the thick disc, a notable fraction are giants and move on more eccentric orbits than their thin disc counterparts---two results that challenge standard disc-evolution models. However, this should be interpreted with caution given detection biases and sample size. We also report TOI-1927 b and TOI-2643 b, the first puffy, low-density giant planets known to orbit thick disc stars, unexpected in old, metal-poor environments where planets should cool and contract efficiently. Together, these findings reveal an early Milky Way far more hospitable to planetary diversity than its harsh conditions would initially suggest.}

\end{abstract}


\section{Introduction}

The Galactic thick disc emerged between roughly $8-12$ billion years ago, during the Milky Way’s dynamically violent and chemically fast-evolving early phase, and now persists as a fossil record of the intense heating and rapid $\alpha-$element enrichment that marked that epoch \citep{1983MNRAS.202.1025G, 1998A&A...338..161F, 2019MNRAS.489..176M}, being distinctly characterised by elevated abundances of {magnesium, silicon, titanium, and calcium}. Stars belonging to this population emerged from low-metallicity molecular clouds during highly efficient star-formation epochs \citep{1997ApJ...477..765C, 2014ARA&A..52..415M}, embedded in environments shaped by strong feedback from core-collapse supernovae \citep{2007MNRAS.376.1465K, 2014MNRAS.445..581H, 2024ApJ...961L..41J}, and recurrent galaxy mergers \citep{2020MNRAS.494.3880B}. Such conditions were expected to truncate or even preclude the persistence of long-lived protoplanetary discs and thereby inhibit the emergence of dynamically cold planetary architectures (e.g., \citealt{2025ApJ...979..120H}). Nevertheless, it was within this feedback-dominated regime that the chemo-dynamical backbone of the present-day thick disc was imprinted, establishing the metallicity distribution function, elemental abundance patterns, and orbital structure that define early Galactic planet formation.

Theoretical models predict that thick disc stars inherited protoplanetary discs of comparatively low mass and short duration, characterised by diminished gas and dust surface densities $\Sigma$, disc masses $M_{\rm disc} \sim 0.01-0.05~M_\star$, and lifetimes of roughly only $\tau_{\rm disc} \sim 1-5~{\rm Myr}$, where intense ultraviolet photo-evaporation from nearby massive stars and frequent dynamical interactions within dense star-forming clusters would have accelerated disc dispersal and truncated their outer regions \citep{2025ApJ...979..120H}. Within the canonical core-accretion paradigm \citep{1996Icar..124...62P, 2004ApJ...604..388I}, the characteristic assembly timescale for a planetary core of mass $M_{\rm core}$ scales as $\tau_{\rm core} \sim (\Sigma_{\rm solids} \Omega)^{-1}$, where $\Omega$ denotes the local Keplerian frequency. For metallicities ${\rm [Fe/H]} \sim -0.5$ to $-1.0$ {dex}, the diminished reservoir of refractory species prolongs $\tau_{\rm core}$ beyond nominal disc dispersal timescales, thereby strongly disfavouring the formation of Jovian-mass planets and instead promoting the emergence of low-mass rocky bodies or subcritical, gas-poor cores \citep{2012A&A...547A.112M}. 

However, the presence of gas giants around even a small number of metal-poor hosts suggests that core accretion may not be the only viable formation pathway in such environments, with the gravitational fragmentation of massive, cold outer discs---the disc-instability mechanism \citep{1997Sci...276.1836B, 2012A&A...543A..89A, Vorobyov25, 2025arXiv251021863N}---providing a metallicity-agnostic route to giant-planet formation, operating on dynamical timescales set by the Safronov-Toomre criterion (${\mathcal{Q}}\propto \left(M_{\rm disc} T^2\right)/\left(\Sigma a^3\right) \lesssim 1$; \citealt{1960AnAp...23..979S, 1964ApJ...139.1217T}) and cooling-time constraints; alternative mechanisms also include streaming instability \citep{2024ApJ...969..130L}, or turbulent concentration \citep{2020ApJ...892..120H}. Since this pathway circumvents the need for rapid solid-core growth, the small number of giant planets around low-metallicity hosts may, in principle, be interpreted as evidence for fragmentation-driven formation, operating only in those exceptional systems whose early discs were sufficiently massive and cold for gravitational instability to occur. Planet formation, in these cases, may have proceeded preferentially within the inner few astronomical units of such discs, where accretion rates were high, orbital periods short, and the ratio $\tau_{\rm core}/\tau_{\rm disc}$ remained favourable to core assembly \citep{2017RSOS....470114E, 2024arXiv241211064A}.

Historically, this theoretical picture was consistent with observational evidence that thick disc stars were {found to host planets in only limited numbers by early surveys}, {broadly consistent with} the planet-metallicity correlation established by previous studies (e.g., \citealt{1997MNRAS.285..403G, 2004A&A...415.1153S, 2005ApJ...622.1102F, 2014Natur.509..593B, 2018AJ....155...89P, 2023AJ....165..262Z, 2024AJ....168..128B, Behmard:2025AJ....170..282B}). This paradigm is now being re-evaluated, as recent studies have significantly increased the number of known thick disc exoplanet hosts. {In particular, \citet{2025ApJS..281...61L} identified 62 thick disc stars hosting transiting planet candidates through combined spectroscopic and kinematic analyses of Kepler and K2 targets, demonstrating that planet occurrence in the thick disc may be more common than previously appreciated. Prior to this work, 32 confirmed exoplanet\footnote{{An object is ideally confirmed as a planet when both mass and radius measurements are available from radial velocity and transit observations, respectively, with both quantities primarily below the deuterium-burning limit \citep[$M_{p} \lesssim 13 M_{\rm{Jup}}$, $R_{p} \lesssim 25 R_{\oplus}$,][]{Spiegel:2011ApJ...727...57S, Hou:2022MNRAS.511.3133H}. Throughout this work, however, we classify an object as a confirmed planet when a dynamical mass measurement is available through radial velocity observations, since this directly measures the companion's gravitational influence on the host star. Objects known only via transit photometry, including statistically validated candidates, are referred to as planet candidates, as no direct mass constraint exists and transit-only validation remains probabilistic. See discussions in \cite{Santerne:2012A&A...545A..76S, Cabrera:2017A&A...606A..75C}}.} and three exoplanet candidates (detected only via transits with no dynamical mass measurement available) have been identified around 12 thick-disc stars \citep{Valenti:2005ApJS..159..141V, Bouchy:2010AA...519A..98B, Jofre:2015AA...574A..50J, Campante:2015ApJ...799..170C, Mortier:2020MNRAS.499.5004M, Lacedelli:2022MNRAS.511.4551L, Dai:2023AJ....166...49D, 2025arXiv250910136G, Behmard:2025AJ....170..282B, 2025ApJS..281...61L, Eschen:2025MNRAS.544.2614E}, representing systems with combined chemical and kinematic thick disc classification and precise orbital characterisation from spectroscopic and photometric analyses. This count is therefore complementary to, rather than in tension with, the broader photometric census of \citet{2025ApJS..281...61L}, with further candidates emerging from ongoing high-precision radial velocity, transit, and astrometric programmes (e.g., \emph{Gaia}: \citealt{2021ApJ...913L...3P, 2021AJ....162..100C, 2025A&A...694A..70M}).} The existence of such systems demonstrates that planet formation is not entirely prohibited in low-metallicity, dynamically active environments, challenging long-standing assumptions about the threshold metallicity for planet formation and the universality of the metallicity-occurrence relation \citep{2019Geosc...9..105A, 2024A&A...683A.118A}.

Beyond individual detections, large-scale surveys have revealed broader connections between planetary systems and their Galactic environment. Thanks to missions such as Kepler \citep{Borucki:2010Sci...327..977B} and TESS \citep{2015JATIS...1a4003R}, which have discovered thousands of exoplanets, and to \textit{Gaia} \citep{2023A&A...674A...1G}, which has mapped precise astrometry for billions of stars, it is now possible to link planetary architectures with stellar kinematics and population membership. Early work showed that the conditions of stellar birth environments leave lasting imprints on planetary outcomes. For instance, \citet{Winter:2024ApJ...972L...9W} and \citet{Kruijssen:2020ApJ...905L..18K} found that planet-hosting stars tend to share more co-moving companions than stars in low-density regions, and \citet{Hamer:2019AJ....158..190H} further connected giant planet occurrence to stellar kinematics, showing that hot Jupiter hosts have smaller Galactic velocity dispersion than main-sequence non-host stars. Since the velocity dispersion is correlated with age, it means that, on average, older populations typically do not have as many hot Jupiters as younger stellar populations. Building on this, \citet{Bashi:2019AJ....158...61B} and \citet{Dai:2021AJ....162...46D} reported correlations between planet occurrence and stellar motion, though interpreting these patterns is challenging because kinematics, age, and chemistry are tightly linked \citep{Gandhi:2019ApJ...880..134G}. Even so, metallicity alone cannot explain all observations. \citet{Zink:2023AJ....165..262Z} showed that small planets are more common around stars farther from the Galactic mid-plane---an effect consistent with an age-dependent mechanism, since older stars typically reach larger vertical excursions--- and supporting this interpretation, \citet{Ballard:2024arXiv240910485B} demonstrated that the observed decline in small-planet occurrence with Galactic height can be reproduced if tightly packed systems form more efficiently around younger stars. Most recently, \citet{Sagear:2025arXiv250923973S} found that single-transiting planets around thick disc stars have systematically higher orbital eccentricities than those around thin disc stars, a difference unlikely to arise from metallicity or giant planet companions alone. Collectively, these results demonstrate that planetary system formation and architecture are governed not only by local disc physics, but also by the broader stellar properties, such as age and chemical composition, as well as the Galactic environment, and its large-scale conditions jointly modulating both the likelihood of planet formation and the types of systems that emerge (see \citealt{2024AJ....168..128B}).

In this first paper of the series, we report the identification and characterisation of new planets orbiting $\alpha-$enhanced, thick disc stars, as well as revisit several systems that were previously known but never recognised as members of this ancient Galactic component. We re-analyse publicly available radial velocity and transit data jointly when both datasets exist, ensuring a consistent treatment of noise, systematics, and parameter inference across heterogeneous literature datasets. Furthermore, we reassess thick disc membership by combining precise \emph{Gaia}-derived kinematics with an independent chemical validation, confirming each star's classification through its characteristic $\alpha-$element enhancement, thereby securing robust population classification. {This approach is complementary to the photometric occurrence-rate methodology of \cite{2025ApJS..281...61L}, whose \emph{Kepler} and \emph{K2} surveys demonstrated that planet hosting in the thick disc may be widespread; where that work established statistical prevalence, the present work provides precise, homogeneous orbital and physical parameters for individual systems.} Through this dual effort, we construct a homogeneous catalogue of planets orbiting thick disc stars, which provides a coherent empirical foundation for future statistical investigations into planet occurrence, architecture, and composition in the chemically and dynamically ancient environments of the early Milky Way. 

{In Section \ref{sec:sample}, we describe the sample selection criteria and initial kinematic screening. Sections \ref{sec:observations} and \ref{sec:datatreat} detail the observations and data reduction. In Sections \ref{sec:fundamentalparams} and \ref{sec:elabuncances}, we derive stellar fundamental parameters and chemical abundances, as well as determine Galactic thin and/or thick membership of our targets in Section \ref{sec:pop}. In Section \ref{sec:rv_transit_analysis}, we describe the radial velocity and transit data analysis, presenting new and updated planetary parameters. Our discussion and conclusions are presented in Sections \ref{sec:discussions} and \ref{sec:conclusions}.}

\section{Sample Selection}\label{sec:sample}
Our planet sample was drawn from the NASA Exoplanet Archive\footnote{\url{https://exoplanetarchive.ipac.caltech.edu/}} \citep{2013PASP..125..989A, 2025PSJ.....6..186C}, which provides detailed properties for over 6,000 confirmed exoplanets as well as 7,700 TESS Objects of Interest (TOIs) by the time of this writing. It also includes stellar parameters for the host stars, inferred from both photometry and spectroscopy. We used this information to apply cuts in stellar and exoplanet parameters: orbital period $P_{p} \leq 300$ d, planet radius $R_{p} \leq 20\ R_{\oplus}$, effective temperature $4000 \leq$ \teff\ $\leq 6500$ K, metallicity $-1.0 \leq$ \feh\ $\leq +0.5$ dex, and surface gravity $4.1 \leq$ \logg\ $\leq 4.6$ dex. These criteria ensure that the stars are on the main sequence, although a few subgiants appear in the sample due to large parameter uncertainties when only photometric analyses are available or when low-quality spectra are used. The resulting sample of confirmed planets and TOIs was then cross-matched with the \textit{Gaia} DR3 \citep{2023A&A...674A...1G} catalogue to obtain precise astrometric data for the host stars, and we subsequently applied additional cuts to avoid binary systems and to focus on relatively bright stars, selecting those with $G < 15$ mag and renormalised unit weight error (RUWE) smaller than $1.2$ \citep{2020MNRAS.496.1922B}. 

Our sample selection also included 14 Kepler stars with thick disc–like kinematics that were observed by TESS \citep{2015JATIS...1a4003R}. Only three of these stars host exoplanets confirmed with radial velocity measurements; the rest are transit-only candidates, which we include as potential ancient systems of interest for future studies (see Subsection \ref{sub:keplerstars}). We also identified two additional thick disc stars from the Inti sample \citep{2021MNRAS.504.1873Y}, both solar-type hosts of newly detected exoplanets, and included them in our total sample.

\subsection{Initial Kinematic Screening}\label{sec:kinscreen}

\begin{figure}
    \centering
    \begin{tabular}{c}
        \includegraphics[width=\columnwidth]{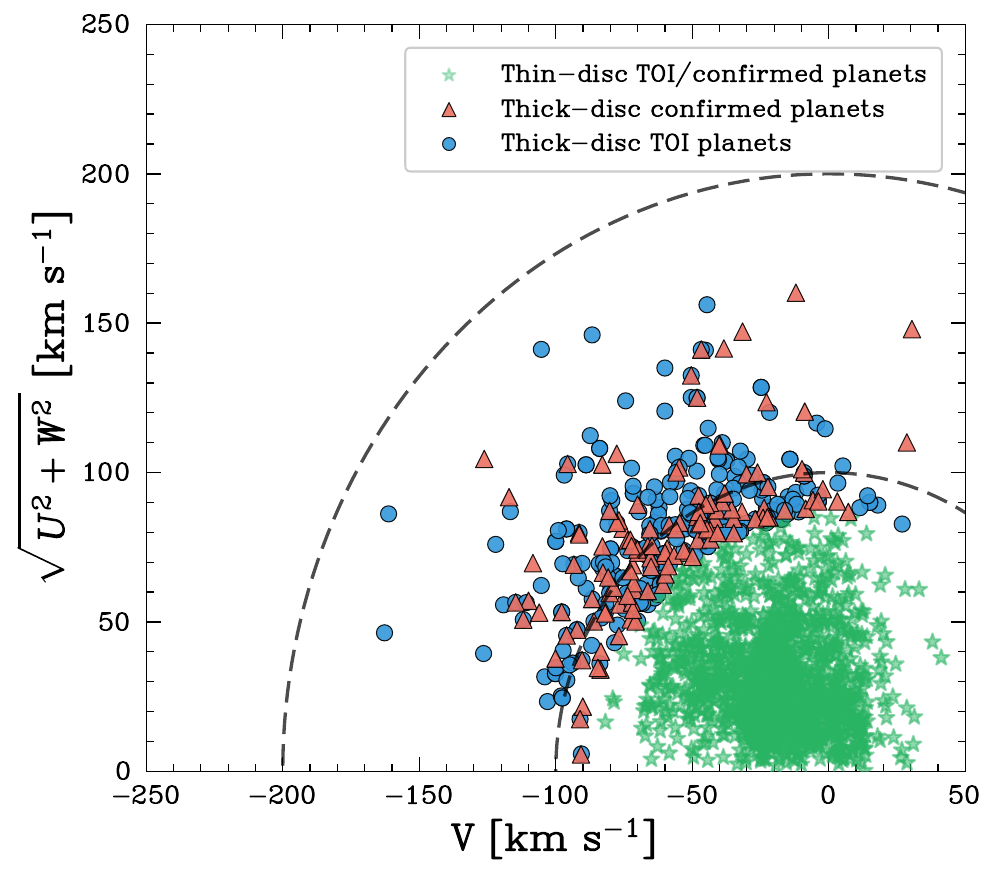}
    \end{tabular}
    \caption{Toomre diagram for our sample of stars with thick disc-like kinematics. Green stars indicate thin disc host stars from the NASA Exoplanet Archive \citep{2013PASP..125..989A}, while blue circles and red triangles represent stars hosting TOIs and confirmed planets, respectively.}
    \label{fig:toomre}
\end{figure}

We first assessed the kinematic heating of the host stars to establish whether they belong to the thick disc population using the {\sc Gala} code\footnote{\url{http://gala.adrian.pw/en/latest/}} \citep{gala, adrian_price_whelan_2020_4159870} in conjunction with \textit{Gaia} DR3 astrometry. The six-dimensional coordinates ($\alpha, \delta, \mu_{\alpha\cos\delta}, \mu_\delta, \overline{\pi}, {\rm RV}$) were transformed to a Galactocentric frame, assuming the Sun is located at ${\rm (x, y, z)} = (-8, 0, 0)$ kpc and has a velocity of $(v_x, v_y, v_z) = (-11.1, 244, 7.25)$ km s$^{-1}$ \citep{2010MNRAS.403.1829S, 2012MNRAS.427..274S}. We adopted a Milky Way potential composed of a spherical Hernquist bulge \citep{1990ApJ...356..359H}, a Miyamoto–Nagai disc \citep{1975PASJ...27..533M}, and a spherical Navarro–Frenk–White (NFW) dark matter halo \citep{1996ApJ...462..563N}, implemented in {\sc Gala} as the {\sc MilkyWayPotential}. A transformation to a non-inertial reference frame was applied to perform numerical orbit integration using the Hamiltonian form of the potential. The orbits were integrated with a time step of 0.5 Myr over 13.8 Gyr. From these, we derived the Galactic space velocities ($U, V, W$), the total velocity $H \equiv \left(U^2 + V^2 + W^2\right)^{1/2}$, eccentricity, azimuthal angle, and orbital energy. Together, these parameters indicate that the stars in our sample are consistent with the thick disc kinematic distribution \citep{2003A&A...410..527B, 2009MNRAS.399.1145S, 2023A&A...671A.136D}, as shown in the Toomre diagram ($V$ vs. $\sqrt{U^2 + W^2}$; see Figure \ref{fig:toomre}). Note that in this work, to distinguish thin from thick disc stars, we adopted a conservative velocity range of $90 \leq v_{\mathrm{tot}} \leq 200~\mathrm{km\ s^{-1}}$, similar to the criteria used in the literature, for comparison, the range $70 \leq v_{\mathrm{tot}} \leq 200~\mathrm{km\ s^{-1}}$ in \citet{Bensby:2004A&A...421..969B}. We identified a total of 145 stars hosting TOIs and 125 stars hosting confirmed exoplanets that show thick disc-like kinematics, although chemical confirmation was still required.

\section{Spectral Observations}\label{sec:observations}
As kinematic information alone is insufficient to confirm thick disc membership, we next verified the classification using chemical diagnostics, specifically the enhancement in $\alpha-$elements. To this end, we used high-resolution spectroscopy for a sample of stars exhibiting thick disc–like kinematics, including systems with confirmed exoplanets and TOIs. Our observations also include Kepler stars. From the TOI sample, we confirmed two exoplanets (see Section \ref{subsec:newplanets}), and work is still in progress to confirm thick disc membership for the remaining stars. The spectra also allow us to derive precise stellar parameters, including ages, masses, and radii (see Section \ref{sec:fundamentalparams}).

\subsection{Las Campanas Observatory}
\subsubsection{MIKE Observations}
We conducted observations with the Magellan Inamori Kyocera Echelle spectrograph (MIKE; \citealp{Bernstein:2003SPIE.4841.1694B, Shectman:2003SPIE.4837..910S}) on the 6.5-m Magellan Clay Telescope at Las Campanas Observatory. Spectra were obtained on 2024 October 22–23 and 2025 January 4, June 2, 21, and 27. We used the narrowest $0.''35\times5''$ slit, $1\times1$ binning, and the standard blue and red grating settings, providing wavelength coverage from 320 to 1,000 nm with resolving powers ($R = \lambda/\Delta\lambda$) of 83,000 and 65,000 in the blue and red arms, respectively. For stars with $G\geq11$ mag, we used the $0.''7\times5''$ slit with the $2\times2$ binning, yielding a resolving power of 35,000 in the blue and 28,000 in the red.

For each observing run, we collected the standard calibration frames, including quartz flats, milky flats, and Thorium-Argon (ThAr) lamp frames. Additionally, for an optimal wavelength solution, we took ThAr lamp exposures every 1-2 hours. We used the latest version of the CarPy\footnote{\url{https://code.obs.carnegiescience.edu/pipelines/carnegie-python-distribution}} software package \citep{Kelson:2000ApJ...531..159K, Kelson:2003PASP..115..688K, Kelson:2014ApJ...783..110K} to reduce the MIKE spectra.

\begin{deluxetable}{lcccccccc} 
\tablecaption{Log of observations for stellar characterisation. $^\alpha$Instruments used to observe each star. \label{tab:instruments}} 
\tablewidth{0pt} 
\tablehead{
\colhead{\textit{Gaia} DR3} &
\colhead{Other ID} & 
\colhead{R.A.} & 
\colhead{Decl.} & 
\colhead{$G$} & 
\colhead{RUWE} &
\colhead{Instrument$^{\alpha}$} & 
\colhead{$R$} & 
\colhead{S/N}  \\ 
\colhead{} &
\colhead{} & 
\colhead{(deg)} & 
\colhead{(deg)} &
\colhead{(mag)} &
\colhead{} &
\colhead{} & 
\colhead{$\lambda/\Delta \lambda$} & 
\colhead{at $6500$\,\AA}
}
\startdata
6643607700111053568 & TOI-3277    & 290.856 & $-54.330$ & 13.35 & 0.996 & MIKE/PFS   & 83,000/110,000 & 200 \\
4354950641755249024 & HD 150433   & 250.284 & $-02.859$ & 7.05  & 0.971 & MIKE/HARPS & 83,000/115,000 & 300 \\
4153637759337630720 & HD 168746   & 275.457 & $-11.922$ & 7.78  & 0.959 & MIKE/HARPS & 83,000/115,000 & 300 \\
4708607629114652032 & HD 4308     & 11.165  & $-65.652$ & 6.38  & 1.036 & MIKE/HARPS & 83,000/115,000 & 350 \\
5855730584310531200 & HD 111232   & 192.215 & $-68.424$ & 7.41  & 1.244 & MIKE/HARPS & 83,000/115,000 & 300 \\
6745589980571162752 & HD 181720   & 290.721 & $-32.920$ & 7.68  & 1.108 & MIKE/HARPS & 83,000/115,000 & 300 \\
5902750168276592256 & TOI-2011    & 230.439 & $-48.318$ & 5.48  & 0.906 & MIKE/HARPS & 83,000/115,000 & 340 \\
6434759761383343744 & HD 175607   & 285.274 & $-66.194$ & 8.42  & 0.920 & MIKE/HARPS & 83,000/115,000 & 250 \\
4847957293278177024 & HD 20794    & 50.000  & $-43.066$ & 4.06  & 1.976 & MIKE/HARPS & 83,000/115,000 & 330 \\
4988158666636598272 & HD 6434     & 16.166  & $-39.490$ & 7.57  & 0.981 & HARPS & 115,000 & 280 \\
4840050395924931968 & HD 27631    & 64.939  & $-41.960$ & 8.09  & 0.959 & MIKE/HARPS & 83,000/115,000 & 240 \\
2817634821194781824 & HD 220197   & 350.494 & $16.632$  & 8.75  & 0.948 & MIKE & 83,000 & 270 \\
6596925769287980416 & TOI-126     & 339.489 & $-35.154$ & 13.19 & 0.971 & MIKE & 83,000 & 200 \\
6489771832811810432 & HD 219077   & 348.532 & $-62.701$ & 5.92  & 1.084 & MIKE/HARPS & 83,000/115,000 & 340 \\
3620325206217720320 & K2-262      & 206.846 & $-6.139$  & 9.90  & 1.051 & MIKE/HARPS & 83,000/115,000 & 235 \\
3596250888028092160 & K2-156      & 182.699 & $-6.294$  & 12.93 & 0.973 & MIKE & 83,000 & 220 \\
3607243530404017280 & K2-190      & 205.363 & $-13.160$ & 10.95 & 1.185 & MIKE & 83,000 & 200 \\
3803004978659671040 & TOI-1927    & 163.788 & $-0.7370$ & 12.80 & 1.064 & MIKE & 83,000 & 190 \\
6257641627989839744 & K2-408      & 227.916 & $-17.684$ & 11.72 & 1.104 & MIKE & 83,000 & 215 \\
2230429435107258752 & HIP 109384  & 332.409 & $+71.313$ & 9.40  & 0.960 & SOPHIE     & 75,000         & 80  \\
2079192121810035840 & Kepler-463  & 298.936 & $+44.857$ & 12.85 & 0.955 & SOPHIE     & 75,000         & 60  \\
2078722600285671936 & Kepler-112  & 296.981 & $+43.209$ & 13.75 & 0.963 & HIRES      & 48,000         & 40  \\
2077409645964537472 & Kepler-517  & 293.675 & $+41.295$ & 12.12 & 0.878 & HIRES      & 48,000         & 40  \\
2099597408349457024 & Kepler-1898 & 287.205 & $+38.498$ & 13.16 & 1.017 & HIRES      & 48,000         & 40  \\
2105921803532722432 & Kepler-1619 & 285.703 & $+44.422$ & 13.29 & 0.926 & HIRES      & 48,000         & 40  \\
2132155017099178624 & Kepler-10   & 285.679 & $+50.241$ & 10.92 & 0.899 & HIRES      & 48,000         & 40  \\
2132156632006693504 & Kepler-1258 & 285.404 & $+49.994$ & 13.77 & 1.005 & HIRES      & 48,000         & 140 \\
39404267653786624   & K2-173      & 60.375  & $15.625$  & 12.17 & 1.144 & TRES & 44,000 & 40 \\
658978451430055936  & K2-337      & 127.761 & $+16.900$ & 12.22 & 0.976 & TRES       & 44,000         & 32  \\
2891248292906892032 & TOI-1926    & 89.5988 & $-30.811$ & 12.90 & 1.017 & HARPS      & 115,000        & 120  \\
3582339626395387264 & TOI-2643    & 182.166 & $-8.7472$ & 11.64 & 1.011 & HARPS      & 115,000        & 150  \\
652007410271734912  & K2-183      & 125.007 & $+14.019$ & 12.66 & 1.032 & HIRES      & 48,000         & 100  \\
5168359139139252352 & HD 21019    & 50.823  & $-7.795$  & 6.02  & 1.716 & HARPS & 115,000 & 220 \\
6180035765135232128 & HD 114729   & 198.183 & $-31.874$ & 6.53  & 0.947 & HARPS & 115,000 & 320 \\
\enddata 
\end{deluxetable}

\subsubsection{PFS Observations}
Specifically for TOI-3277, we obtained spectra on 2025 July 3, 5, 8, 9, 16, and 17 using the Carnegie Planet Finder Spectrograph \citep[PFS;][]{Crane:2006SPIE.6269E..31C, Crane:2008SPIE.7014E..79C, Crane:2010SPIE.7735E..53C}. PFS is a high-resolution echelle spectrograph covering the wavelength range $\sim$391–734 nm. We used the $0.''3$ slit with $3 \times 3$ detector binning, achieving a resolving power of $R \sim 110,000$. A total of 21 spectra were obtained through the iodine gas cell to model the instrument profile, along with one iodine-free template taken on 2025 July 9. The data reduction and radial velocity extraction were performed using a custom IDL pipeline \citep{Butler:1996PASP..108..500B}. The derived velocities were used to constrain the orbital parameters of TOI-3277 b (Y. Gaibor et al., in preparation). The iodine-free template spectrum was also used to refine the stellar parameters of the host star.

\subsection{La Silla Observatory}
We used the European Southern Observatory (ESO) Science Archive to search for publicly available spectra of our sample stars, identifying High Accuracy Radial velocity Planet Searcher (HARPS; \citealt{Mayor:2003Msngr.114...20M}) spectra from 82 different ESO programs through the Phase 3 data access interface\footnote{\url{https://archive.eso.org/wdb/wdb/adp/phase3_spectral/form}}. HARPS is mounted on the ESO 3.6-m telescope at La Silla Observatory and functions as a high-resolution ($R \sim 115{,}000$) echelle spectrograph covering the wavelength range $\sim$378–691 nm; observations were acquired between 2010 and 2024, with each exposure automatically reduced by the instrument pipeline to ensure consistent and high-quality spectral data.

\subsection{The W.M. Keck Observatory}
Kepler and K2 stars (Kepler-112, Kepler-517, Kepler-1619, Kepler-10, Kepler-1258, Kepler-1898, and K2-183) were observed with the High Resolution Echelle Spectrometer \citep[HIRES;][]{Vogt:1994SPIE.2198..362V} on the 10 m Keck I telescope. The nearly continuous 400–840 nm coverage is assembled from three detectors (blue, green, and red), each sampling a distinct part of the wavelength range. The HIRES spectra were downloaded from the Keck Observatory Archive\footnote{\url{https://koa.ipac.caltech.edu/cgi-bin/KOA/nph-KOAlogin}} using the Program IDs: U023Hr\_2011 (PI: Marcy), A092Hr\_2011 (PI: Smith), N204Hr\_2011 (PI: Borucki), U040Hr\_2012 (PI: Marcy), N137Hr\_2013 (PI: Payne), C235Hb\_2016 (PI: Wang), and U082Hr (PI: Hansen). In total, we retrieved seven spectra with spectral resolution $R \sim 48,000$. The Keck-MAKEE pipeline was used for standard echelle spectral reduction, including bias subtraction, flat-fielding, scattered-light subtraction, spectral extraction, and wavelength calibration. 

\subsection{L'Observatoire de Haute-Provence}
We collected five spectra of HIP 109384 and six of Kepler-463 from the online archive\footnote{\url{http://atlas.obs-hp.fr/sophie/}}
 of the Spectrographe pour l’Observation des Phénomènes des Intérieurs stellaires et des Exoplanètes (SOPHIE). SOPHIE is a high-resolution, fibre-fed, cross-dispersed échelle spectrograph mounted on the 1.93 m telescope at the Observatoire de Haute-Provence (OHP; \citealt{Perruchot2008, Bouchy2013}). It covers wavelengths from $\sim$387 to 694 nm. We obtained calibrated spectra reduced with version 0.50 of the SOPHIE Data Reduction Software \citep[DRS;][]{Bouchy2009}. These observations were taken in high-resolution mode (HR mode, $R = 75{,}000$): for HIP 109384 in 2021 October and November, and for Kepler-463 in 2011 July, September, and November. The peak S/N per spectral element ranges from approximately 100 to 200.

\subsection{Fred Lawrence Whipple Observatory}
We obtained spectra of K2-173 and K2-337 with the Tillinghast Reflector Echelle Spectrograph \citep[TRES;][]{Szentgyorgyi:2007RMxAC..28..129S}, mounted on the 1.5 m Tillinghast Reflector at the Fred Lawrence Whipple Observatory on Mt. Hopkins, Arizona. TRES delivers a resolving power of $R = 44{,}000$ and covers wavelengths from 390 to 910 nm, and has been used in recent studies of stellar properties and chemical abundances of exoplanet hosts \citep{2026ApJ..1004...91R}. The spectra available in the TRES archive\footnote{\url{http://tdc-www.harvard.edu/instruments/tres/tresearch.html}} are processed with the standard reduction pipeline \citep{Buchhave:2010ApJ...720.1118B}, which performs flat-fielding, cosmic-ray removal, order extraction, and wavelength calibration using ThAr exposures. A TRES spectrum of K2-337 was obtained in November 2015, and one of K2-173 in October 2015.

\section{Data Treatment}\label{sec:datatreat}
To determine stellar parameters, we used the method of spectroscopic equilibrium within an absolute abundance analysis. Because solar abundances are required to derive abundances relative to the Sun, we measured our own solar abundances for MIKE, HARPS, SOPHIE, and HIRES (see Subsection \ref{sec:sp} for details). For the remaining instruments, including PFS and TRES, we adopted the solar abundances from \citet{asplund2021}. We obtained solar reference spectra with MIKE using sunlight reflected from the asteroid Vesta. Similarly, HARPS spectra for the Sun are also available in the ESO archive, and the solar spectra from HIRES and SOPHIE are available through their respective instrument archives. On average, the solar spectra from all instruments have a signal-to-noise ratio (S/N) of $\sim$300-400 per pixel at $\sim$650 nm. 

The MIKE, HIRES, SOPHIE, and TRES spectra were corrected for both radial and barycentric velocities using \textsc{iSpec}\footnote{\url{https://blancocuaresma.com/s/iSpec}}
 \citep{Blanco:2014A&A...569A.111B, Blanco:2019MNRAS.486.2075B}. For HARPS and PFS, we adopted the radial velocities provided by their respective pipelines. All spectra were shifted to the rest frame and normalised in pixel space using cubic spline fits (typically of third degree) using the \textsc{dopcor} and \textsc{continuum} tasks of \textsc{IRAF\footnote{\textsc{IRAF} is distributed by the National Optical Astronomy Observatory, operated by the Association of Universities for Research in Astronomy, Inc., under a cooperative agreement with the National Science Foundation; \url{https://iraf-community.github.io}}}, respectively. For cooler stars ($T_{\rm eff} < 5000$ K), we carefully avoided molecular bands blueward of 500 nm, where continuum placement becomes uncertain, and restricted the analysis to longer wavelengths. The normalised spectra were then co-added using the IRAF \texttt{scombine} task with an average combination method. The resulting S/N ratios and spectral resolutions for each target are listed in Table \ref{tab:instruments}. Note that in this work, we report only the stars that we confirm as thick disc members. The remaining stars, observed with the same instruments, will be presented in Paper II (Yana Galarza et al., in preparation).

\section{Fundamental Parameters}\label{sec:fundamentalparams}

\subsection{Equivalent Widths}
We measured equivalent widths (EWs) of \ion{Fe}{1} and \ion{Fe}{2} lines by selecting pseudo-continuum regions of 6~\AA\ around each feature. Gaussian profiles were fitted to the lines using the \textsc{KAPTEYN}\footnote{\url{https://www.astro.rug.nl/software/kapteyn/index.html}} \texttt{kmpfit} package \citep{KapteynPackage}. We adopted our own updated version of the line list {from \citet{Ramirez:2009A&A...508L..17R}, \citet{Melendez:2014ApJ...791...14M}, and \citet{Ramirez:2015ApJ...808...13R}}, which provides atomic data, including statistical weights and oscillator strengths. All measurements were performed manually, line by line, to ensure maximum precision. To avoid saturation effects in the determination of spectroscopic stellar parameters, we only considered iron lines with EWs $<$ 130 m\AA.

\subsection{Spectroscopic Stellar Parameters}
\label{sec:sp}
The spectroscopic stellar parameters—effective temperature (\teff), surface gravity (\logg), metallicity (\feh), and microturbulent velocity (\vmic)—were derived by enforcing spectroscopic equilibrium, which involves fulfilling three conditions: (i) excitation balance, such that the derived iron abundances show no trend with excitation potential, to obtain \teff; (ii) ionization balance, requiring consistent abundances from \ion{Fe}{1} and \ion{Fe}{2} lines, to determine \logg; and (iii) the absence of correlation between abundance and reduced equivalent width ($\log(\mathrm{EW}/\lambda)$), to infer \vmic. Iron abundances were calculated using the curve-of-growth method implemented in the \textsc{abfind} driver of \textsc{MOOG}\footnote{\url{https://www.as.utexas.edu/~chris/moog.html}}. 

\startlongtable
\begin{deluxetable*}{lcccccccc} 
\tablecaption{{Stellar parameters inferred in this work (upper section) and for thick disc stars hosting confirmed planets in the literature (lower section). We include systems for which mass and radius measurements are available, or a mass measurement from radial velocity observations alone, consistent with our planet confirmation criterion. $^\alpha$ Stellar parameters inferred in this work: age, mass, and radius.} \label{tab:sp}} 
\tablewidth{0pt}
\setlength{\tabcolsep}{4pt}
\tablehead{
\colhead{ID} &
\colhead{\teff} & 
\colhead{\logg} & 
\colhead{\feh} & 
\colhead{\vmic} & 
\colhead{Age} &
\colhead{Mass} &  
\colhead{Radius} & 
\colhead{Reference} \\ 
\colhead{} &
\colhead{(K)} & 
\colhead{(dex)} & 
\colhead{(dex)} &
\colhead{(km s$^{-1}$)} &
\colhead{(Gyr)} &
\colhead{(\sm)} & 
\colhead{(\sr)} &
\colhead{}
}
\startdata
TOI-3277    & $4958\pm66$  & $4.27\pm0.17$ & $-0.345\pm0.033$ & $0.87\pm0.15$ & $12.99\pm0.84$ & $0.77\pm0.01$ & $0.805\pm0.01$ & This work \\
HD 150433   & $5656\pm13$  & $4.40\pm0.08$ & $-0.342\pm0.017$ & $1.00\pm0.02$ & $13.43\pm0.11$ & $0.84\pm0.01$ & $1.042\pm0.01$ & This work \\
HD 168746   & $5593\pm15$  & $4.39\pm0.10$ & $-0.046\pm0.018$ & $0.93\pm0.03$ & $12.19\pm0.42$ & $0.90\pm0.01$ & $1.094\pm0.01$ & This work \\
HD 4308     & $5667\pm14$  & $4.39\pm0.09$ & $-0.303\pm0.018$ & $1.01\pm0.02$ & $13.26\pm0.30$ & $0.84\pm0.01$ & $1.034\pm0.01$ & This work \\
HD 111232   & $5480\pm15$  & $4.46\pm0.09$ & $-0.382\pm0.019$ & $0.86\pm0.02$ & $13.43\pm0.11$ & $0.79\pm0.01$ & $0.891\pm0.00$ & This work \\
HD 181720   & $5698\pm12$  & $4.03\pm0.08$ & $-0.561\pm0.018$ & $1.18\pm0.02$ & $12.29\pm0.29$ & $0.86\pm0.01$ & $1.528\pm0.01$ & This work \\
TOI-2011    & $5658\pm14$  & $4.36\pm0.09$ & $-0.335\pm0.018$ & $1.00\pm0.02$ & $13.40\pm0.16$ & $0.84\pm0.01$ & $1.044\pm0.01$ & This work \\
HD 175607   & $5388\pm13$  & $4.46\pm0.09$ & $-0.573\pm0.020$ & $0.83\pm0.02$ & $13.44\pm0.09$ & $0.75\pm0.01$ & $0.809\pm0.00$ & This work \\
HD 20794    & $5419\pm15$  & $4.47\pm0.10$ & $-0.351\pm0.022$ & $0.82\pm0.03$ & $13.45\pm0.08$ & $0.79\pm0.01$ & $0.885\pm0.00$ & This work \\
HD 6434     & $5725\pm13$  & $4.28\pm0.08$ & $-0.593\pm0.019$ & $1.08\pm0.02$ & $13.47\pm0.04$ & $0.82\pm0.01$ & $1.091\pm0.01$ & This work \\
HD 27631    & $5697\pm15$  & $4.40\pm0.10$ & $-0.100\pm0.018$ & $1.03\pm0.03$ & $10.84\pm0.43$ & $0.90\pm0.01$ & $1.079\pm0.01$ & This work \\
HD 220197   & $5656\pm17$  & $4.38\pm0.12$ & $-0.448\pm0.027$ & $0.92\pm0.03$ & $13.36\pm0.22$ & $0.82\pm0.01$ & $1.015\pm0.01$ & This work \\
TOI-126     & $5753\pm40$  & $4.34\pm0.25$ & $-0.541\pm0.047$ & $1.21\pm0.07$ & $13.13\pm0.57$ & $0.83\pm0.01$ & $1.056\pm0.01$ & This work \\
HD 219077   & $5335\pm16$  & $3.96\pm0.10$ & $-0.148\pm0.018$ & $1.01\pm0.03$ & $9.32\pm0.33$  & $1.01\pm0.01$ & $1.950\pm0.03$ & This work \\
TOI-1927    & $5395\pm43$  & $4.24\pm0.11$ & $-0.328\pm0.035$ & $0.37\pm0.18$ & $12.56\pm1.37$ & $0.80\pm0.01$ & $0.867\pm0.01$ & This work \\
K2-408      & $5244\pm28$  & $4.37\pm0.09$ & $-0.631\pm0.021$ & $0.23\pm0.24$ & $13.33\pm0.29$ & $0.73\pm0.01$ & $0.772\pm0.01$ & This work \\
HIP 109384  & $5168\pm22$  & $4.37\pm0.18$ & $-0.296\pm0.043$ & $1.04\pm0.05$ & $13.28\pm0.34$ & $0.77\pm0.01$ & $0.827\pm0.01$ & This work \\
TOI-1926    & $5129\pm18$  & $4.38\pm0.17$ & $-0.200\pm0.044$ & $0.50\pm0.04$ & $11.50\pm2.47$ & $0.80\pm0.02$ & $0.826\pm0.01$ & This work \\
TOI-2643    & $4701\pm62$  & $3.02\pm0.13$ & $-0.010\pm0.020$ & $0.74\pm0.11$ & $7.12\pm3.12$  & $1.12\pm0.13$ & $3.900\pm0.08$ & This work \\
HD 21019    & $5373\pm27$  & $3.71\pm0.07$ & $-0.533\pm0.021$ & $1.01\pm0.04$ & $12.28\pm1.11$ & $0.82\pm0.01$ & $3.900\pm0.08$ & This work \\
HD 114729   & $5752\pm28$  & $4.05\pm0.06$ & $-0.322\pm0.020$ & $1.13\pm0.04$ & $12.28\pm1.11$ & $0.82\pm0.01$ & $0.942\pm0.01$ & This work \\
\hline
\multicolumn{9}{c}{{Known thick disc planet host stars by the time of this writting}} \\
\hline
HD 39194$^\alpha$  & $5219\pm10$  & $4.47\pm0.04$ & $-0.560\pm0.010$ & $0.62\pm0.04$ & $13.74\pm1.22$ & $0.71\pm0.01$ & $0.71\pm0.01$  & (1) \\
HD 37124        & $5500\pm44$  & $4.60\pm0.06$ & $-0.440\pm0.010$ & ...           & $11.70\pm3.30$ & $1.47\pm0.23$ & $1.00\pm0.04$  & (2) \\
Kepler-444      & $5046\pm74$  & $4.60\pm0.06$ & $-0.550\pm0.030$ & ...           & $11.23\pm0.95$ & $0.75\pm0.04$ & $0.75\pm0.01$  & (3) \\
TOI-561         & $5372\pm70$  & $4.50\pm0.12$ & $-0.400\pm0.050$ & ...           & $11.00\pm3.15$ & $0.81\pm0.04$ & $0.84\pm0.01$  & (4) \\
TOI-2018        & $4174\pm38$  & $4.62\pm0.03$ & $-0.580\pm0.180$ & ...           & ...            & ...           & ...            & (5) \\
TOI-5963        & $5800\pm100$ & $4.20\pm0.01$ & $-0.460\pm0.110$ & ...           & $12\pm5$       & $1.01\pm0.03$ & $1.06\pm0.04$  & (6) \\
K2-111          & $5775\pm60$  & $4.25\pm0.15$ & $-0.460\pm0.050$ & $1.02\pm0.05$ & $13.5\pm0.7$   & $0.84\pm0.02$ & $1.25\pm0.02$  & (7) \\
TOI-1203$^\alpha$  & $5741\pm10$  & $4.29\pm0.03$ & $-0.410\pm0.010$ & $1.12\pm0.01$ & $11.50\pm0.42$ & $0.91\pm0.01$ & $1.17\pm0.01$  & (8) \\
Kepler-10$^\alpha$ & $5630\pm28$  & $4.27\pm0.07$ & $-0.121\pm0.022$ & $0.71\pm0.07$ & $11.50\pm0.85$ & $0.90\pm0.01$ & $1.089\pm0.01$ & (9) \\
K2-262$^\alpha$    & $4679\pm18$  & $4.51\pm0.24$ & $-0.401\pm0.071$ & $0.35\pm0.05$ & $13.06\pm0.86$ & $0.66\pm0.00$ & $0.663\pm0.01$ & (10) \\
\hline
\multicolumn{9}{c}{{Stars hosting exoplanet candidates detected only via the transit method}} \\
\hline
K2-173      & $5552\pm62$  & $4.17\pm0.15$ & $-0.344\pm0.034$ & $0.96\pm0.12$ & $12.66\pm1.20$ & $0.82\pm0.01$ & $0.947\pm0.01$ & This work \\
K2-156      & $4717\pm78$  & $4.18\pm0.19$ & $-0.240\pm0.054$ & $1.05\pm0.15$ & $7.04\pm4.45$  & $0.69\pm0.01$ & $0.663\pm0.00$ & This work \\
Kepler-463  & $5720\pm43$  & $4.30\pm0.33$ & $-0.275\pm0.058$ & $0.32\pm0.11$ & $12.53\pm1.29$ & $0.85\pm0.01$ & $1.059\pm0.01$ & This work \\
Kepler-112  & $5526\pm35$  & $4.35\pm0.09$ & $-0.479\pm0.027$ & $0.66\pm0.08$ & $13.20\pm0.47$ & $0.79\pm0.01$ & $0.899\pm0.01$ & This work \\
Kepler-517  & $5575\pm30$  & $4.27\pm0.08$ & $-0.146\pm0.025$ & $0.68\pm0.07$ & $12.51\pm0.94$ & $0.87\pm0.01$ & $1.021\pm0.01$ & This work \\
Kepler-1898 & $5333\pm36$  & $4.23\pm0.09$ & $-0.461\pm0.028$ & $0.50\pm0.10$ & $13.23\pm0.46$ & $0.77\pm0.01$ & $0.835\pm0.01$ & This work \\
Kepler-1619 & $5676\pm41$  & $4.09\pm0.09$ & $-0.269\pm0.028$ & $0.85\pm0.07$ & $12.02\pm0.59$ & $0.91\pm0.01$ & $1.457\pm0.02$ & This work \\
Kepler-1258 & $5670\pm41$  & $4.19\pm0.09$ & $-0.314\pm0.028$ & $0.76\pm0.08$ & $13.13\pm0.56$ & $0.86\pm0.01$ & $1.111\pm0.02$ & This work \\
K2-337      & $5735\pm139$ & $3.91\pm0.42$ & $-0.318\pm0.055$ & $0.97\pm0.12$ & $12.85\pm0.87$ & $0.87\pm0.01$ & $1.342\pm0.02$ & This work \\
K2-183      & $5313\pm62$  & $4.07\pm0.11$ & $+0.113\pm0.040$ & $0.56\pm0.12$ & $10.33\pm2.05$ & $0.89\pm0.03$ & $0.942\pm0.01$ & This work \\
K2-190      & $5621\pm22$  & $4.40\pm0.07$ & $-0.242\pm0.020$ & $0.73\pm0.06$ & $12.60\pm0.59$ & $0.85\pm0.01$ & $1.002\pm0.01$ & This work \\
\enddata 
\tablenotetext{}{(1) \citet{Jofre:2015AA...574A..50J}; (2) \citet{Valenti:2005ApJS..159..141V}; (3) \citet{Campante:2015ApJ...799..170C}; (4) \citet{Lacedelli:2022MNRAS.511.4551L}; (5) \citet{Dai:2023AJ....166...49D}; (6) \citet{Bouchy:2010AA...519A..98B}; (7) \citet{Mortier:2020MNRAS.499.5004M}; (8) \citet{2025arXiv250910136G}; (9) \citet{Behmard:2025AJ....170..282B}; (10) \citet{Loaiza:2024ApJ...970...53L}.}
\end{deluxetable*}

The spectroscopic equilibrium was determined using the \textsc{XIRU}\footnote{\url{https://github.com/arthur-puls/xiru}} \citep{Alencastro_thesis:1885-288051} package, a Python code configured to employ the Kurucz ODFNEW model atmospheres \citep{Castelli:2003IAUS..210P.A20C} together with the 2019 version of the LTE code \textsc{MOOG} \citep{Sneden:1973PhDT.......180S}. To achieve equilibrium, \textsc{XIRU} implements Broyden’s method, an iterative algorithm that efficiently solves systems of nonlinear equations by updating an approximate Jacobian until convergence is reached. To compute abundances, we adopted the solar values reported by \citet{asplund2021}. For the MIKE and HARPS measurements, we instead derived solar abundances directly from their corresponding solar spectra. Our iron measurements ($A(\mathrm{Fe})_{\odot,\mathrm{MIKE}} = 7.43 \pm 0.2$ and $A(\mathrm{Fe})_{\odot,\mathrm{HARPS}} = 7.41 \pm 0.2$) agree well with the value $A(\mathrm{Fe})_{\odot} = 7.46 \pm 0.4$ reported by \citet{asplund2021}. {Table \ref{tab:sp} lists the stellar parameters for the newly identified thick disc stars hosting confirmed and candidate exoplanets and those already reported in the literature.}

\subsection{Uncertainty Estimations}
\label{sec:uncertainties}
{Uncertainties in the stellar parameters were estimated following the procedure described in \citet{Epstein:2010ApJ...709..447E} and \citet{Bensby:2014A&A...562A..71B}. Briefly, the uncertainty in \teff\ and \vmic\ are propagated from the null-slope fits imposed during spectroscopic equilibrium, while the uncertainties in \logg\ and \feh\ are derived from the standard errors of the mean of the neutral and ionised iron line abundances, respectively.} 

{For the $\alpha$-element abundances, uncertainties are estimated by propagating the contributions from the equivalent width measurement errors and the uncertainties in the stellar parameters \teff, \logg, \feh, and \vmic. The sensitivity of each abundance to variations in the stellar parameters is evaluated by perturbing each parameter by its associated uncertainty while keeping the others fixed, and the contributions are added in quadrature.}

\subsection{Stellar Ages, Masses, and Radii}
We estimated stellar ages, masses, and radii through an isochrone fitting method, anchored by the extremely precise differential spectroscopic photospheric parameters. We apply the same methodology described in detail by \citet{Reggiani:2022AJ....163..159R, Reggiani:2022AJ....163..252R, Reggiani:2024AJ....167...45R}. While the classical spectroscopic approach simultaneously minimizes line-by-line iron abundance inference difference between \ion{Fe}{1} \& \ion{Fe}{2}-based abundances as well as their dependencies on excitation potential and reduced equivalent widths to infer T$_{\text{eff}}$, log$g$, and [Fe/H], isochrones fitting is needed to infer fundamental (mass, luminosity, and radius) stellar parameters that are consistent with the differentially determined photospheric parameters.  

For our isochrone fitting we use high-quality multiwavelength photometry: \emph{Gaia} DR3 $G$ \citep{gaia2016,gaia2018,arenou2018,evans2018,hambly2018,riello2018, gaia2021,fabricius2021,lindegren2021a,lindegren2021b,torra2021}, J, H, and Ks bands from the Two Micron All Sky Survey (2MASS) All-Sky Point Source Catalog \citep[PSC,][]{skrutskie2006}, and W1 and W2 bands from the Wide-field Infrared Survey Explorer (WISE) AllWISE mid-infrared data \citep{wright2010,mainzer2011}, SkyMapper DR4  \textit{u, v, g, r, i, and z} magnitudes \citep{Onken:2024PASA...41...61O}, Sloan Digital Sky Survey (SDSS) DR16 \citep{SDSS16:2020ApJS..249....3A}, and the Panoramic Survey Telescope \& Rapid Response System (Pan-STARRS) DR2 \citep{PS:2016arXiv161205560C, PS:2020ApJS..251....7F}. The included data are based on availability and on data flag-based cuts, as detailed in \cite{Nataf:2024ApJ...976...87N}. We also include the \emph{Gaia} DR3-based \textit{geophotometric} distances from \citep{bailer-jones2021} of our targets in our priors.  Finally, we include extinction $A_V$ inferences based on either \textit{Bayestar} extinction \citep{Green:2019ApJ...887...93G} or, when not available, we use the SFD maps \citep{Schlegel:1998ApJ...500..525S}. In both cases, we employed the {\tt dustmaps\footnote{\url{https://dustmaps.readthedocs.io/en/latest/index.html}}} code to interpolate the extinction of our stars \citep{Green:2018JOSS....3..695M}.

We use the \texttt{isochrones} package\footnote{\url{https://github.com/timothydmorton/isochrones}} \citep{morton2015} to fit the MESA Isochrones and Stellar Tracks \cite[MIST;][]{dotter2016,choi2016,paxton2011,paxton2013,paxton2015,paxton2018,paxton2019} library to our photospheric stellar parameters as well as our input multiwavelength photometry, parallax, and extinction data using \texttt{MultiNest}\footnote{\url{https://ccpforge.cse.rl.ac.uk/gf/project/multinest/}} \citep{feroz2008,feroz2009,feroz2019} via \texttt{PyMultinest} \citep{buchner2014}.

Our adopted stellar parameters (Table \ref{tab:sp}) are used both in the likelihood and as part of the priors of the isochrone analysis, when appropriate. We highlight that the quoted uncertainties for the stellar masses, radii, and ages only include random uncertainties, i.e., derived under the unlikely assumption that the MIST isochrone grid we use in our analyses perfectly reproduces all stellar properties.

\section{{Elemental Abundances}}\label{sec:elabuncances}
{To identify thick disc stars, we measured the $\alpha-$elements \ion{Mg}{1}, \ion{Si}{1}, \ion{Ca}{1}, \ion{Ti}{1}, and \ion{Ti}{2}, which are the primary chemical tracers used to separate the thin and thick disc populations (Section~\ref{sec:pop}). Abundances were determined from equivalent width measurements following the same methodology as in our previous works \citep{Winnick:2025arXiv250816513W, Yana:2025ApJ...983...70Y, Reggiani:2024AJ....167...45R}. For this task, we used the Qoyllur-quipu ($\mathrm{q}^{2}$) Python code \citep{Ramirez:2014A&A...572A..48R}, configured to use the Kurucz ODFNEW model atmospheres and the 2019 version of \textsc{moog}, the same grid adopted for the stellar parameter determination with \textsc{XIRU}, ensuring internal consistency.}

{The adopted line lists are from \citet[][Table 1]{Ramirez:2009A&A...508L..17R}, \citet[][Table 1]{Melendez:2014ApJ...791...14M}, and \citet[][Table 3]{Ramirez:2015ApJ...808...13R}, which provide atomic data including excitation potentials, statistical weights, and oscillator strengths for each transition. Solar reference abundances were derived directly from our solar spectra obtained with MIKE and HARPS, with \citet{asplund2021} adopted for the remaining instruments. For stars observed with multiple instruments, we find no significant systematic offsets in the derived $\alpha$-element abundances, with differences always within 0.02 dex.}

\section{Galactic Population Membership}
\label{sec:pop}
The Milky Way is composed of distinct stellar populations forming the thin disc, the thick disc, and the halo, with each population differing significantly in both kinematics and chemical composition. {Historically, the thin disc is defined as the Galaxy’s innermost plane, where $0.0 \leq |Z| \leq 0.5$ (kpc), hosting predominantly iron-rich young stars (on average $\sim$6 Gyr), with a peak of the metallicity distribution function (MDF) at  [Fe/H] $= -0.06$ dex \citep{Casagrande:2011A&A...530A.138C}, $\alpha$-poor ([$\alpha$/Fe] $\leq$ 0.1 dex), and characterised by low total space velocities stars ($v_{\rm{tot}} \leq 50$ km s$^{-1}$) \citep{Nissen:2004oee..symp..154N, Sharma:2019MNRAS.490.5335S}. The thick disc occupies regions above and below the Galactic plane and is composed primarily of older stars ($\gtrsim$8 Gyr). The peak of the thick disc MDF is between $-0.7 \leq [\rm{Fe/H]} \ \leq -0.5$ dex \citep{Bensby:2004A&A...421..969B}, $\alpha-$enhanced ([$\alpha$/Fe] $\geq 0.2$ dex), and exhibits higher total space velocities compared to their thin disc counterparts ($70 \leq v_{\mathrm{tot}} \leq 200$ km s$^{-1}$). Although this is a somewhat simple description of these galactic components, it is still a good representation of them. For example, if we include additional variables, as height from the galactic plane (Z), or radial distribution, the MDF of the high and low-$\alpha$ populations have peaks at [Fe/H]$\sim-0.1$ and [Fe/H]$\sim -0.5$ for stars at the solar radius (and different heights), with peak ages of $\tau\sim6$ Gyrs and $\tau\sim10$ Gyrs \citep{Imig:2023ApJ...954..124I}, corroborating the description presented above.} The halo represents the outermost component of the Milky Way and is composed of stars that are even more metal-poor and have significantly higher space velocities than their thick and thin disc counterparts.

{Traditionally, these populations were classified either exclusively from kinematics or exclusively by chemistry, from direct observations of small samples of stars. However, large surveys made it possible to combine the two classification methods for a more robust determination of thin/thick discs membership. While Hipparcos and \textit{Gaia} have provided astrometric data for over a billion stars, spectroscopic surveys like GALAH and APOGEE \citep{2025PASA...42...51B, 2017AJ....154...94M, 2026ApJS..285....9S} have provided large scale abundances of $\alpha-$elements (Mg, Si, Ca, and Ti - although Ti is not an actual $\alpha$-element, it is used as a tracer as it behaves similarly on a Galactic scale) for millions of stars. This allowed us to combine [$\alpha$/Fe]\footnote{[$\alpha$/Fe] = ([Mg/Fe] + [Si/Fe] + [Ca/Fe] + [Ti/Fe])/4} chemical abundances with kinematic information to provide robust classifications even for thick disc stars that have thin disc-like kinematics \citep{Kovalev:2019A&A...628A..54K, Chen:2021ApJ...909..115C}}

\begin{figure*}
    \centering
    \includegraphics[width = 0.49\linewidth]{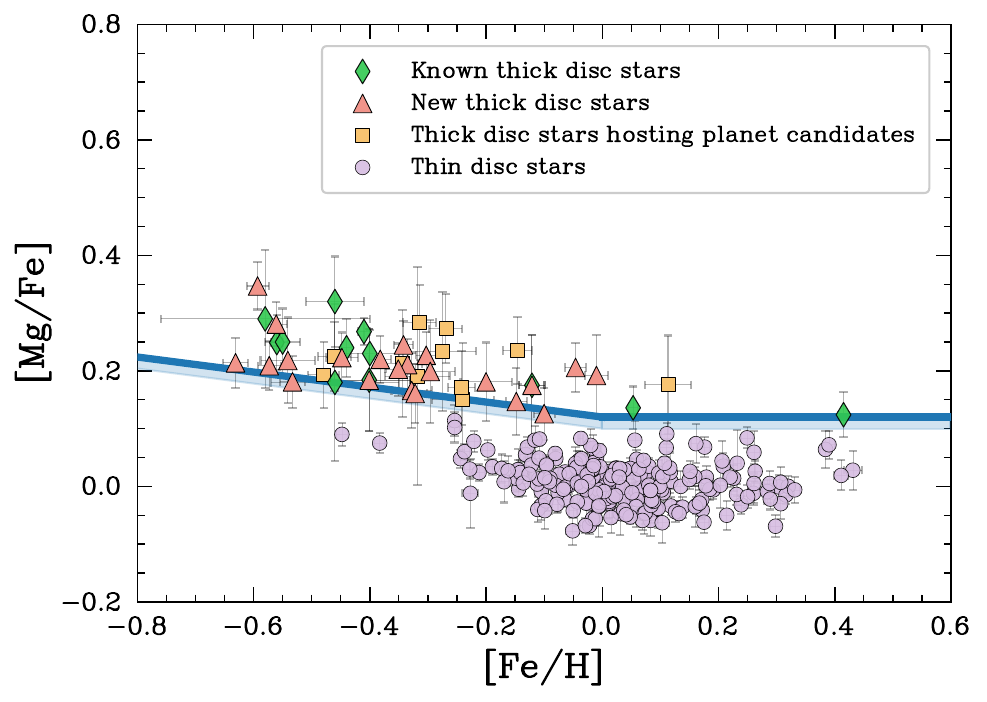} 
    \includegraphics[width = 0.49\linewidth]{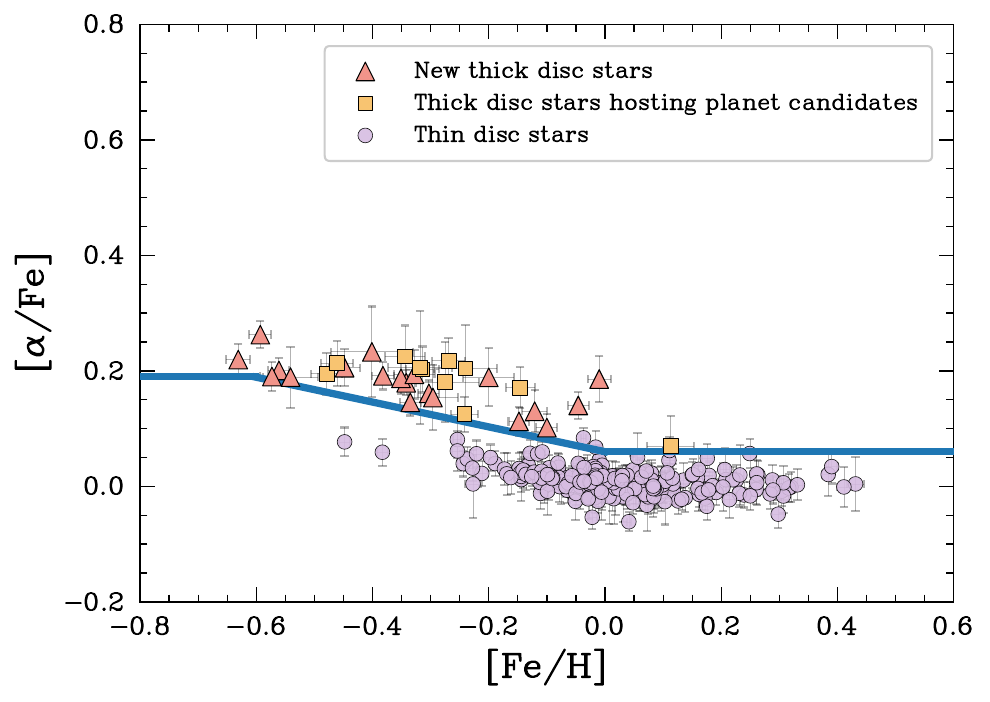}
    \includegraphics[width = \linewidth]{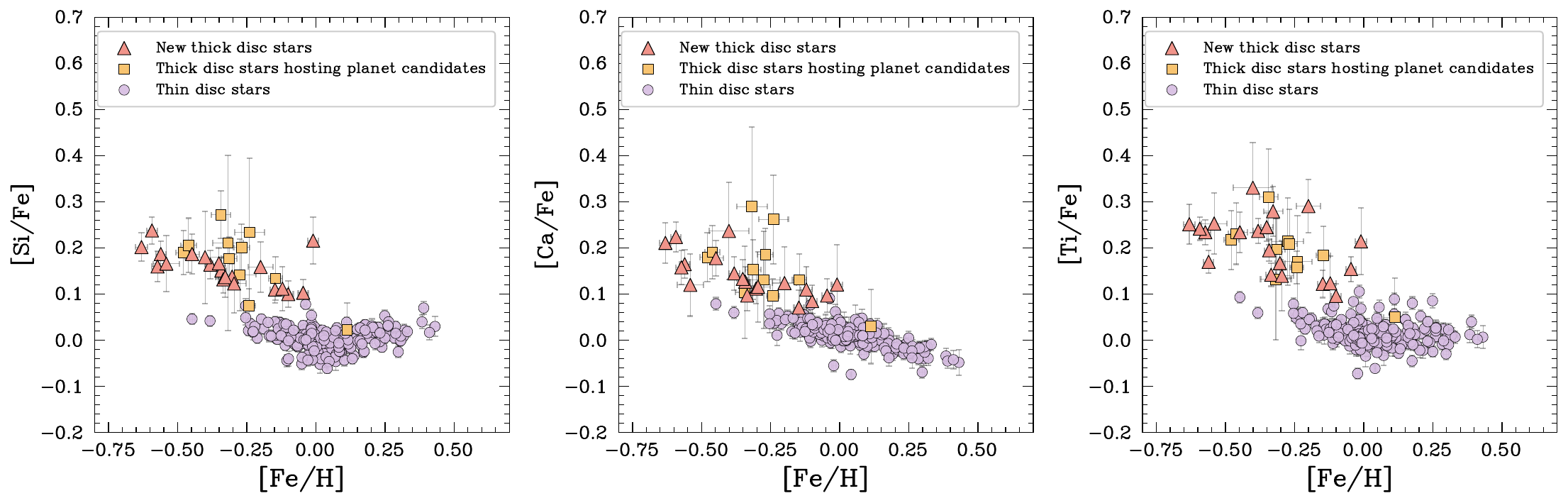}
    \caption{{\it Upper left panel}: [Mg/Fe] as a function of [Fe/H] for our confirmed thick disc stars (triangles {and squares}). Thin disc stars from our Inti sample \citep{2021MNRAS.504.1873Y} are shown as circles, while the known thick disc stars are represented by diamonds. The blue line marks the thin/thick disc boundary defined by \citet{Griffith:2019ApJ...886...84G}, {while the blue shaded region shows the $+0.02$ dex offset between our measurements and those from GALAH DR4 \citep{2025PASA...42...51B}}. {\it Upper right panel}: Same as the left panel, but with the blue line representing the thin/thick disc boundary defined by \citet{Adibekyan:2011A&A...535L..11A} as a function of [$\alpha$/Fe]. {\it Bottom panel}: {Individual $\alpha$-element abundances [Si/Fe], [Ca/Fe], and [Ti/Fe] as a function of [Fe/H] for our new thick disc stars (triangles and squares) and thin disc comparison sample (circles). All elements consistently show the expected $\alpha-$enhancement for thick disc stars across the full metallicity range of our sample, except K2-183, which appears enhanced in [Mg/Fe] but not in the other $\alpha-$elements (see text).}}
    \label{fig:membership}
\end{figure*}

In this work, we use the high resolution and the high–to–moderate signal-to-noise ratio of our spectra to infer precise chemical abundances of $\alpha-$elements and confirm the population membership of our sample. To remain consistent with large spectroscopic surveys such as GALAH in classifying thick disc stars, we adopted magnesium (\ion{Mg}{1}) as the representative $\alpha$-element in our initial analysis, before later considering the full [$\alpha$/Fe] ratio. {The \ion{Mg}{1} abundance calculation is explained in Section \ref{sec:elabuncances}}. To classify the thick disc stars, we adopted the division defined by \citet[][see their Equation 1]{Griffith:2019ApJ...886...84G} with data from GALAH, which separates the high-$\alpha$ and low-$\alpha$ stellar populations using cuts in [Mg/Fe]. {Since a systematic offset could exist between our abundances and those from GALAH DR4 \citep{2025PASA...42...51B}, we performed a consistency test by determining stellar parameters and Mg abundances for 18 stars in common (see Figure \ref{app:abscomp}) between this work and our upcoming paper (Yana Galarza et al. 2026, in preparation). We found negligible mean offsets of $-0.01$ dex in \feh\ and $-25$ K in \teff. We detect a small offset of $+0.02$ dex in [Mg/Fe] relative to our measurements, which we applied to the adopted division (blue shaded region in the upper left panel of Figure \ref{fig:membership}). However, this shift is minor and comparable to the abundance uncertainties in this work. The offset in [Mg/Fe] is likely driven by the increased scatter observed at [Mg/Fe] $\gtrsim +0.1$ dex, the regime occupied by thick disc stars, where lower effective temperatures broaden the wings of the magnesium lines and molecular blends affect the pseudo-continuum placement. The agreement is considerably better for thin disc stars with [Mg/Fe] $\lesssim$ 0.1 dex, as seen in the middle panel of Figure \ref{app:abscomp}.}

To confirm the thick disc membership of our sample, we also used the [$\alpha$/Fe] ratio (see upper right panel of Figure \ref{fig:membership}). As with \ion{Mg}{1}, the abundances of \ion{Si}{1}, \ion{Ca}{1}, and both \ion{Ti}{1} and \ion{Ti}{2} were derived from EW measurements, and the separation used to classify thin and thick disc stars follows \citet{Adibekyan:2011A&A...535L..11A}. {As shown in the upper right panel of Figure \ref{fig:membership}, the [$\alpha$/Fe] enhancement of our thick disc stars separates cleanly from our thin disc comparison sample, and the \citet{Adibekyan:2011A&A...535L..11A} boundary naturally separates both populations when applied directly to our abundances, without requiring any offset correction.}

We established thick disc membership for 22 stars with confirmed and candidate exoplanets observed with MIKE and HARPS. Abundances were always derived from the highest-quality spectra available and supplemented with spectra from other instruments only when lines were missing or affected by artefacts. For the remaining 12 stars, we used publicly available spectra from the SOPHIE, HARPS, and TRES archives. {Our results using the divisions defined by \citet{Griffith:2019ApJ...886...84G} and \citet{Adibekyan:2011A&A...535L..11A}
are shown in the top panels of Figure \ref{fig:membership}, where the separation between our sample of stars with confirmed planets (triangles and diamonds), exoplanet candidates (squares),  and the thin disc population is clearly visible. Further support for the thick disc classification is provided by the individual $\alpha$-element ratios shown in the bottom panels of Figure \ref{fig:membership}, where [Si/Fe], [Ca/Fe], and [Ti/Fe] are all consistently enhanced in our thick disc stars relative to the thin disc comparison sample, reinforcing the thick disc membership assignment.} Table \ref{app:alpha} lists the [Mg/Fe] and [$\alpha$/Fe] ratios for our sample.

{In our sample of thick disc exoplanet candidates, we note that K2-183 appears enhanced in [Mg/Fe] but not in the other individual $\alpha-$elements, and its combined [$\alpha$/Fe] ratio places it in the transition region between the two populations. Given the moderate S/N ($\sim100$) and resolution ($48,000$) of the available spectrum, which may limit the precision of the other $\alpha$-element measurements, we retain K2-183 as a thick disc member. Higher quality spectra will be needed to confirm this classification definitively.}

For comparison between the two populations, we also used MIKE- and HARPS-based abundances from our Inti sample \citep{2021MNRAS.504.1873Y} of thin disc stars (purple circles in Figure \ref{fig:membership}), which will be presented in Paper II (Yana Galarza et al., in preparation). These abundances were measured in the same way as for the thick disc stars reported here, so no systematic differences are expected {between the two comparison samples. The complete set of chemical abundances for 22 elements for the stars analysed in this work, including the individual $\alpha$-element ratios, will also be presented in Paper II.}

\subsection{Thick Disc Stars in the Kepler Survey}\label{sub:keplerstars}
The Kepler and K2 surveys are rich not only in photometric data but also because many spectroscopic surveys (e.g., GALAH) have targeted these stars to improve their stellar parameters and, consequently, the properties of their exoplanets. Among available datasets, the KOA and ExoFOP offer the largest publicly accessible HIRES spectra for Kepler and K2 stars, enabling precise characterisation of planets and their host stars. \citet{Loaiza:2024ApJ...970...53L, 2025ApJS..281...61L} used ExoFOP archival data to homogeneously determine stellar parameters for 109 K2 stars, and \citet{Loaiza:2024ApJ...970...53L} later extended this analysis to 301 K2 stars using both HIRES and TRES spectra. By measuring magnesium abundances and performing a kinematic analysis, they identified 62 thick disc stars (see their Figures 5 and 6), demonstrating that exoplanets around thick disc stars may be common. {Their sample, however, mostly includes transiting exoplanets, with only one planet (K2-262 $b$) having mass inferred from radial velocities.}

Our sample of thick disc candidates includes seven Kepler stars and seven K2 stars, seven overlapping with \citet{Loaiza:2024ApJ...970...53L}. Two stars of our sample, K2-262 and K2-408, host confirmed exoplanets, while the rest are candidates identified via transits. Here, we focus on the exoplanet candidates, deriving stellar parameters, age, mass, radius (Table \ref{tab:sp}), and $\alpha$-element abundances (Table \ref{app:alpha}) from MIKE, HIRES, SOPHIE, and TRES spectra. We confirmed thick disc membership for these stars (see Figure \ref{fig:membership}), which, together with the \citet{Loaiza:2024ApJ...970...53L} sample, represent bona fide thick disc hosts suitable for radial velocity studies to investigate planet formation in metal-poor environments. 

\section{Joint Keplerian Analysis}\label{sec:rv_transit_analysis}

We derived the orbital and physical properties of confirmed and candidate planets through a homogeneous and simultaneous re-analysis of the available transit photometric and, when available, radial velocity datasets using {\sc pyaneti}\footnote{\url{https://github.com/oscaribv/pyaneti}} \citep{2022MNRAS.509..866B}. {Where previously orbital and physical solutions exist, we compare the results of our re-analysis against parameters retrieved from the NASA Exoplanet Archive Planetary Systems catalogue \citep{ps}, accessed in October--November 2025}. For each system (see Appendix \ref{app:lcs}), all publicly available transit light curves were retrieved from the {\sc exo.MAST}\footnote{\url{https://exo.mast.stsci.edu}} archive, and come from space-based missions such as Kepler, K2, and TESS, which were processed by the Science Processing Operations Centre (SPOC; \citealt{2016SPIE.9913E..3EJ}) standard aperture pipeline. These photometric data were combined, when available, with archival radial velocity (RV) time series from high-resolution spectrographs retrieved from either the KOA, the ESO Science Archive, or the system's discovery papers. Only epochs exceeding conservative signal-to-noise thresholds (${\rm S/N}_{\rm RV} > 20$, ${\rm S/N}_{\rm phot} > 20$) were retained to ensure high-quality data for the re-analysis, with all time-series subsequently standardised to barycentric Julian dates and corrected for secular acceleration as well as instrument-specific zero-point offsets.

For each orbiting companion, the set of fitted parameters depends on the available observables, with all free parameters sampled using uniform priors. For radial velocity-only systems, we fitted the orbital period $P$, time of periastron passage $T_{\rm p}$ (or equivalently, the mean anomaly at a reference epoch), velocity semi-amplitude $K$, and the eccentricity vector components $\sqrt{e}\sin\omega$ and $\sqrt{e}\cos\omega$, which jointly define the orbital eccentricity $e$ and argument of periastron $\omega$. For transiting systems, additional parameters were inferred, including the epoch of mid-transit $T_0$, impact parameter $b$, planet-to-star radius ratio $R_{\rm p}/R_\star$, scaled semi-major axis $a/R_\star$, orbital inclination $i$, and equilibrium temperature at zero albedo $T_{\rm eq}$. In cases where both radial velocity and transit datasets were available, all parameters were jointly fitted, allowing the photometric and spectroscopic data to inform each other through shared orbital elements and yielding self-consistent masses, radii, and orbital geometries. For each radial velocity dataset, we additionally included an additive, instrument-based, white-noise term $\sigma_{\rm jit}$ to account for scatter not captured by formal uncertainties, being simultaneously fitted with uniform priors. Quadratic limb-darkening coefficients for transiting systems were included following the prescriptions of \citet{2013MNRAS.435.2152K} and \citet{2016MNRAS.457.3573E}. Moreover, stellar parameters (mass $M_\star$, radius $R_\star$, and effective temperature $T_{\rm eff}$) were fixed to values derived in Section~\ref{sec:fundamentalparams}. 

Sampling was performed with an ensemble of $100-700$ independent chains, with the number of chains scaling according to the model complexity, the number of free parameters, and the number of orbiting companions, with simpler single-planet models requiring fewer chains ($\sim100$), whereas multi-planet systems with correlated noise demanded larger ensembles ($\sim700$). Each chain was evolved for $\mathcal{O}(10^4)$ iterations after burn-in, and convergence was verified via the Gelman-Rubin statistic, requiring $\hat{R} < 1.01$ for all parameters. Posterior distributions were subsequently marginalised to derive credible intervals for all fitted and derived quantities (see \citealt{2019MNRAS.482.1017B, 2022MNRAS.509..866B}).

Such a homogeneous re-analysis tries to mitigate the inconsistencies introduced by different fitting strategies in the literature, yielding parameters and uncertainties that are directly comparable across the sample {(see, e.g., \citealt{2010MNRAS.408.1689S, 2012ApJ...757..161T, 2021A&A...656A..53S, 2022A&A...663A.161M})}. The resulting catalogue therefore offers a statistically coherent reference for population-level studies of planet formation and evolution in the chemically and dynamically ancient thick disc, and provides a quantitative basis for examining links between planetary architectures and the chemistry, kinematics, and ages of their host stars, as discussed in Section \ref{sec:conclusions}. Results are presented in Appendix \ref{app:lcs} and \ref{app:gaussian_process}. 

\begin{figure*}
    \centering
    \includegraphics[width = 0.49\linewidth]{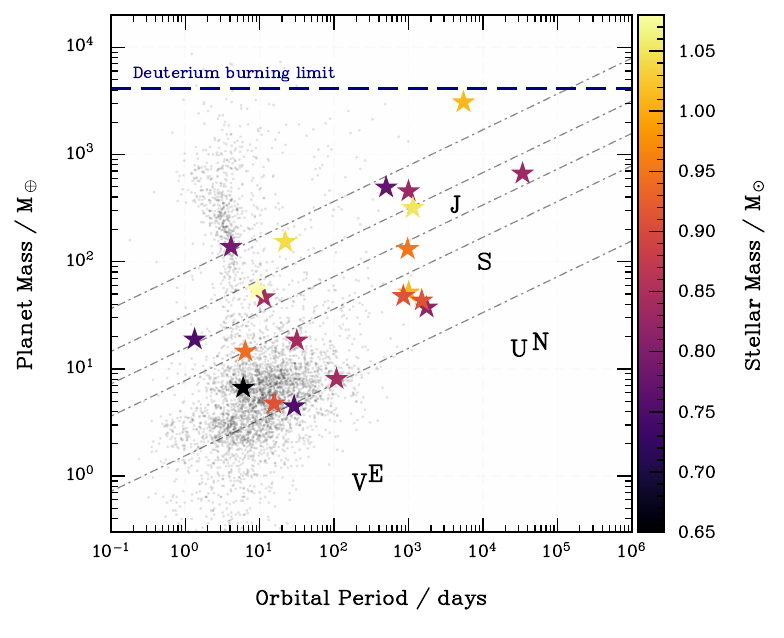}
    \includegraphics[width = 0.49\linewidth]{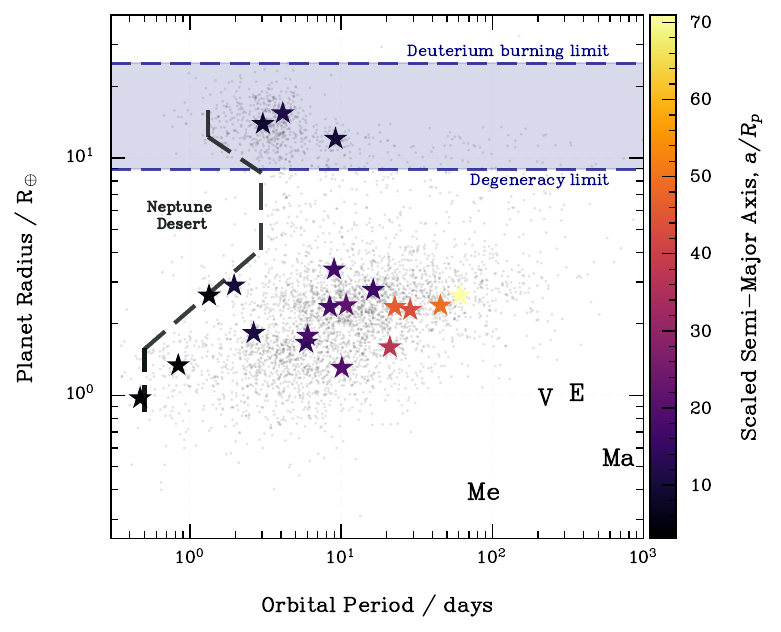}
    \caption{Period-Mass (left) and Period-Radius (right) for thick disc exoplanets. Grey points represent the full NASA Exoplanet Archive population. Thick disc planets are shown as star symbols, colour-coded by the stellar mass (left) and scaled semi-major axis (right). Solar System planets are labelled with their initial letters. The dashed blue line indicates the brown dwarf mass/radius limit ($13~M_{\rm Jup}$ / $25~R_{\oplus}$; \citealt{2011ApJ...727...57S, 2022MNRAS.511.3133H}), and the shaded degenerate zone highlights where massive planets and brown dwarfs are indistinguishable by radius alone, further requiring mass measurements for proper classification. In the left panel, dashed lines of constant semi-amplitude $K = 1,~5,~10,~20,~{\rm and}~50$ m s$^{-1}$ are indicated assuming $M_\star = 1~M_\odot$. In the right panel, the dashed black lines delineate the Neptune Desert boundaries following \cite{2024A&A...689A.250C}.}
    \label{fig:thickdiscmassradius}
\end{figure*}

Figure \ref{fig:thickdiscmassradius} shows the mass–radius–period distribution of the analysed thick disc planets, placed in context with those reported in the NASA Exoplanet Archive. {In the period-mass diagram (left panel), our thick disc planets span a diverse range of masses from sub-Neptunes ($\sim$10 M$_\oplus$) to super-Jovian scales ($>300~M_{\rm Jup}$), with orbital periods ranging from $\sim0.5$ to $\sim600$ days. Notably, the thick disc sample includes several massive planets with $M_p > 100~M_\oplus$ at intermediate periods ($10-100$ days), a region that is relatively sparsely populated in the general exoplanet population.} {The period-radius distribution (right panel) also shows that thick disc planets occupy diverse regions of parameter space, with no clear avoidance of any particular radius range. Interestingly, none of our thick disc planets falls within the Neptune desert boundaries, consistent with the general rarity of planets in this region. The thick disc sample includes both compact, high-density planets and more inflated worlds, suggesting diverse atmospheric and internal structure properties.} 

{When compared to the broader exoplanet population (grey points), thick disc planets do not appear to preferentially occupy any specific region of period-mass-radius space, indicating that the fundamental planet formation processes operating in thick disc environments may produce outcomes broadly similar to those in the thin disc. However, the presence of several massive, short-period planets ($M_p > 100~M_\oplus$, $P<10$ days) may reflect the different stellar encounter rates and dynamical histories characteristic of thick disc stellar populations, potentially facilitating high-eccentricity migration pathways that bring massive planets to close-in orbits.} {Furthermore, the color-coding by stellar mass (left) and scaled semi-major axis (right) reveals that our more massive planets tend to orbit more massive stars, consistent with the known correlation between stellar and planetary mass, and the scaled semi-major axis distribution indicates that several planets experience significant stellar irradiation ($a/R_p < 10$), placing them in regimes where atmospheric escape and tidal effects may be important for their long-term evolution and is worthwhile investigation in future follow-ups.}

\subsection{Gaussian Process Regression}\label{sec:gaussian_process}

Radial velocity signals can be also modelled as the sum of Keplerian components superposed on Gaussian Processes (GPs) terms that accounts for correlated stellar activity and additional stochasticity, thereby providing a flexible, non-parametric framework to capture residual astrophysical signals that are not strictly periodic nor stationary, as even old, magnetically evolved stars may retain low-amplitude yet temporally correlated surface variability arising from long-lived active regions, rotational modulation, or granulation-driven processes, which cannot be adequately captured by white-noise models alone (see Appendix \ref{app:gaussian_process}). These motions at the stellar surface induce quasi-periodic radial velocity variations that are described as

\begin{equation}
    {\rm RV}(t_i) = A_0 \, \gamma_{\rm MQ-P}(t_i, t_j) + A_1 \, \frac{\mathrm{d}}{\mathrm{d}t} \gamma_{\rm MQ-P}(t_i, t_j), \label{eq:1}
\end{equation}  

where $\gamma_{\rm MQ-P}(t_i, t_j)$ is a Matérn quasi-periodic covariance function evaluated between epochs $t_i$ and $t_j$, defined by

\begin{eqnarray}
    \gamma_{\rm MQ-P}(t_i, t_j) &\equiv& A_0^2 \, k_{\mathrm{Mat\acute{e}rn}, \nu=1/2}(|t_i - t_j|; \lambda_e) \notag \\ 
    &\times& \exp\left[-\frac{\sin^2\left(\pi \frac{t_i - t_j}{P_{\rm GP}}\right)}{2 \lambda_p^2}\right],
\end{eqnarray}  

with the Matérn kernel of order $\nu=1/2$ (dubbed exponential kernel) given by  

\begin{equation}
    k_{\mathrm{Mat\acute{e}rn}, \nu=1/2}(\tau; \lambda_e) = \exp\left(-\frac{\tau}{\lambda_e}\right), \label{eq:3}
\end{equation}  

where $\tau = \vert t_i - t_j\vert$ is the time lag between observations \citep{2019dmde.book.....H, 2023ARA&A..61..329A}. The hyper-parameters $A_0$ and $A_1$ represent the amplitude of the activity-induced variations and the contributions from the derivative of the GP, respectively, allowing the model to capture correlations not only in the RV values but also in their temporal gradients, $P_{\rm GP}$ corresponds to the characteristic recurrence timescale of active regions, typically associated to the stellar rotation period \citep{2023ARA&A..61..329A}, while $\lambda_p$ and $\lambda_e$ control the smoothness of the periodic component and the evolutionary timescale of active regions, respectively \citep{2020MNRAS.495L..61L, 2021AJ....162..160N}. 

\begin{figure*}[t]
    \centering
    \includegraphics[width = 0.49\linewidth]{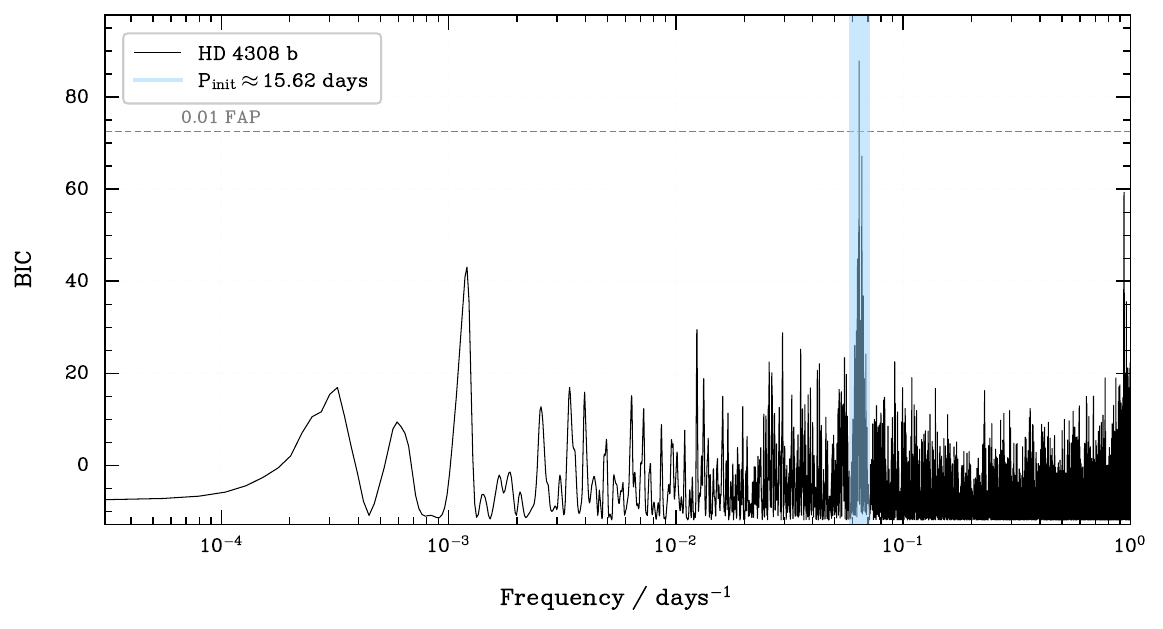}
    \includegraphics[width = 0.49\linewidth]{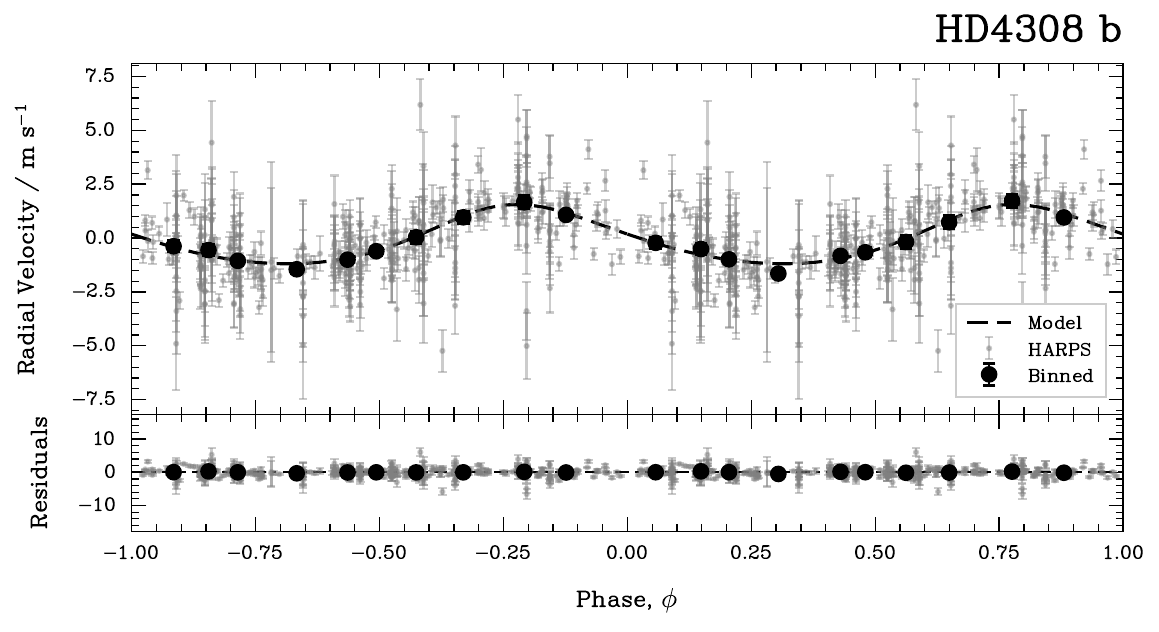}\\
    \includegraphics[width = 0.49\linewidth]{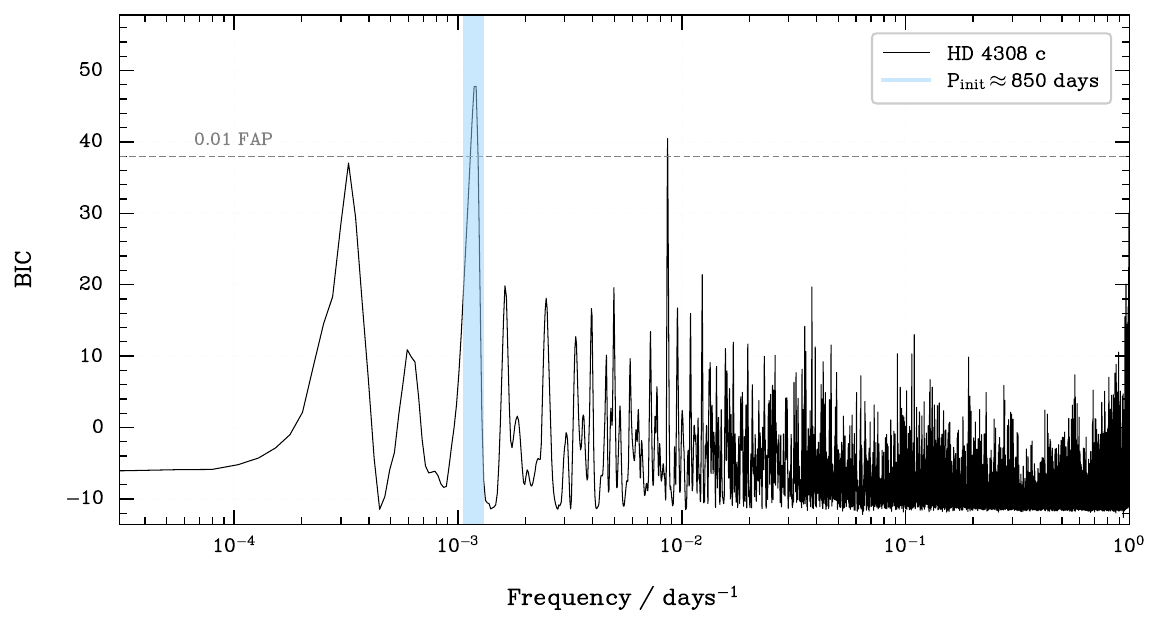}
    \includegraphics[width = 0.49\linewidth]{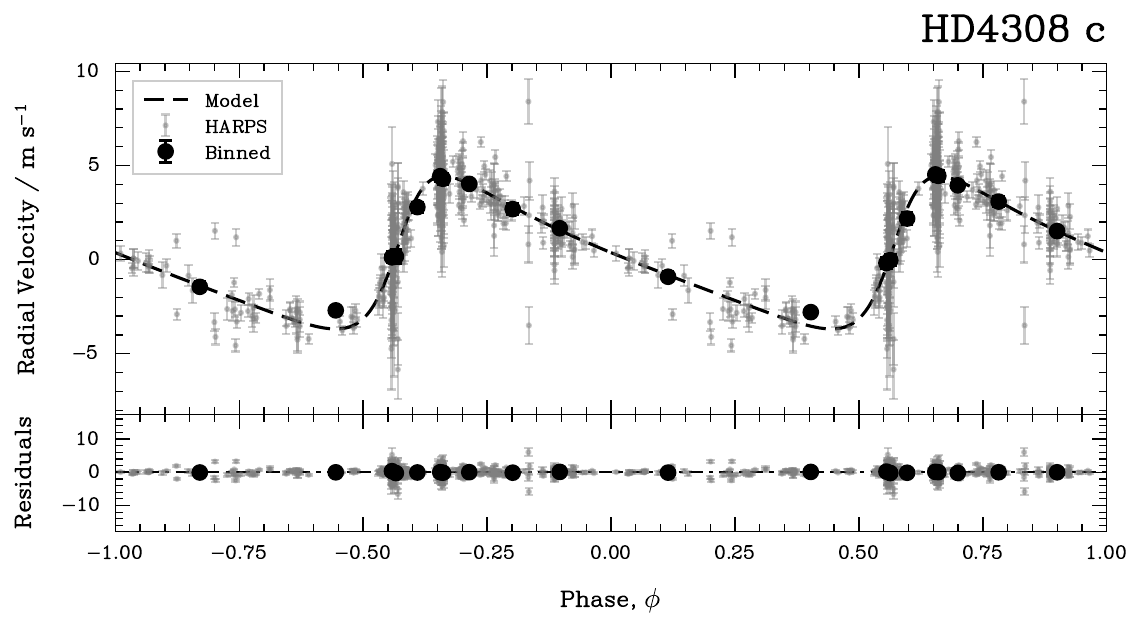}\\
    \includegraphics[width = 0.49\linewidth]{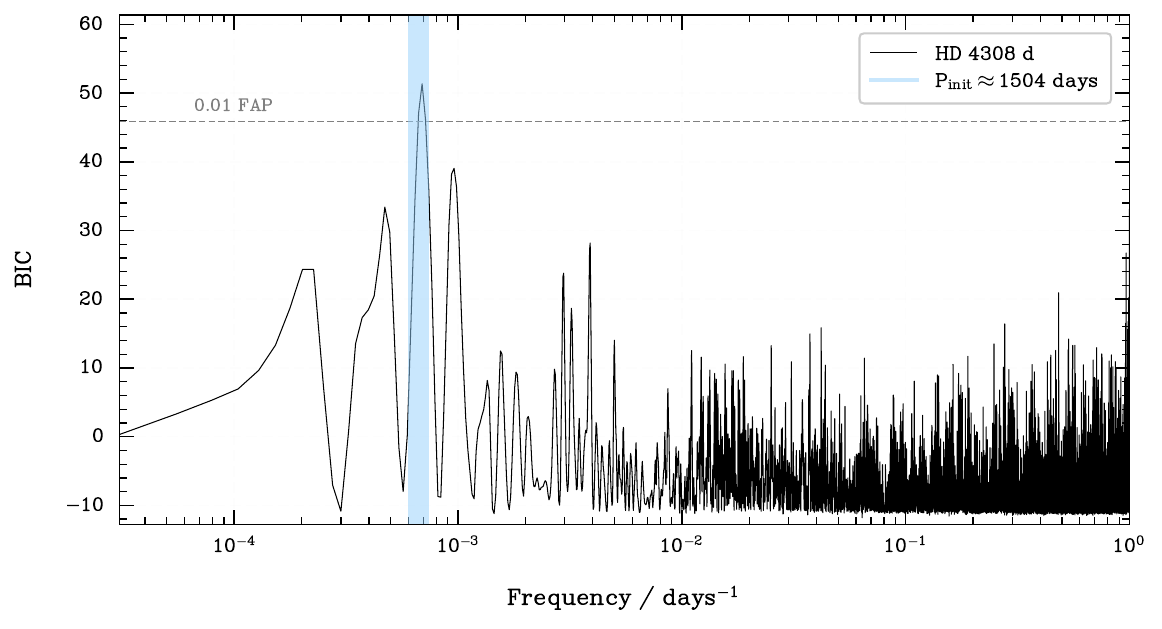}
    \includegraphics[width = 0.49\linewidth]{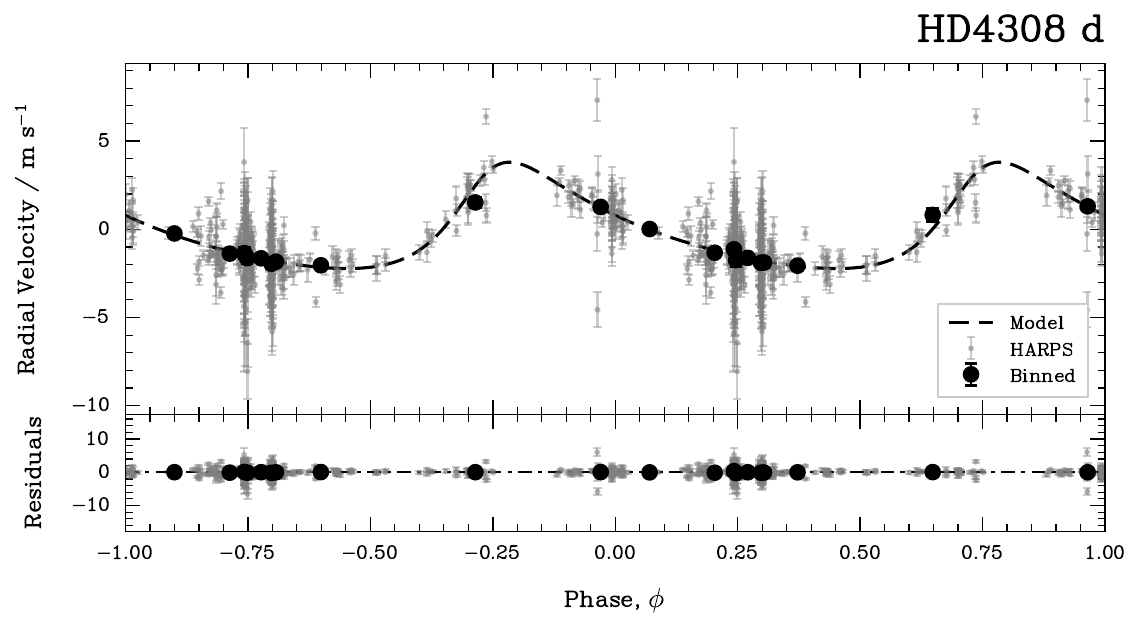} \\
    \includegraphics[width = \linewidth]{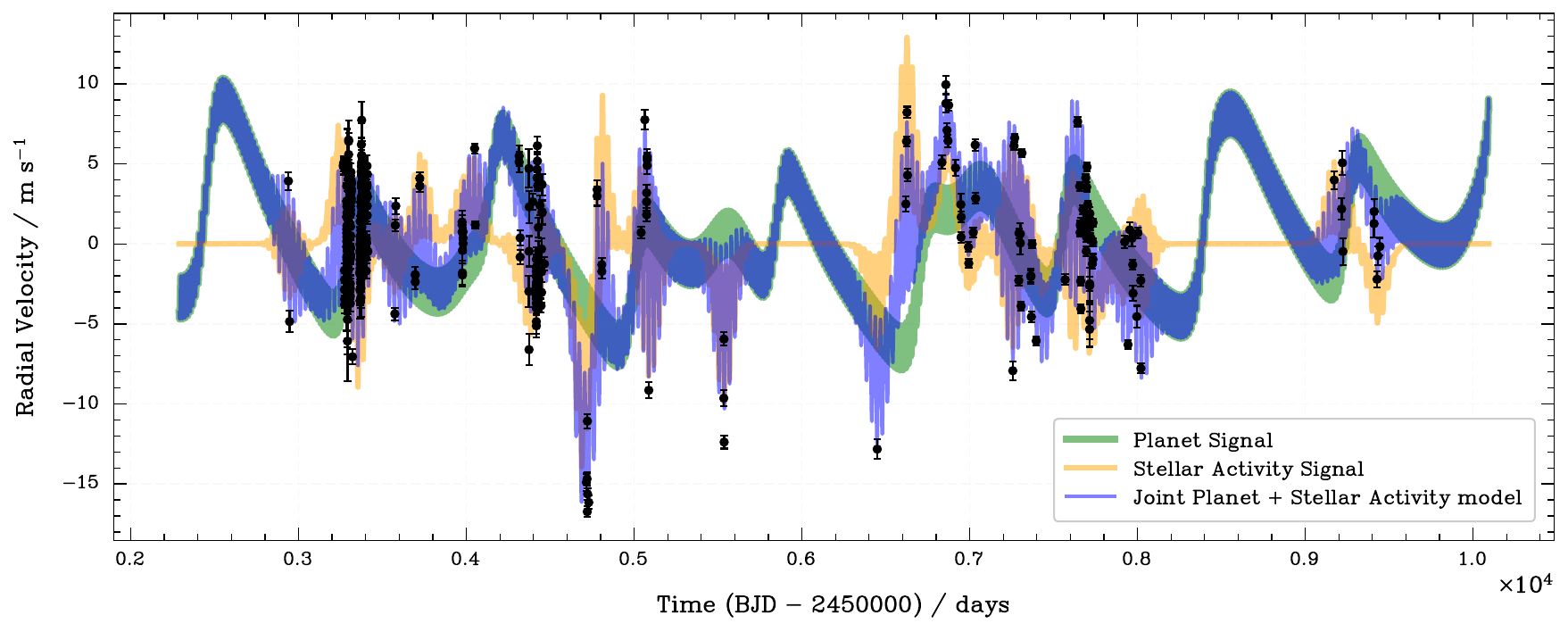}
    \caption{Periodograms and best-fit models over HARPS radial velocities for the three planets: HD 4308 $b$ (first panel), HD 4308 $c$ (second panel), and HD 4308 $d$. The binned data correspond to 20 uniformly spaced bins (also applied to all phase diagrams presented in this work). Residuals for each fit are displayed beneath the corresponding panel. In the last panel, the planet's signal is represented in green, stellar activity modulations in yellow, and the joint model in blue.}
    \label{fig:HD4308}
\end{figure*}

\subsection{New Exoplanet Hosts}\label{subsec:newplanets}

\subsubsection{HD 4308}

The HD 4308 system hosts three planets (Figure \ref{fig:HD4308}) discovered through radial velocity modelling spanning over $\sim17$ years. The innermost planet, HD 4308 $b$, orbits at $P \sim 15.62 \pm 0.49$ days on a moderately eccentric orbit with $e = 0.146 \pm 0.316$, has a minimum mass of $m\sin{i} = 4.75 \pm 4.79~M_\oplus$, and was first reported by \citet{2006A&A...447..361U} {from a shorter observational baseline; our homogeneous reanalysis of the full $\sim17$-year dataset yields a revised minimum mass and non-zero eccentricity for this planet, consistent with the additional constraints provided by the extended temporal coverage \citep{2009ApJ...701.1732O}.} The middle planet, HD 4308 $c$, has $P \sim 850.1 \pm 25.3$ days, $e = 0.467 \pm 0.335$, and $m\sin{i} = 47.8 \pm 21.5~M_\oplus$, while the outer planet, HD 4308 $d$, orbits at $P \sim 1504.0 \pm 55.6$ days with $e = 0.367 \pm 0.305$ and $m\sin{i} = 43.5 \pm 25.8~M_\oplus$. Both {HD 4308 $c$ and HD 4308 $d$} are newly discovered exoplanets, {undetected in prior analyses owing to the more limited baseline than available.} No additional signals indicative of further companions were detected in the residuals.

\subsubsection{HD 150433}

\begin{figure*}[h]
    \centering
    \includegraphics[width = \linewidth]{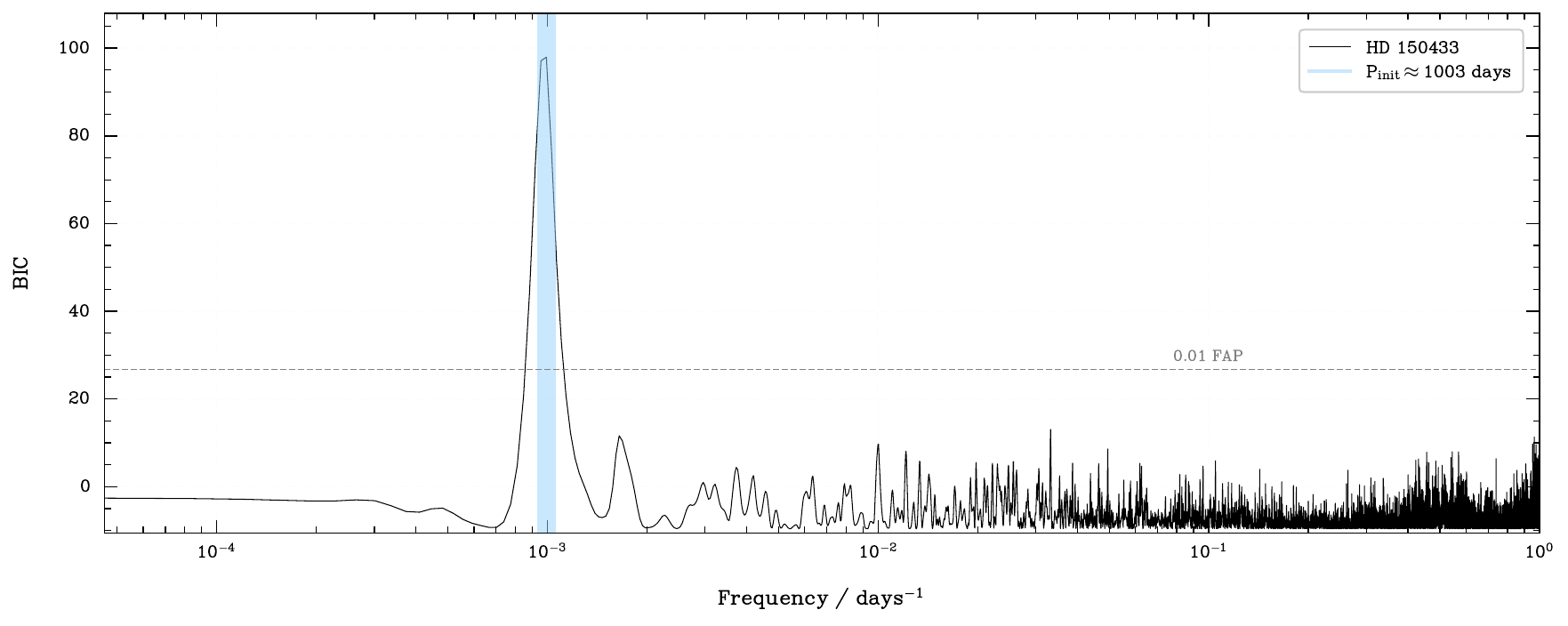} \\
    \includegraphics[width = \linewidth]{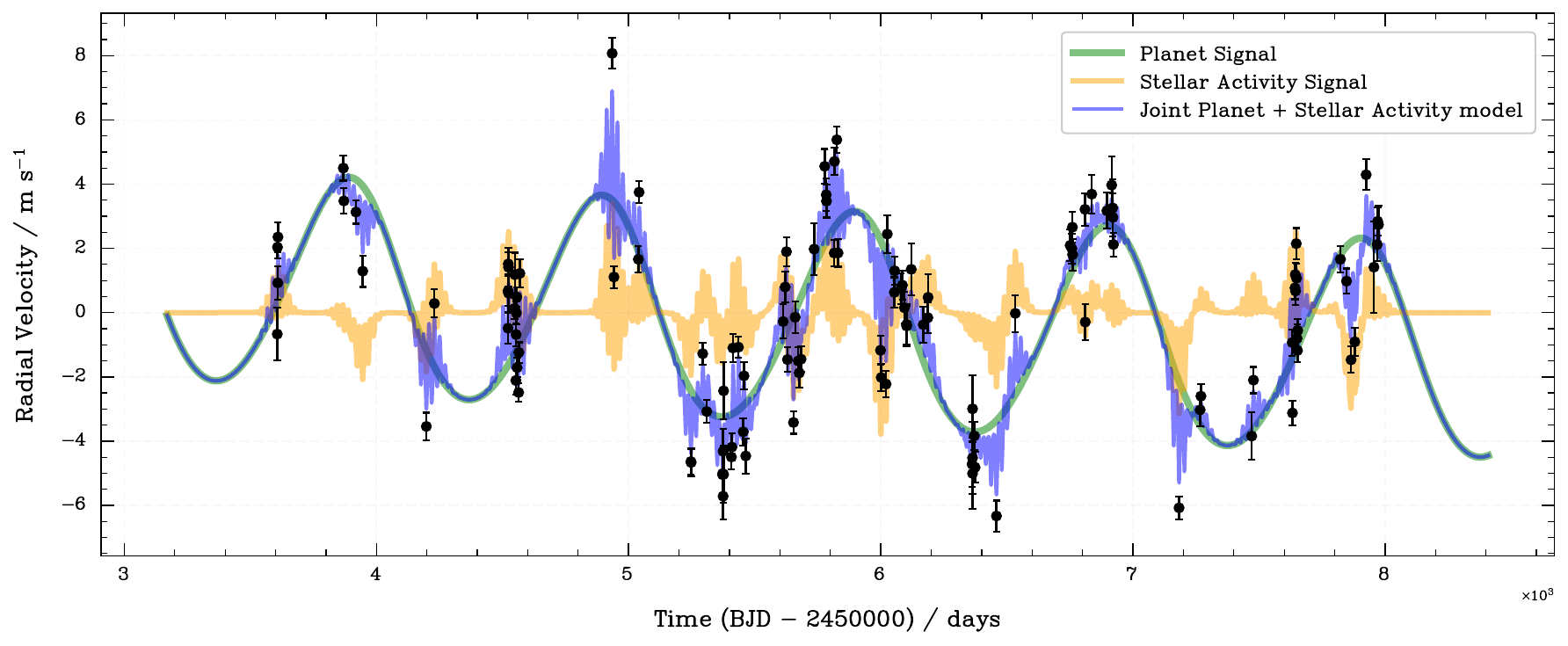} \\
    \includegraphics[width = \linewidth]{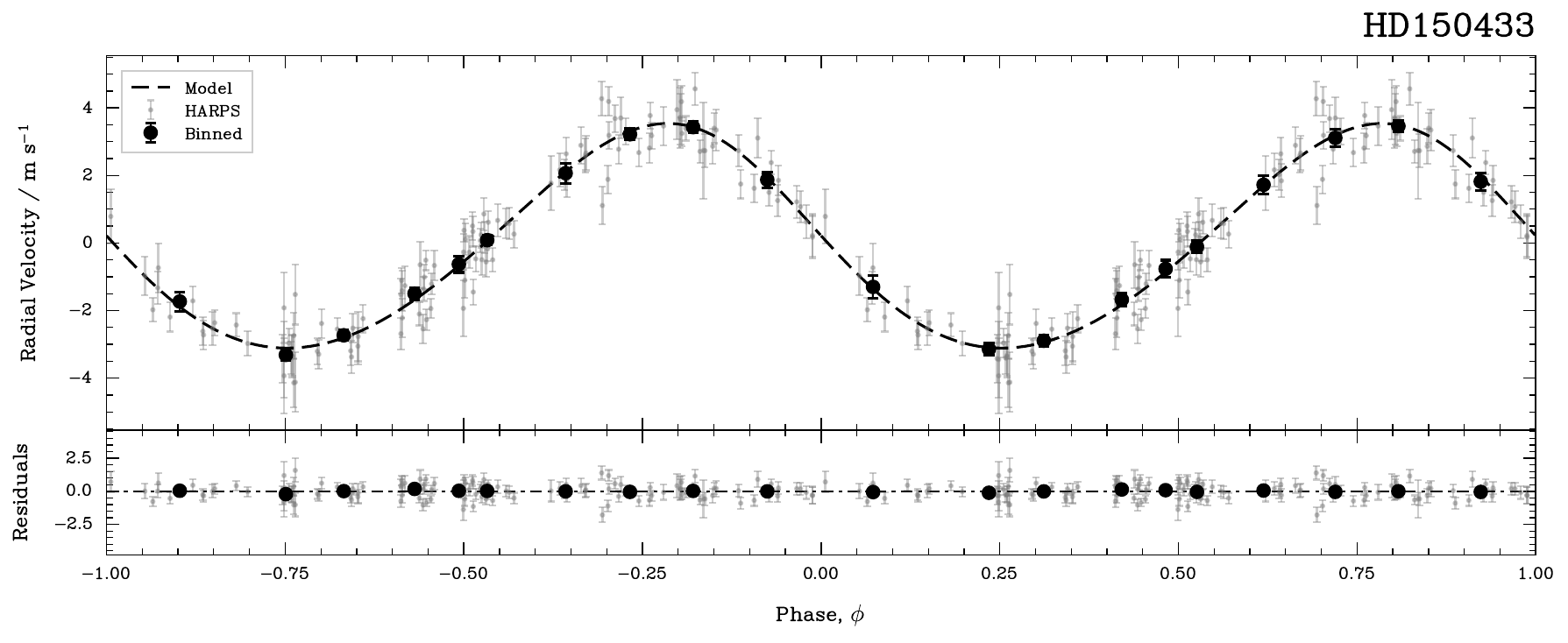}
    \caption{Periodogram (first panel) and best-fit model over HARPS radial velocities for HD 150433 $b$ (second and third panels). The dashed horizontal grey line in the first panel indicates 0.01 false-alarm probability levels, and the periodogram peak is indicated by the blue horizontal line. In the second panel, the planet's signal is represented in green, stellar activity modulations in yellow, and the joint model in blue. Residuals in the third panel are shown in the lower panel.}
    \label{fig:HD150433}
\end{figure*}

{HD 150433 $b$ is a newly discovered planet detected through radial velocity modelling of HARPS spectra} spanning $\sim$11 years, orbiting at $P \sim 1003 \pm 18$ days on a near-circular orbit ($e = 0.082 \pm 0.076$), and it has a minimum mass of $m \sin{i} = 0.16 \pm 0.02$ M$_{\rm Jup}$ (Figure \ref{fig:HD150433}).  No evidence for additional companions was found in the residuals, nor were there indications of barycentric motion attributable to long-period, unseen companions. \bigskip

\subsubsection{TOI-126}

\begin{figure}[h]
    \centering
    \includegraphics[width = \linewidth]{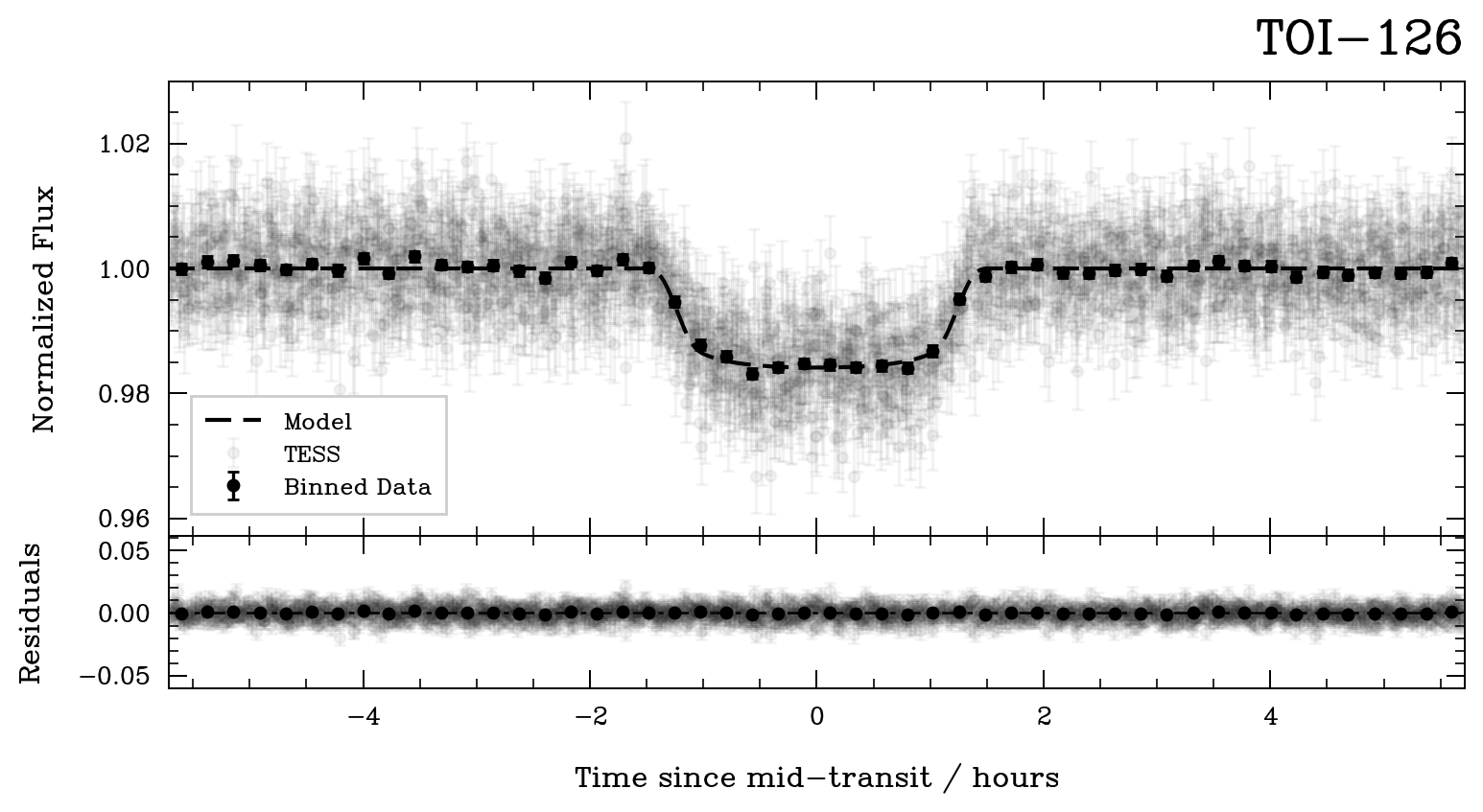}
    \caption{TESS transit light curve of TOI-126 $b$ with the best-fit transit model (black line). Residuals are displayed in the lower panel.}
    \label{fig:TOI126}
\end{figure}

TOI-126 b was detected using TESS transit photometry, orbiting an old, Sun-like thick disc star at a period of $P\sim3.0353\pm0.0002$ days, assuming a circular orbit (Figure \ref{fig:TOI126}). This planet was listed as a candidate in the NASA Exoplanet Archive (TOI-126.01; $R_p = 14.13~R_\oplus$) with no published mass or formal confirmation; our homogeneous reanalysis yields a refined planetary radius of $R_p = 13.92\pm0.04~R_\oplus$ and a planetary mass of $m{\rm sin}i = 0.88\pm0.12~M_{\rm Jup}$ from CORALIE radial velocities \citep{Anderson:2014arXiv1410.3449A, Queloz:2000A&A...354...99Q}, thereby confirming its planetary nature. The semi-major axis is $a = 0.0428\pm0.0039$ au, placing it in the hot Jupiter regime. The transit duration is $T_{\rm tot} = 2.858_{-0.063}^{+0.069}$ hours, and the orbital inclination was estimated at $i = 87.75\pm1.72^\circ$, corresponding to a scaled semi-major axis $a/R_\star = 8.69\pm0.67$. Furthermore, its estimated equilibrium temperature is $T_{\rm eq} \sim 1435$ K, receiving an insolation of $S_{\rm p}\sim2.86~S_\oplus$

\subsubsection{TOI-3277}

\begin{figure}[h]
    \centering
    \includegraphics[width = \linewidth]{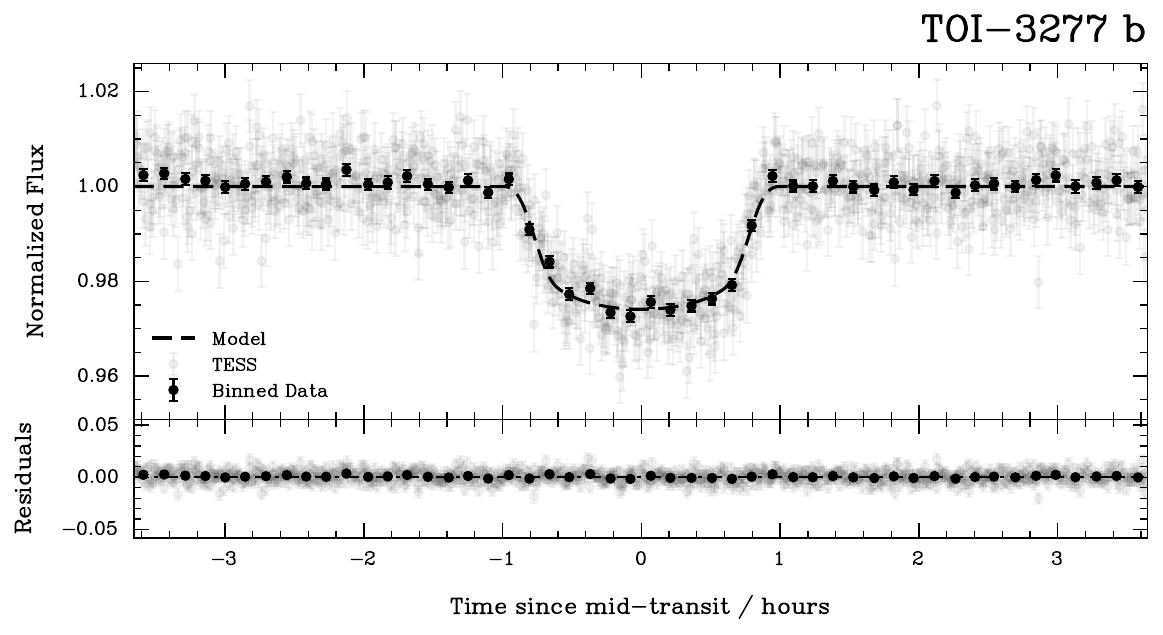}
    \includegraphics[width = \linewidth]{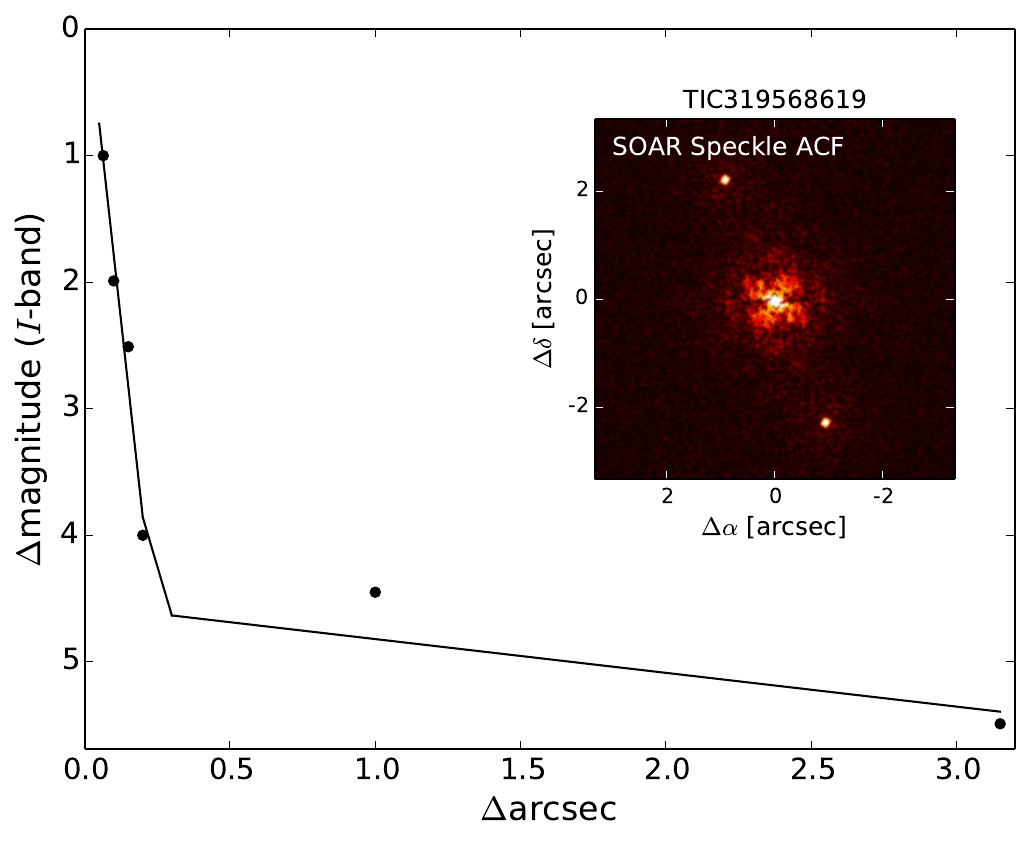}
    \caption{{\it First panel}: TESS transit light curve of TOI-3277 $b$ with the best-fit transit model (black line). Residuals are displayed in the lower panel. {\it Second panel}: I band speckle image and sensitivity curve for TOI-3277 from the SOAR telescope in the 880 nm band. The inset image is the primary target showing a companion (TOI-3277 B) at 2.2\arcsec.}
    \label{fig:TOI3277}
\end{figure}

TOI-3277 $b$ was discovered using TESS transit photometry, with follow-up radial velocity measurements obtained using PFS; a dedicated analysis of these observations will be presented in Y. Gaibor et al. (in preparation). The planet orbits a cool star at a period of $P \sim 1.52$ days on a nearly circular orbit ($e = 0.033\pm0.002$; see lower panel of Figure \ref{fig:TOI3277}). The transit depth corresponds to a planetary radius of {$R_p = 1.163\pm0.043$ R$_{\rm Jup}$}, while the radial velocity semi-amplitude of {$K = 357\pm8$ m/s yields a planetary mass of $M_p = 1.705\pm0.042$ M$_{\rm Jup}$}. The orbital inclination is $i = 87.22\pm1.81^\circ$, with a scaled semi-major axis of {$a/R_\star = 7.00\pm0.45$, and a semi-major axis of $a = 0.0262\pm0.0018$ au}. The resulting equilibrium temperature is {$T_{\rm eq} = 1327\pm49$ K, and the planet receives an insolation of $S \sim 517~S_\oplus$}. The transit duration is {$T_{\rm tot} = 1.869\pm0.043$ hours, and the planet has a bulk density of $\rho_p = 1.34_{-0.16}^{+0.14}$ g/cm$^3$.} 

Additionally, we find that TOI-3277 is a binary system whose companion, TOI-3277 B, lacks spectroscopic characterisation; the only available photometric ($G = 16.27$) and astrometric constraints are provided by \textit{Gaia}. This absence of spectroscopic information precludes a robust determination of the companion's physical properties, including its atmospheric parameters, chemical composition, and radial velocity, all of which are essential for assessing its nature and evolutionary state, as well as its dynamical relationship to the primary. High-angular-resolution observations are therefore required to spatially resolve the system and enable future dedicated spectroscopic follow-up of TOI-3277 B. As part of this effort, we searched for stellar companions to TOI-3277 using speckle imaging obtained with the 4.1-m Southern Astrophysical Research (SOAR) telescope \citep{Tokovinin:2018PASP..130c5002T} on 2021 July 14 UT, observing in the Cousins $I$ band, which closely matches the visible bandpass of TESS. These observations were sensitive to companions up to 4.5 magnitudes fainter than the primary at an angular separation of 1\arcsec. Further details of the observations within the SOAR TESS survey are provided in \citet{Ziegler:2020AJ....159...19Z}. The $5\sigma$ detection sensitivity and speckle auto-correlation functions are shown in the second panel of Figure~\ref{fig:TOI3277}. TOI-3277 B is clearly detected at a projected separation of 2.2\arcsec\ from the primary, confirming the binarity and establishing the feasibility of future spatially resolved spectroscopic observations.

\subsection{TOI-1927 b and TOI-2643 b: Two Puffy Planets in the Thick Disc}\label{sec:puffy}

\begin{figure*}
    \centering
    \includegraphics[width = \linewidth]{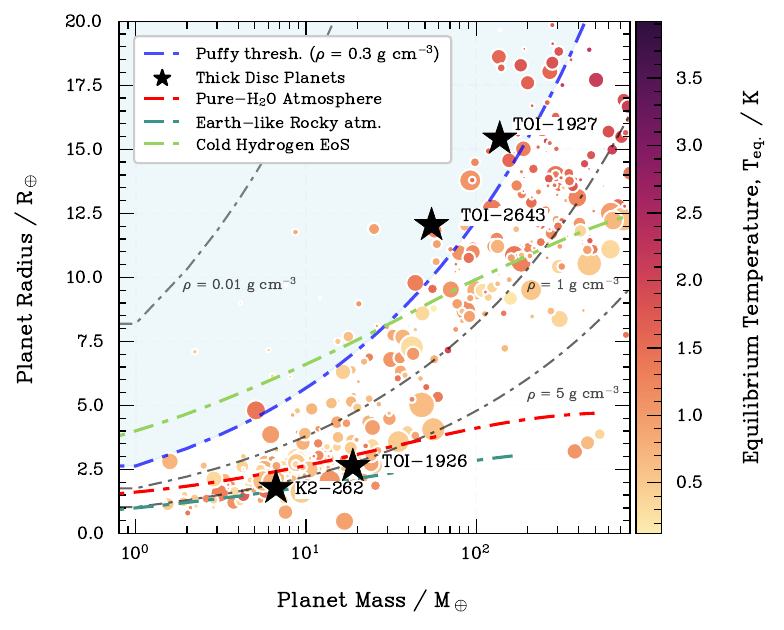}
    \caption{Mass-Radius diagram (black stars) shown in the context of the known exoplanet population (coloured circles by equilibrium temperature and sized by their orbital eccentricity). Overplotted are theoretical mass-radius relations from \cite{2019PNAS..116.9723Z} for different interior compositions: Cold Hydrogen Equation of State \citep{2014ApJS..215...21B}, Earth-like rocky composition, and pure-H$_2$O atmospheres. Dashed lines indicate constant density contours at $\rho$ = 0.01, 1, and 5 g cm$^{-3}$. The thick blue dashed line marks the {\it puffy planet} threshold at $\rho = 0.3$ g cm$^{-3}$, above which planets are considered inflated.}
    \label{fig:massradius}
\end{figure*}

\begin{figure}
    \centering
    \includegraphics[width = \linewidth]{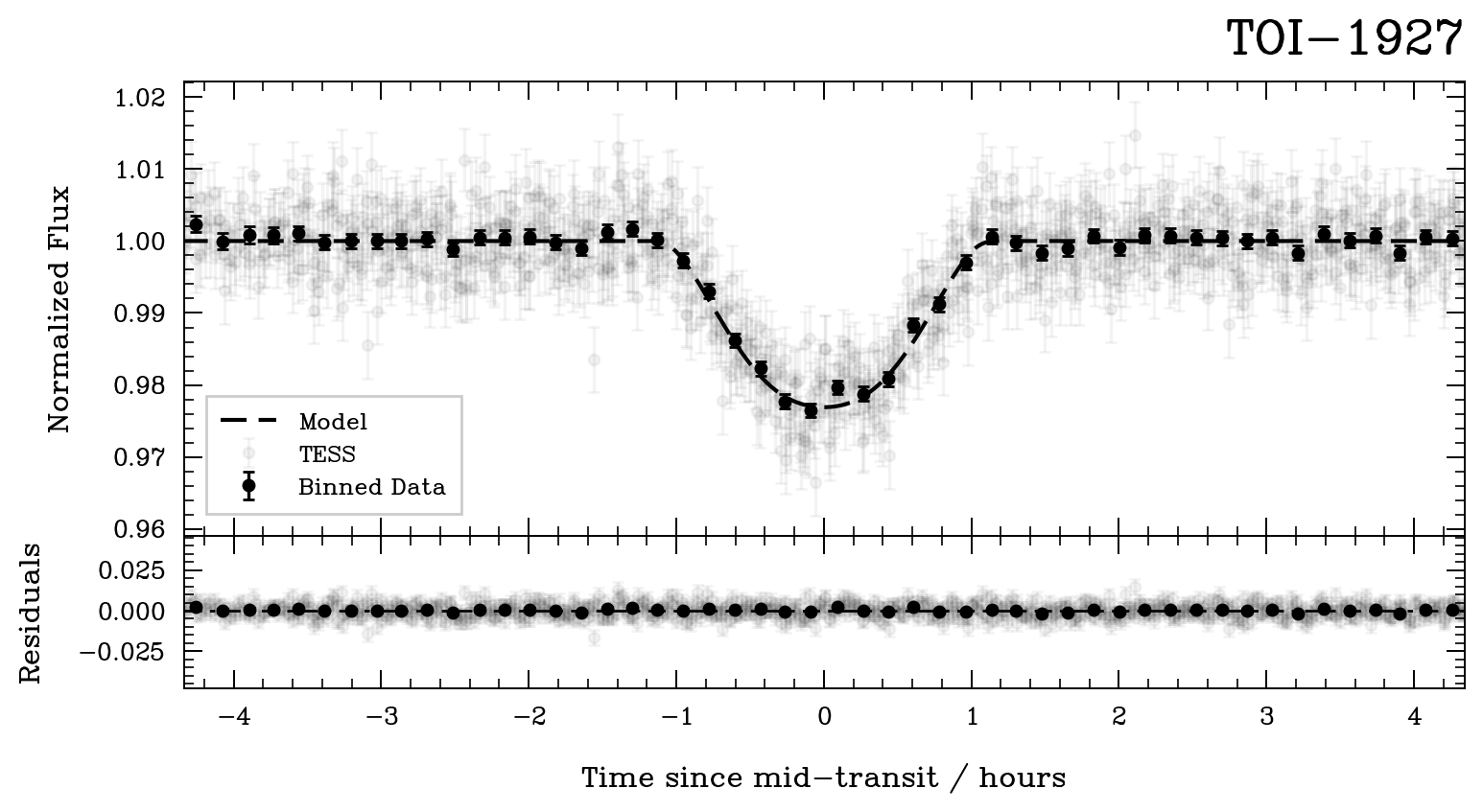}
    \includegraphics[width = \linewidth]{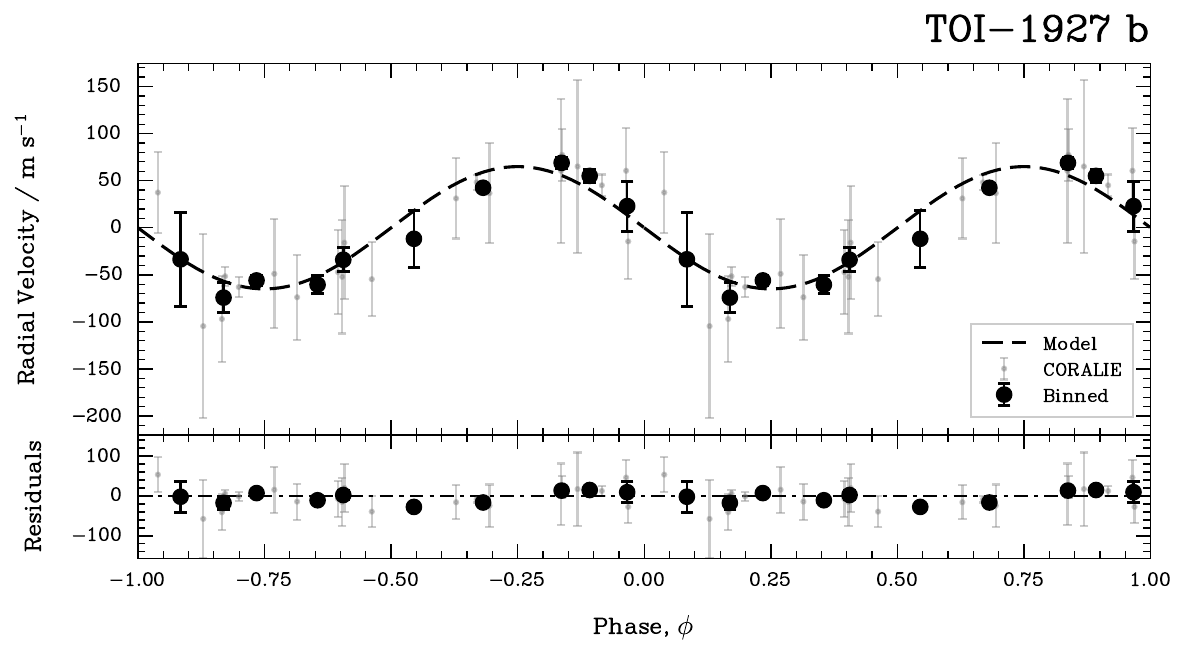} \\
    \includegraphics[width = \linewidth]{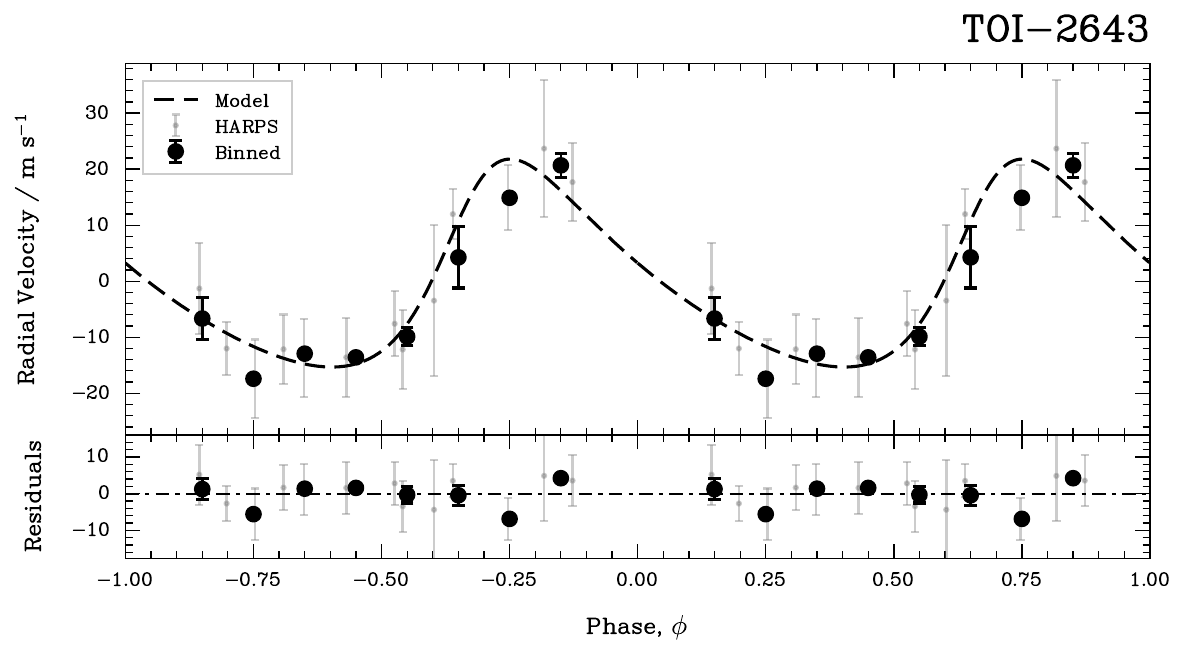}
    \includegraphics[width = \linewidth]{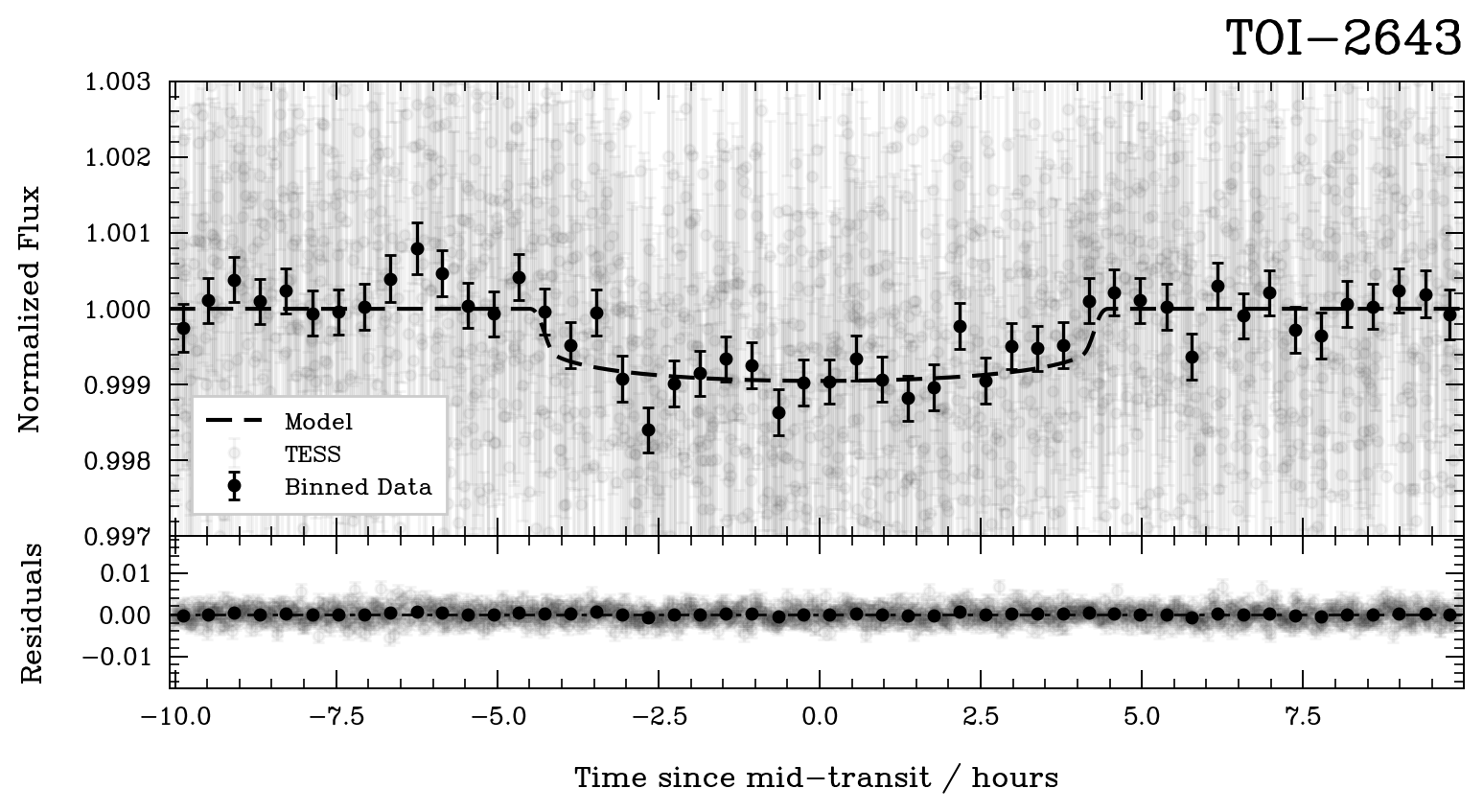}
    \caption{Best-fit models over CORALIE and HARPS radial velocities, respectively, plus TESS transit light curve for TOI-1927 b (first panel) and TOI-2643 b (second panel). Residuals for each fit are displayed beneath the corresponding panel.}
    \label{fig:puffy}
\end{figure}

Within our sample of thick disc planetary systems, we identified two planets with anomalously low bulk densities: {TOI-1927~$b$ (WASP-183~$b$)} with $P = 4.11$ days, $M_p = 0.43~M_{\rm Jup}$, $R_p = 1.37~R_{\rm Jup}$, and $\rho_p = 0.19~{\rm g~cm^{-3}}$, {consistent with the discovery solution of \citet{2019MNRAS.485.5790T} ($M_p = 0.502 \pm 0.047~M_{\rm Jup}$, $R_p = 1.47^{+0.94}_{-0.33}~R_{\rm Jup}$),} and {TOI-2643~$b$ (K2-132~$b$)} with $P = 9.17$ days, $M_p = 0.17^{+0.19}_{-0.13}~M_{\rm Jup}$, $R_p = 1.07 \pm 0.07~R_{\rm Jup}$, and $\rho_p = 0.17^{+0.20}_{-0.13}~{\rm g~cm^{-3}}${; the latter mass is uncertain owing to the limited RV semi-amplitude relative to the stellar activity level of this giant star host, modelled here via a GP, and is consistent within $2\sigma$ with the value reported by \citet{2017AJ....154..254G}.} Both planets occupy the extreme low-density tail of the giant-planet population ($\rho_{\rm p} \lesssim 0.3~{\rm g~cm^{-3}}$; Figure~\ref{fig:massradius}; see also \citealt{2017AJ....154..254G, 2019MNRAS.485.5790T}), placing them among the highly-inflated, or puffy, planets---{the first ones reported orbiting thick disc stars, a context neither \citet{2019MNRAS.485.5790T} nor \citet{2017AJ....154..254G} identified for these systems.} The Keplerian fits for both puffy planets are displayed in Figure~\ref{fig:puffy}. These objects are intrinsically uncommon in old, metal-poor environments, where long cooling timescales and reduced atmospheric opacities are generally expected to favour contraction rather than inflation \citep{2012Natur.486..375B, 2014ApJ...792....1L, 2016ApJ...831...64T, 2016ApJ...825...29G}. Their presence in our sample therefore adds two valuable data points to an otherwise sparsely populated region of parameter space.

Radius inflation at billion-year ages requires either sustained heating or inhibited cooling, particularly at high equilibrium temperatures ($T_{\rm eq} \gtrsim 1000$ K; for context $T_{\rm eq} \approx 1089~{\rm K}$ for TOI-1927 b and $\approx 1110~{\rm K}$ for TOI-2643 b), where the planet would otherwise cool and contract efficiently, and several mechanisms have been proposed to explain anomalously large radii (see e.g, \citealt{2024AJ....168...91Y}), including: (i) irradiation-driven heating, where stellar flux modifies the deep atmospheric temperature gradient and delays contraction (e.g., \citealt{2010ApJ...714L.238B, 2017ApJ...846...47P, 2018AJ....155..214T}; (ii) H/He envelope contraction regulated by thermal cooling \citep{2014ApJ...792....1L}; and, (iii) atmospheric energy deposition from photoionisation-driven winds, which generate extended upper atmospheres \citep{2013ApJ...775..105O, 2024AJ....167...79K}. Additionally, photochemical hazes may increase the apparent radius by altering the scattering profile of the upper atmosphere \citep{2021ApJ...920..124O}.

The thick disc context of these companions also places stringent constraints on the physics of radius inflation since these stars exhibit ages $\gtrsim 8-9~{\rm Gyr}$, lower metallicities, and distinct formation conditions. {If the planets formed with a reduced inventory of high-Z material (silicates, ices, and iron-bearing compounds) the lower envelope and atmospheric opacities would enhance radiative energy transport}, increasing the intrinsic luminosity $L_{\rm int}$ and thereby shortening the Kelvin--Helmholtz timescale $\tau_{\rm KH} \propto L_{\rm int}(\kappa)^{-1}$, where $L_{\rm int}(\kappa)$ explicitly depends on the opacity $\kappa$ \citep{2005AREPS..33..493G, 2007ApJ...659.1661F}. A shorter $\tau_{\rm KH}$ corresponds to faster cooling and contraction, contrary to the observed inflated radii. Alternatively, variations in core mass or primordial entropy alter $L_{\rm int}$ and the radius evolution, modifying the energy budget required to sustain inflation. The persistence of large radii in TOI-1927 $b$ and TOI-2643 $b$ therefore indicates that one or more energy sources must remain effective over billion-year timescales, for example: irradiation-driven deposition, $L_{\rm dep} = \varepsilon_{\rm dep} \,\pi R_{\rm p}^{2} F_{\rm inc}$ (e.g., \citealt{2010ApJ...716.1323B}, tidal dissipation at non-zero eccentricity, $\dot{e}_{\rm tide} \propto \mathcal{Q}^{-1}$ \citep{2008ApJ...678.1396J, 2012arXiv1209.5724S}, or even magnetically mediated atmospheric heating \citep{2014ApJ...794..132R}. Whether sustained heating, inhibited interior cooling, or extended mass-loss envelopes dominate remains unconstrained, making these systems especially diagnostic of the metallicity dependence and long-term efficacy of inflation mechanisms. 

When it comes to observational constraints, the low surface gravities of puffy planets enhance scale heights and produce favourable conditions for atmospheric characterisation. Transmission spectroscopy---particularly at ${\rm Ly}-\alpha$ at $\lambda1215~{\rm \AA}$, H$\alpha$ at $\lambda6562~{\rm \AA}$, and the metastable He I triplet at $\lambda10833~{\rm \AA}$---can help quantify mass-loss efficiencies, ionisation states, and the vertical structure of the upper atmosphere \citep{2018Natur.557...68S, 2018ApJ...855L..11O, 2019A&A...623A.131K}. Moreover, multi-band transmission and thermal emission spectra also allow constraining atmospheric metallicities, temperature–pressure profiles, potential hazes, and disequilibrium chemistry, while simultaneously, precise orbital modelling (eccentricities, obliquities, and tidal quality factors) and improved stellar chronologies can evaluate whether residual tidal heating remains energetically viable at these ages, which will be the subject of future follow-up studies.

\section{Discussions}\label{sec:discussions}

\subsection{{Giant Planet Formation Mechanisms in the Galactic Thick Disc}}\label{sec:demo}

\begin{figure}[h]
    \centering
    \includegraphics[width = \linewidth]{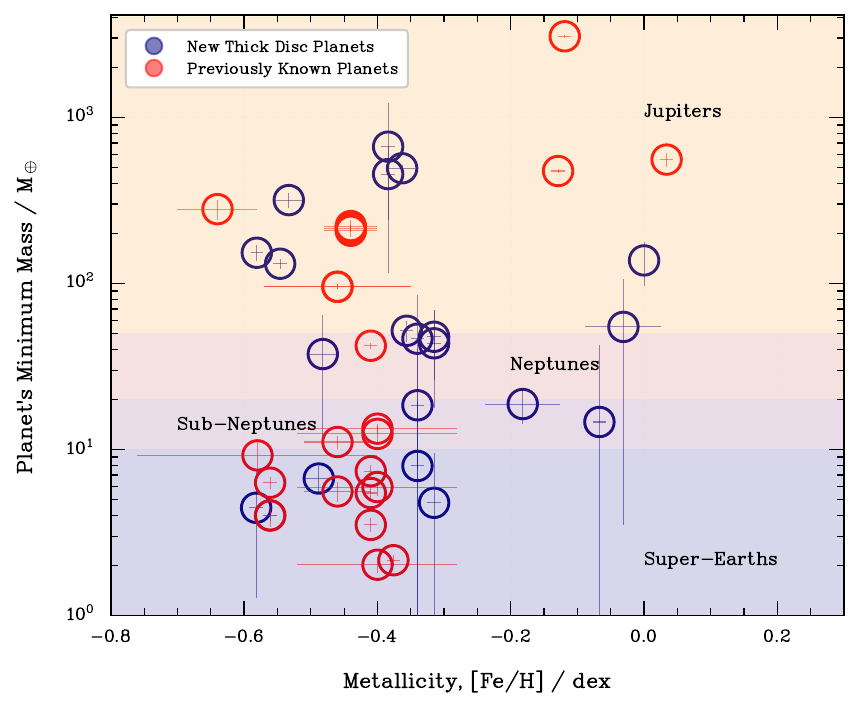}
    \caption{Planet minimum mass vs. host star metallicity [Fe/H] for planets previously known to be thick disc (red) and planets with thick disc membership confirmed in this work (blue; see Table \ref{tab:sp}). Shaded regions roughly delineate Super-Earths ($M_p < 10~M_\oplus$), Sub-Neptunes ($10 < M_p < 20~M_\oplus$), Neptunes ($20 < M_p < 50~M_\oplus$), and Jupiters ($M > 50~M_\oplus$; see \citealt{2007ApJ...656..545V, 2014ApJ...792....1L, 2016ApJ...825...19W, 2020A&A...634A..43O}).}
    \label{fig:mass_feh}
\end{figure}

{Among the 27 thick disc systems with measured planet masses and ${\rm [Fe/H]} < 0$ dex characterised, 14 (51.9\%) host only gas giants, 10 (37\%) host only small planets (super-Earths and sub-Neptunes), and 3 (11.1\%) host both populations (see Figure \ref{fig:mass_feh}), which indicates a relatively high incidence of gas-giant–hosting systems within this metal-poor sample. The well-established planet-metallicity correlation in the thin disc \citep{Fischer:2005ApJ...622.1102F, Wang:2015AJ....149...14W, Petigura:2018AJ....155...89P} predicts a strong suppression of giant planet formation at ${\rm [Fe/H]} < 0$ dex, where lower solid surface densities may slow core growth and make the $\sim 10~M_\oplus$ threshold required for runaway gas accretion increasingly difficult to reach \citep{1996Icar..124...62P, 2012ApJ...751...81J}. That exoplanet occurrence rates in the thick disc appear suppressed by roughly $\sim 50\%$ relative to the thin disc \citep{Bashi:2022MNRAS.510.3449B, Zink:2023AJ....165..262Z} is consistent with this picture; yet the conditional incidence of gas giants among planet-hosting systems remains high, suggesting that while metal-poor environments hinder planet formation overall, those systems that do form planets may preferentially channel their solid mass budget into giant planet cores.}

{Two interpretations here, therefore, merit consideration. First,} disc instability---which is less sensitive to metallicity than core accretion \citep{1997Sci...276.1836B}---{may operate more efficiently in thick disc protoplanetary discs, bypassing the solid-assembly bottleneck entirely. Second, the $\alpha-$enhanced composition of thick disc hosts may modulate the metallicity dependence of core accretion in ways not fully captured by [Fe/H] alone, i.e., variations in [Mg/Si] and [Fe/Mg]} {ratios alter the silicate-to-metal fraction of forming solids} \citep{2015A&A...577A..83D, 2018A&A...612A..46M, 2021PSJ.....2..113S}, {potentially concentrating the available refractory material into denser, more massive cores despite the overall metal deficit} \citep{2010ApJ...716....1B, 2014ApJ...793..124U}. {We caution, however, that} detection biases {favour} gas giants {in both transit and radial velocity surveys, and our sample remains relatively small; the near-parity, in that sense, should therefore be regarded as suggestive rather than conclusive.}

{The interpretation may be further complicated by the fact that classifying these objects themselves is uncertain at low-metallicity regimes}. For instance, {the canonical mass-based boundary between gas giants and brown dwarfs---set near the deuterium-burning threshold at $\sim11-16~M_{\rm Jup}$ \citep{2011ApJ...727...57S}---becomes increasingly unreliable in metal-poor regimes, where} post-formation cooling erases formation imprints, rendering mass, radius, or luminosity alone {ambiguous} discriminants \citep{2019A&A...624A..20M, 2019A&A...623A..85L}. Properties that clearly separate massive Jovian planets from low-mass brown dwarfs in solar- and super-solar metallicities may no longer apply, suggesting that robust classification in ancient, metal-poor environments requires more consideration of formation pathways---core accretion vs. gravitational instability as dominant processes---{through differences in high-Z (rock- and ice-forming) enrichment \citep{2007ApJ...659.1661F, 2008Icar..195..863H, 2012A&A...541A..97M, 2017MNRAS.470.2387N}, as well as the dynamical imprint of migration or past scattering phenomena \citep{1997Sci...276.1836B, 2005MNRAS.364L..56R, 2007ApJ...662.1282M, 2009ApJ...695L..53B}}. {An independent and powerful avenue to break this degeneracy is atmospheric characterisation, with the C/O ratio and atmospheric metallicity of a planet potentially encoding its formation location and accretion history, with gravitational instability products expected to reflect the bulk disc composition while core accretion planets are systematically enriched in high-Z material through solid accretion \citep{2019ApJ...887L..20W, 2023Natur.618...43B}. In the thick disc context---where host star metallicities are intrinsically sub-solar---disentangling planetary atmospheric enrichment from the underlying disc composition will require transmission or emission spectroscopy with the \emph{James Webb Space Telescope} (JWST; \citealt{2024RvMG...90..411K, 2025ARA&A..63...83S}) or \emph{Atmospheric Remote-sensing Infrared Exoplanet Large-survey} (ARIEL; \citealt{2018SPIE10698E..0HP}), making these systems high-priority targets for atmospheric follow-up.} {Enlarging the confirmed thick disc sample---and potentially extending it toward halo star hosts} (e.g., \citealt{2024A&A...692A.150B})---{will be essential for testing, therefore, whether the near-parity reported in this work persists and for disentangling formation pathways from observational biases.}

\subsection{Orbital Eccentricities and Tidal Effects}\label{sec:ecc_tidal}

Planets orbiting thick disc stars have persisted for $\gtrsim8$ Gyr, indicating that any primordial dynamical instabilities were either intrinsically modest or efficiently damped during the early phases of system assembly. Yet, such longevity does not clearly guarantee dynamical quiescence, where weak but persistent perturbations---e.g., from undetected long-period companions, episodic stellar encounters, or even the cumulative action of Galactic tides---can, over billion-year timescales, drive secular excitation of eccentricities or gradual erosion of planetary multiplicity \citep{1998ApJ...508L.171L, 2004AJ....128..869Z}. Consequently, measurable deviations in eccentricity distributions relative to thin disc systems would provide direct evidence for dynamical heating shaped by the broader Galactic environment, rather than solely by internal system architecture. 

The current sample of exoplanets orbiting thick disc stars remains, however, far too limited to establish whether such signatures are present, and existing eccentricity estimates suffer from short temporal baselines and modest precision. Long-term, high-cadence radial velocity monitoring---together with forthcoming \emph{Gaia} astrometric measurements---will be essential for resolving low-amplitude orbital motion and identifying distant or inclined companions that may contribute to the observed dichotomy in orbital eccentricity distributions.

\begin{figure}
    \centering
    \includegraphics[width = \linewidth]{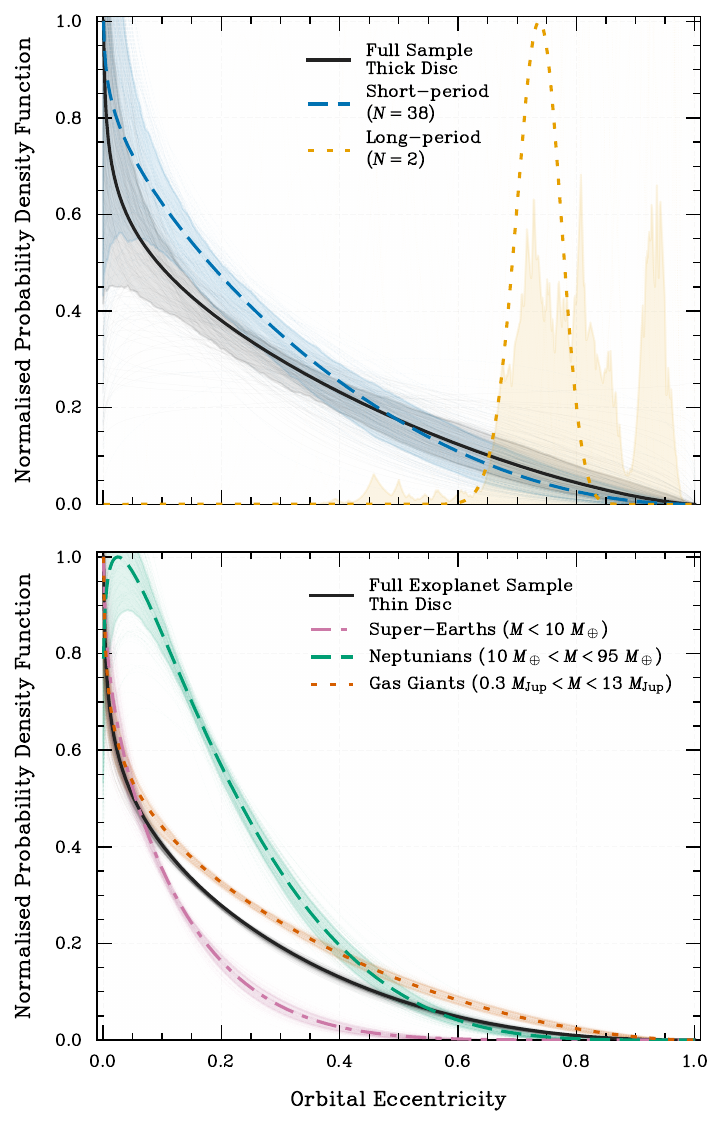}
    \caption{{Normalised probability density function of orbital eccentricities for planets around thick (top) and {\it kinematically-consistent} thin (bottom) disc stars, fitted using hierarchical Bayesian modelling with an underlying beta distribution. The solid line in both panels show the full sample, with the thick disc population further split into short- and long-period populations, and the thin disc split into Super-Earths ($M < 10~M_\oplus$), Neptunians ($10~M_\oplus < M < 0.3~M_{\rm Jup}$), and Gas Giants ($0.3~M_{\rm Jup} < M < 13~M_{\rm Jup}$), respectively. Thick lines indicate median fits, shaded regions show $68\%$ credible intervals, and lighter overlapping lines display posterior samples illustrating the uncertainty in the fitted distributions.}}
    \label{fig:eccdistribution}
\end{figure}

Nevertheless, {following the framework introduced in \cite{2026arXiv260702682F}}, the eccentricity distribution of thick disc planetary systems can be characterised more robustly by adopting a hierarchical Bayesian model in which the population-level distribution is represented by an underlying Beta function $\mathcal{B}(\alpha,\beta)$, as demonstrated in Figure \ref{fig:eccdistribution}. An initial Kolmogorov-Smirnov test scanning period thresholds across the full orbital range of our sample identifies a tentative separation at $P \approx 3000$ days, yielding $D = 0.97$ and p-value $p = 0.003$. Although the division is significant, we emphasise that it depends on an extremely limited sample ($N = 40$ planets with well-constrained eccentricities, partitioned into $N_{\rm SP} = 38$ and $N_{\rm LP} = 2$), and the inferred threshold may, and will, shift substantially as additional thick disc planets are discovered and characterised in future investigations. With this critical caveat in mind, we proceed with separate HBM fits for short-period (SP; $P < 3000$ d) and long-period (LP; $P \geq 3000$ d) subsamples, alongside a fit to the full population for comparison. The resulting HBM analysis yields Beta distributions of the form {$\mathcal{B}_{\rm Full\,Sample}\left(0.88^{+0.18}_{-0.16}, 2.42^{+0.60}_{-0.51}\right)$ for the full sample with population mean $\langle e \rangle_{\rm Full~Sample} = 0.26\pm0.07$. For the short-period subsample, we obtain $\mathcal{B}_{\rm SP}\left(0.94^{+0.22}_{-0.18}, 3.02^{+0.83}_{-0.68}\right)$, with $\hat{e}_{\rm SP} \rightarrow 0$ and $\langle e\rangle_{\rm SP} = 0.23\pm0.06$, and in striking contrast, the long-period subsample---though comprising only two planets---exhibits a markedly different distribution, $\mathcal{B}_{\rm LP}\left(103^{+260}_{-92}, 37^{+101}_{-33}\right)$, peaking at $\hat{e}_{\rm LP} \approx 0.73$ and yielding $\langle e\rangle_{\rm LP} \leq 0.73$.} The large uncertainties on the long-period parameters clearly reflect the severe small-number limitations, yet the qualitative distinction---whereby outer planets appear systematically more eccentric than their inner counterparts---is visually evident.

{Comparing the full thick disc and thin disc populations directly, we find $\langle e \rangle_{\rm Thick} = 0.26\pm0.07$ versus $\langle e \rangle_{\rm Thin} = 0.20\pm0.01$, indicating that thick disc planetary systems are, on average, modestly more eccentric than their thin disc counterparts, albeit with substantially larger uncertainty owing to the relatively small thick disc sample size ($N = 40$) with reliably measured eccentricities. We further partition the thin disc sample by planetary mass, adopting the same framework presented in Section \ref{sec:kinscreen} and drawing planets from the NASA Exoplanet Archive \citep{2025PSJ.....6..186C}---restricted to confirmed planets with measured mass and radius, excluding controversial entries, and using the archive's default parameter set finding out that the full sample follows $\mathcal{B}_{\rm FS,~Thin} = 0.86^{+0.02}_{-0.03}, 3.30^{+0.13}_{-0.11}$; $N = 1546$ and $\langle e \rangle_{\rm FS,~Thin} = 0.20\pm0.01$. Separate Beta-distribution fits yield: (a) Super-Earth ($M < 10~M_\oplus$) with $\mathcal{B}_{\rm SE} = 0.91^{+0.01}_{-0.09}, 7.05^{+0.33}_{-1.32}$; $N = 300$, and $\langle e \rangle_{\rm SE} = 0.11\pm0.01$, (b) Neptunians ($10~M_\oplus < M < 0.3~M_{\rm Jup}$) with $\mathcal{B}_{\rm N} = 1.10^{+0.08}_{-0.07}, 4.95^{+0.41}_{-0.38}$; $N = 345$, and $\langle e \rangle_{\rm N} = 0.18\pm0.01$, and (c) Gas Giant ($0.3~M_{\rm Jup} < M < 13~M_{\rm Jup}$) with $\mathcal{B}_{\rm G} = 0.86^{+0.04}_{-0.03}, 2.76^{+0.14}_{-0.14}$; $N = 899$ with $\langle e \rangle_{\rm G} = 0.23\pm0.01$ (Figure \ref{fig:eccdistribution}), all consistent with or below the thick disc mean, with only the most massive gas giants approaching comparable eccentricities. This suggests the thick disc's elevated eccentricity may not be simply a by-product of hosting more massive planets, but may instead reflect the dynamically hotter formation or migration environment associated with thick disc kinematics.}

Such inferred posteriors indicate that the population of exoplanets orbiting thick disc stars exhibit a period-dependent eccentricity structure, with short-period planets occupying a moderately eccentric regime while long-period companions display markedly higher eccentricities. However, given the exceedingly small sample sizes---particularly for the long-period sub-sample---this interpretation remains highly tentative and must be regarded as preliminary until confirmed by new additions to the current sample. The observed bi-modality is qualitatively consistent, however, with scenarios in which inner planets have undergone partial tidal circularisation or dynamical damping, while outer planets retain primordial excitation or have experienced ongoing secular perturbations \citep{2025arXiv250923973S}, i.e., thick disc companions orbit on higher eccentric orbits compared to thin disc ones (see their Figure 5). 

\begin{figure}[h]
    \centering
    \includegraphics[width = \linewidth]{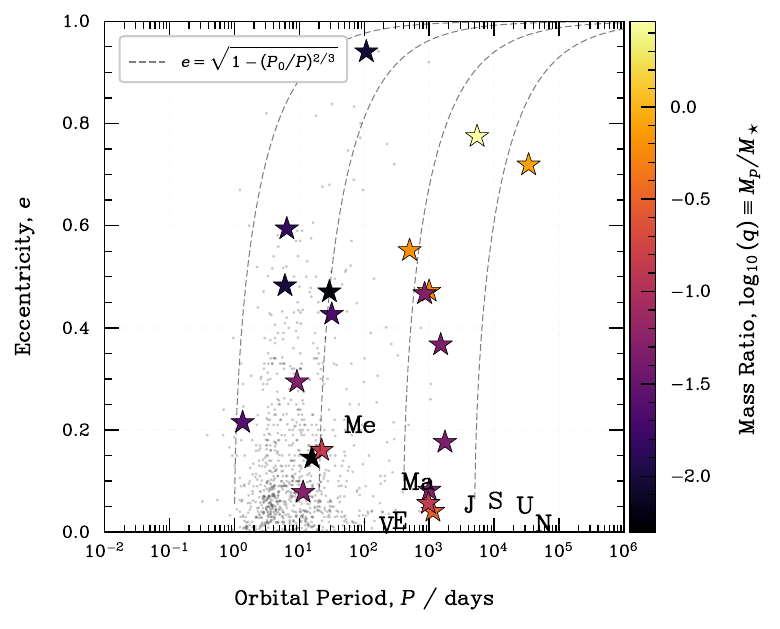}
    \caption{Orbital period and eccentricity relation for thick disc planets (coloured by planet-to-stellar mass ratio $q$) compared to the NASA Exoplanet Archive (grey) and Solar System planets (labelled). Dashed curves show angular momentum conservation tracks $e = \sqrt{1 - (P_0/P)^{2/3}}$ for reference initial periods $P_0 = 1,~20,~400,~{\rm and}~5000$ days.}
    \label{fig:period_ecc}
\end{figure}

Figure \ref{fig:period_ecc} further presents the orbital period-eccentricity plane for thick disc planets, overlaid with angular momentum conservation tracks $e = \sqrt{1 - (P_0/P)^{2/3}}$ for initial periods of $P_0 = 1,~20,~400,~{\rm and}~5000$ days \citep{2016ApJ...829...34S}. Statistical comparison via Mann--Whitney--Wilcoxon $U$ tests \citep{wilcoxon45, mann1947test} reveals that thick disc planets are systematically more eccentric than \emph{field}/thin disc systems across multiple period bins: at $P < 10$ days, for instance, we find $\langle e_{\rm thick}\rangle = 0.396$ (N = 4) versus $\langle e_{\rm field}\rangle = 0.132$ (N = 547), with $p = 0.0047$; at $10 < P < 100$ days, $\langle e_{\rm thick}\rangle = 0.256$ (N = 5) versus 
$\langle e_{\rm field}\rangle = 0.167$ (N = 316), with $p = 0.1366$; at $100 < P < 1000$ days, $\langle e_{\rm thick}\rangle = 0.497$ (N = 5) versus $\langle e_{\rm field}\rangle = 0.202$ (N = 46), with $p = 0.0255$; while the excess becomes statistically insignificant at $P > 1000$ days, where $\langle e_{\rm thick}\rangle = 0.360$ (N = 6) versus 
$\langle e_{\rm field}\rangle = 0.203$ (N = 3), with $p = 0.7143$, owing to the small sample sizes.

Strikingly, two planets lie near constant-angular momentum tracks---TOI-1926 $b$ ($P \approx 1.34$ days, $e \approx 0.215$) and HD 136352 $d$ ($P \approx 107.5$ days, $e \approx 0.940$)--the latter requiring either extremely weak tidal dissipation ($\mathcal{Q}_{\rm p} \gg 10^6$; following \citealp{2008ApJ...678.1396J}) or ongoing dynamical pumping from von Zeipel-Kozai-Lidov cycles \citep{2007ApJ...669.1298F}, secular interactions with unseen companions, or past stellar encounters \citep{2006ApJ...640.1086F}. We estimated that tidal circularisation timescales---$\tau_{\rm circ} \propto P \cdot\mathcal{Q}_{\rm p}/\kappa_2 \cdot (M_\star/M_{\rm p}) \cdot (a/R_{\rm p})^5$ following \cite{1966Icar....5..375G, 1981A&A....99..126H, 2008Icar..193..637W}---exceed 8 Gyr for all companions with $e > 0.15$, consistent with long-term survival and suggestive that either initially dynamically hot architectures---such as planet–planet scattering or primordial excitation by binaries \citep{2008ApJ...686..580C, 2008ApJ...678..498N}---ongoing secular perturbations from undetected companions, or weak Galactic tidal influences \citep{2013Natur.493..381K} have shaped their present configurations. All of these considerations imply and reinforce the view that thick disc planetary systems may have undergone distinct formation pathways and dynamical evolutions relative to their thin disc counterparts \citep{2012A&A...545A..32A}.

\section{Conclusions}\label{sec:conclusions}

{We report the discovery of new exoplanets orbiting thick disc stars, together with the confirmation of thick disc membership for several previously known planet-hosting systems through combined chemical and kinematic analyses. When supplemented with previously reported thick disc planet-hosting systems (Table \ref{tab:sp}; \citealt{Jofre:2015AA...574A..50J, Valenti:2005ApJS..159..141V, Campante:2015ApJ...799..170C, Lacedelli:2022MNRAS.511.4551L, Dai:2023AJ....166...49D, Bouchy:2010AA...519A..98B, Mortier:2020MNRAS.499.5004M,2025arXiv250910136G, Behmard:2025AJ....170..282B, 2025ApJS..281...61L}), the sample expands to a total of 66 confirmed exoplanets.} By employing high-precision radial velocities, transit photometry, and a uniform spectroscopic framework, we derived precise stellar ages, masses, radii, and orbital parameters for all systems considered. These results carry significant implications for planet formation and evolution during the early Milky Way, where feedback-dominated conditions are generally thought to have limited efficient planet assembly. In probing ancient, $\alpha-$enhanced environments, we establish a link between exoplanet demographics and Galactic archaeology, with forthcoming studies (Papers II and III) set to investigate population-level trends and detailed chemical signatures.

{Our analysis provides tentative evidence for an elevated giant-planet fraction around thick disc stars, which is consistent with previous studies that reported a relatively high incidence of giant planets among metal-poor, $\alpha-$enhanced thick disc stars despite their generally low metallicities \citep{2009MNRAS.399L.103G, 2009ApJ...698L...1H, 2012A&A...547A..36A}. However, given the limited sample size and potential selection effects, larger homogeneous samples with well-characterised systems are required to establish the statistical significance of this trend and to determine whether it is linked to thick disc membership itself or to enhanced $\alpha-$element abundance of these stars.} We also identify two unusually low-density `puffy' planets (TOI-1927 b and TOI-2643 b)---the first detected around thick disc stars---with densities as low as $0.17~{\rm g~cm^{-3}}$, offering new constraints on atmospheric evolution across billion-year timescales. Furthermore, thick disc planets may exhibit systematically higher orbital eccentricities than their thin disc counterparts {(see Figure \ref{fig:eccdistribution}}), hinting at distinct dynamical pathways influenced by Galactic tides or primordial excitation. A larger sample will, however, be essential to secure these emerging trends statistically.

\begin{acknowledgments} 

    T.F. acknowledges support from Yale Graduate School of Arts and Sciences. J.Y.G. acknowledges support from a Carnegie Fellowship. H.R. acknowledges support from NOIRLab, which is managed by the Association of Universities for Research in Astronomy (AURA) under a cooperative agreement with the National Science Foundation. E.M. acknowledges funding from FAPEMIG under project number APQ-02493-22 and a research productivity grant number 309829/2022-4 awarded by the CNPq, Brazil. V.L.T. acknowledges support from the CNPq through the Postdoctoral Junior (PDJ) fellowship, process No. 152242/2024-4.

    We gratefully acknowledge Keerthi Vasan for conducting MIKE observations for some of the stars included in this work. We also thank Diego Lorenzo-Oliveira for his comments, which helped improve the paper.

    The observations were carried out within the framework of the Subaru-Keck/Subaru-Gemini time exchange program, which is operated by the National Astronomical Observatory of Japan. We are honoured and grateful for the opportunity of observing the Universe from Maunakea, which has cultural, historical and natural significance in Hawaii. We are also deeply thankful to the Likan Antai of the Atacama Desert, who have observed the dark constellations of the Andean skies for millennia.

    This work has made use of data from the European Space Agency (ESA) mission Gaia (\url{https://www.cosmos.esa.int/gaia}), processed by the Gaia Data Processing and Analysis Consortium (DPAC, \url{https://www.cosmos.esa.int/web/gaia/dpac/consortium}). The ID ESO programs used in this work are: 072.C-0488(E), 198.C-0836(A), 183.C-0972(A), 192.C-0852(A), 196.C-1006(A), 192.C-0852(G), 091.C-0936(A), 077.C-0364(E), 0103.C-0548(A), 0101.C-0232(A), 097.C-0090(A), 0101.C-0232(B), 0100.C-0414(B), 0102.C-0338(A), 105.20AK.002, 083.D-0549(B), 081.D-0066(A), 080.D-0408(A), 086.D-0078(D), 080.D-0047(A), 106.215E.002, 082.D-0499(A), 097.C-0021(A), 084.D-0591(C), 082.D-0499(B), 091.D-0469(A), 106.215E.004, 0102.C-0584(A), 0104.C-0090(A), 192.C-0852(H), 0103.C-0206(A), 196.C-0042(D), 087.C-0831(A), 089.C-0732(A), 0100.C-0097(A), 091.C-0034(A), 082.C-0212(B), 099.C-0458(A), 190.C-0027(A), 085.C-0019(A), 085.C-0063(A), 086.C-0284(A), 082.C-0212(A), 111.2506.001, 60.A-9036(A), 090.C-0849(A), 084.C-0229(A), 087.C-0990(A), 089.C-0050(A), 088.C-0011(A), 096.C-0053(A), 083.C-1001(A), 092.C-0579(A), 095.C-0040(A), 086.C-0230(A), 094.C-0797(A), 093.C-0062(A), 60.A-9709(G), 096.D-0402(A), 0101.C-0232(C), 0100.C-0414(A), 0102.C-0338(B), 60.A-9700(G), 094.C-0901(A), 112.25YG.010, 0104.C-0090(B), 112.25YG.011, 112.25YG.003, 112.25YG.004, 108.22KV.004, 108.22KV.005, 108.22KV.001, 108.22KV.002, 112.25YG.002, 108.22KV.003, 112.25YG.009, 0100.C-0847(A), 0100.C-0474(A), 0101.C-0623(A), 093.C-0919(A), 099.C-0374(A).    

    Funding for the DPAC has been provided by national institutions, in particular the institutions participating in the Gaia Multilateral Agreement. 

    This research has made use of the Exoplanet Follow-up Observation Program (ExoFOP; DOI:10.26134/ExoFOP5) website, which is operated by the California Institute of Technology, under contract with the National Aeronautics and Space Administration under the Exoplanet Exploration Program.

\end{acknowledgments}

\software{
\textsc{numpy} \citep{van_der_Walt:2011CSE....13b..22V}, 
\textsc{matplotlib} \citep{Hunter:4160265}, 
\textsc{pandas} \citep{mckinney-proc-scipy-2010}, 
\textsc{astroquery} \citep{Ginsburg:2019AJ....157...98G}, 
\textsc{iraf} \citep{Tody:1986SPIE..627..733T}, 
\textsc{iSpec} \citep{Blanco:2014A&A...569A.111B, Blanco:2019MNRAS.486.2075B}, 
\textsc{Kapteyn} Package \citep{KapteynPackage}, 
\textsc{moog} \citep{Sneden:1973PhDT.......180S}, 
\textsc{Gala} \citep{gala}, 
\textsc{pyaneti} \citep{2019MNRAS.482.1017B, 2022MNRAS.509..866B}, 
\textsc{smplotlib} \citep{jiaxuan_li_2023_8126529},
\textsc{scipy} \citep{2020NatMe..17..261V}
} 

\facilities{Magellan:Clay/MIKE, ESO:3.6m/HARPS, Keck:I/HIRES, OHP:SOPHIE, FLWO:1.5m/TRES}

\bibliography{bib}
\bibliographystyle{aasjournalv7}

\appendix

\section{Comparison Between GALAH Abundances With Our Measurements}\label{app:abscomp}

\begin{figure*}[h]
    \centering
    \includegraphics[width = 0.32\linewidth]{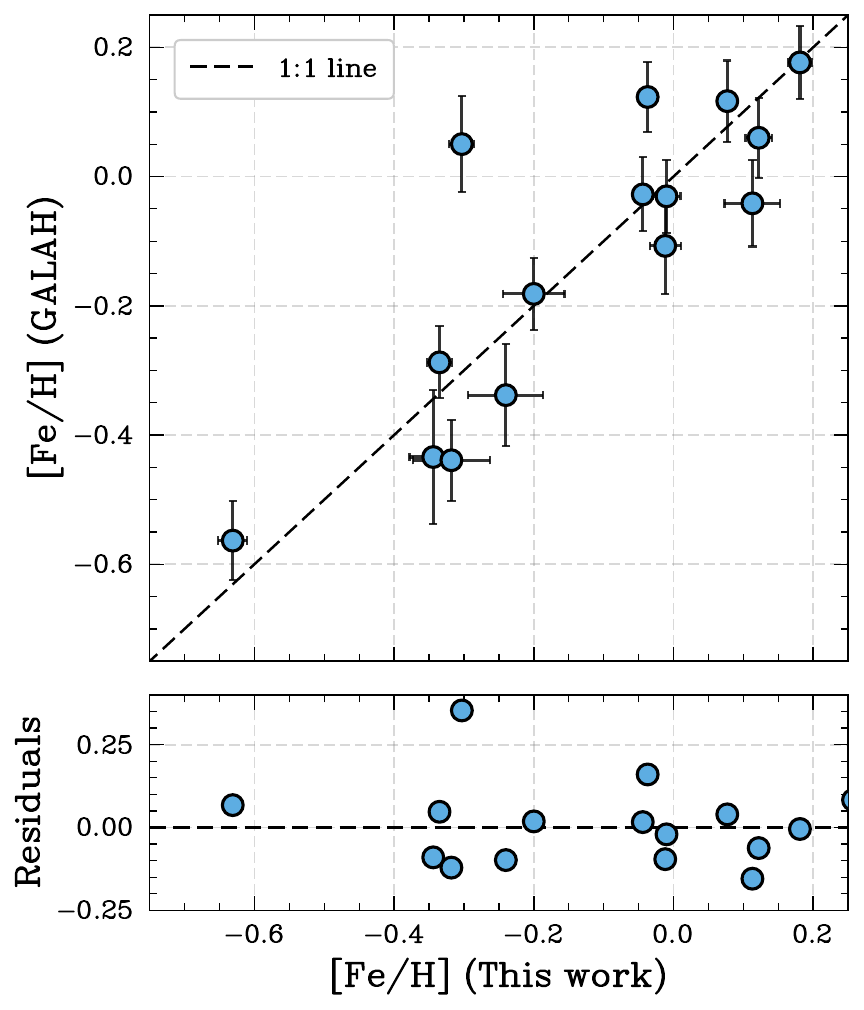} 
    \includegraphics[width = 0.32\linewidth]{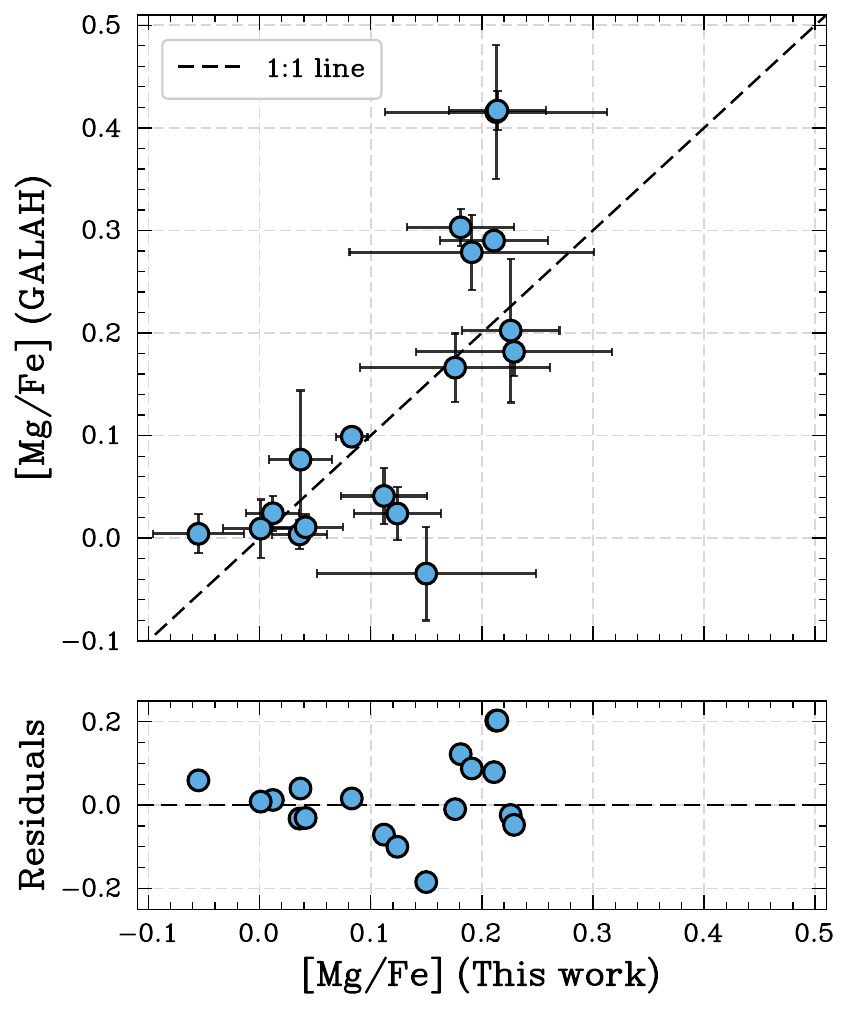}
    \includegraphics[width = 0.32\linewidth]{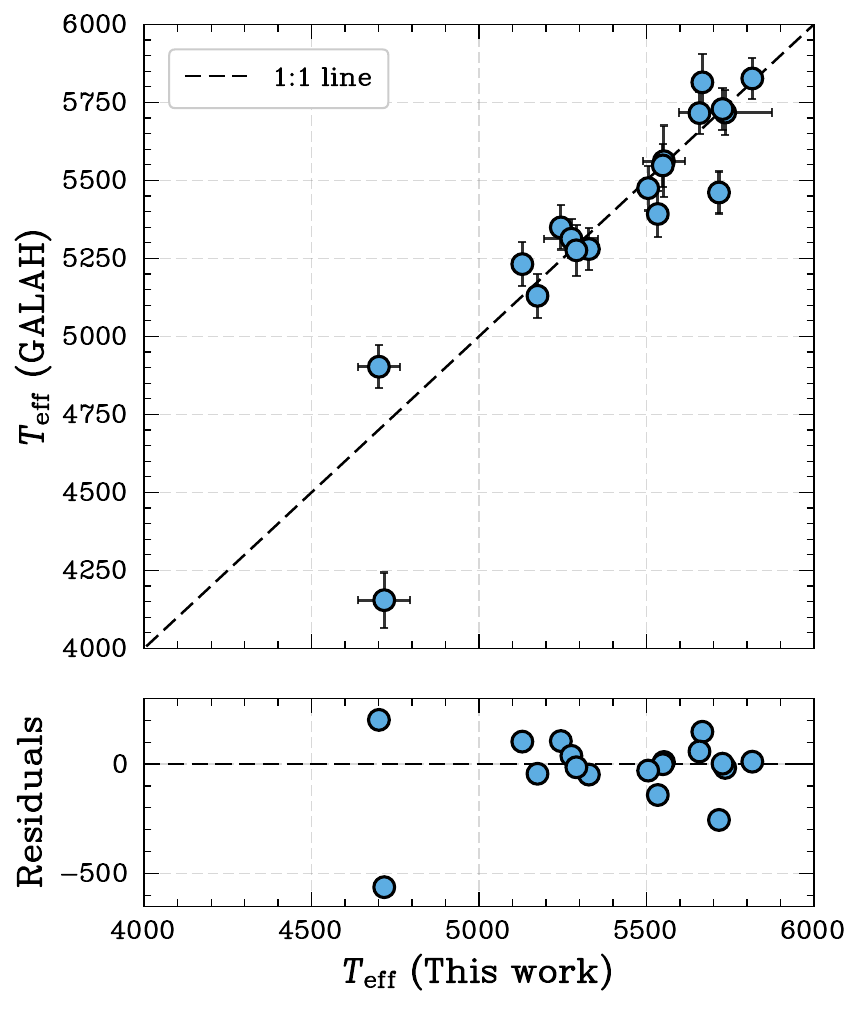}
    \caption{{Comparison between our inferred \feh\ (top left), [Mg/Fe] (top right), and \teff\ (bottom) and those reported in GALAH DR4 \citep{2025PASA...42...51B}. The dashed lines indicate the 1:1 relation. The lower sub-panels show the residuals, which reveal mean offsets of $-0.01$ dex in \feh, $+0.02$ dex in [Mg/Fe], and $-25$ K in \teff.}}
\end{figure*}

\section{$\alpha-$Element Abundances}\label{app:alpha}

\begin{deluxetable*}{lccc} 
\tablecaption{Abundance ratios of [Mg/Fe] and [$\alpha$/Fe] for our sample. A planet is listed as {\it Confirmed} if it was detected via radial velocities or through a combination of radial velocities and transits. {\it Transit} indicates that the planet was detected only via the transit method.} 
\tablewidth{0pt}
\setlength{\tabcolsep}{4pt}
\tablehead{
\colhead{ID} &
\colhead{[Mg/Fe] (dex)} &
\colhead{[$\alpha$/Fe] (dex)} &
\colhead{Exoplanet status}
}
\startdata
HD 150433   & $0.245\pm0.046$ & $0.179\pm0.030$ & Confirmed \\
HD 168746   & $0.206\pm0.047$ & $0.125\pm0.032$ & Confirmed \\
HD 4308     & $0.226\pm0.044$ & $0.160\pm0.028$ & Confirmed \\
HD 111232   & $0.219\pm0.048$ & $0.190\pm0.035$ & Confirmed \\
HD 181720   & $0.281\pm0.048$ & $0.201\pm0.036$ & Confirmed \\
TOI-2011    & $0.211\pm0.048$ & $0.146\pm0.030$ & Confirmed \\
HD 175607   & $0.209\pm0.047$ & $0.190\pm0.032$ & Confirmed \\
HD 20794    & $0.203\pm0.049$ & $0.187\pm0.032$ & Confirmed \\
HD 6434     & $0.347\pm0.045$ & $0.263\pm0.030$ & Confirmed \\
HD 27631    & $0.126\pm0.050$ & $0.079\pm0.032$ & Confirmed \\
HD 220197   & $0.223\pm0.043$ & $0.206\pm0.031$ & Confirmed \\
TOI-126     & $0.218\pm0.074$ & $0.189\pm0.053$ & Confirmed \\
HD 219077   & $0.147\pm0.058$ & $0.108\pm0.025$ & Confirmed \\
K2-262      & $0.184\pm0.089$ & $0.233\pm0.079$ & Confirmed \\
TOI-1927    & $0.166\pm0.065$ & $0.228\pm0.074$ & Confirmed \\
K2-408      & $0.214\pm0.044$ & $0.220\pm0.027$ & Confirmed \\
HIP 109384  & $0.199\pm0.089$ & $0.164\pm0.098$ & Confirmed \\
Kepler-10   & $0.175\pm0.087$ & $0.125\pm0.037$ & Confirmed \\
TOI-1926    & $0.181\pm0.068$ & $0.189\pm0.050$ & Confirmed \\
TOI-2643    & $0.229\pm0.088$ & $0.223\pm0.065$ & Confirmed \\
K2-183      & $0.176\pm0.085$ & $0.070\pm0.052$ & Transit \\
K2-173      & $0.213\pm0.100$ & $0.207\pm0.065$ & Transit \\
K2-156      & $0.150\pm0.099$ & $0.254\pm0.075$ & Transit \\
K2-190      & $0.171\pm0.064$ & $0.125\pm0.030$ & Transit \\
Kepler-463  & $0.234\pm0.103$ & $0.180\pm0.069$ & Transit \\
Kepler-112  & $0.193\pm0.057$ & $0.195\pm0.036$ & Transit \\
Kepler-517  & $0.235\pm0.071$ & $0.171\pm0.053$ & Transit \\
Kepler-1898 & $0.225\pm0.072$ & $0.213\pm0.056$ & Transit \\
Kepler-1619 & $0.273\pm0.060$ & $0.217\pm0.039$ & Transit \\
Kepler-1258 & $0.284\pm0.065$ & $0.203\pm0.039$ & Transit \\
K2-337      & $0.191\pm0.189$ & $0.206\pm0.098$ & Transit 
\enddata
\end{deluxetable*}

\section{Phase Diagrams, Joint Priors and Orbital Parameters for the Re-Analysed Systems}\label{app:lcs}

\begin{figure*}[!h]
    \centering
    \label{fig:XXX}
    \includegraphics[width = 0.32\linewidth]{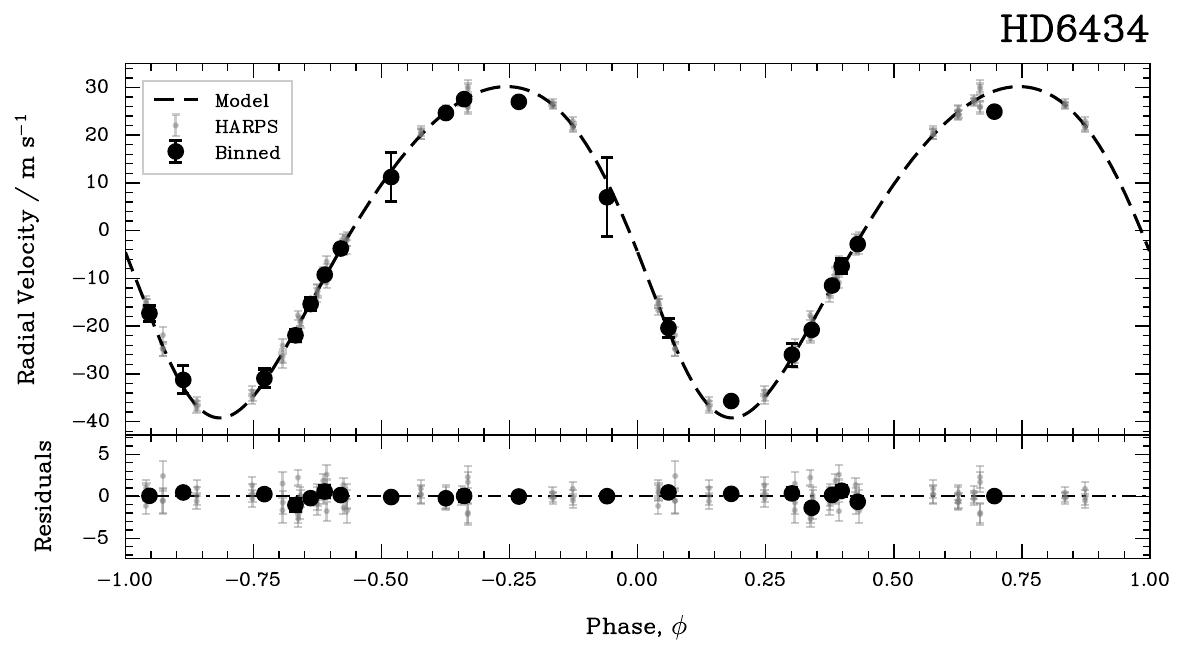}
    \includegraphics[width = 0.32\linewidth]{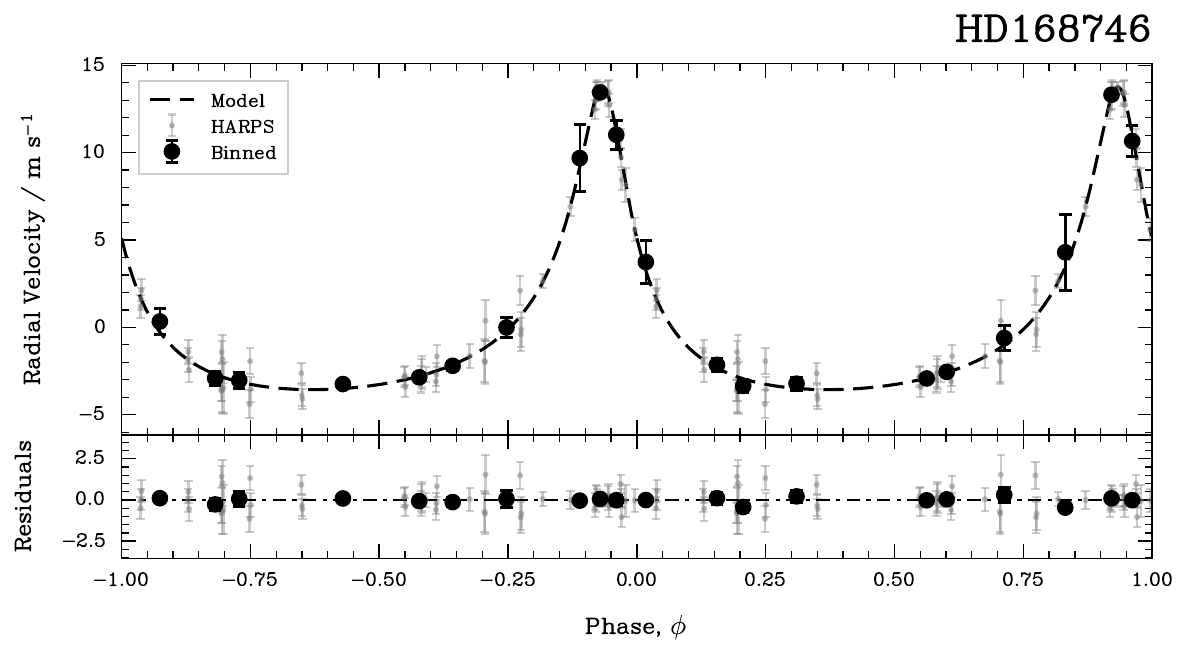}
    \includegraphics[width = 0.32\linewidth]{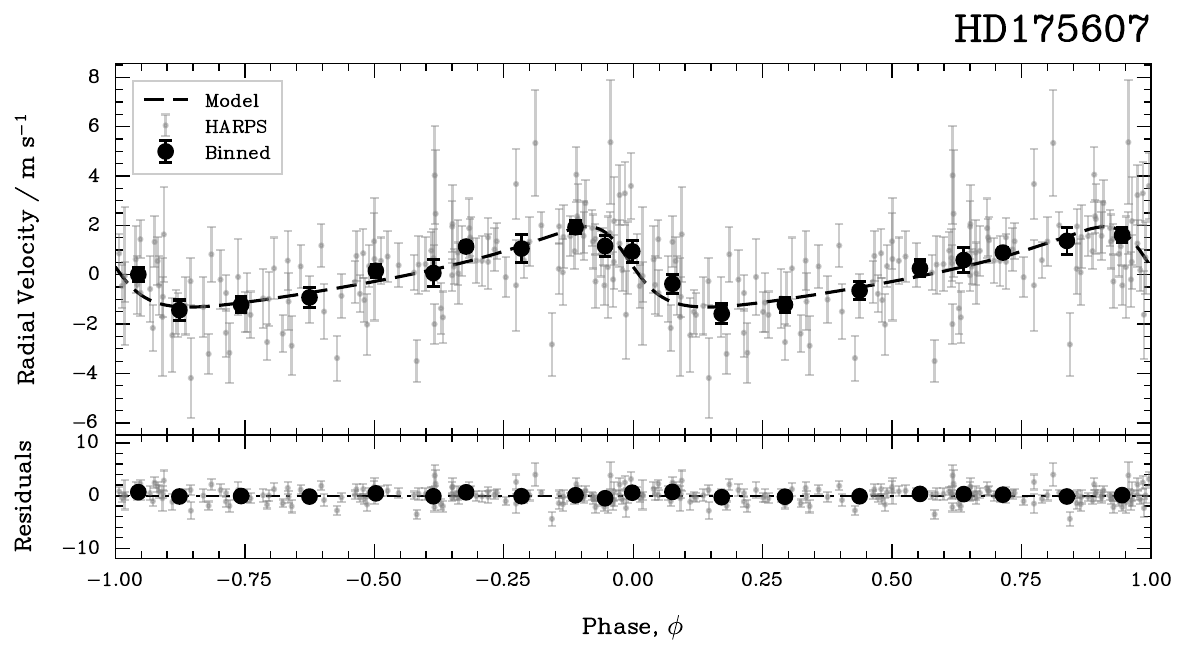}\\
    \includegraphics[width = 0.32\linewidth]{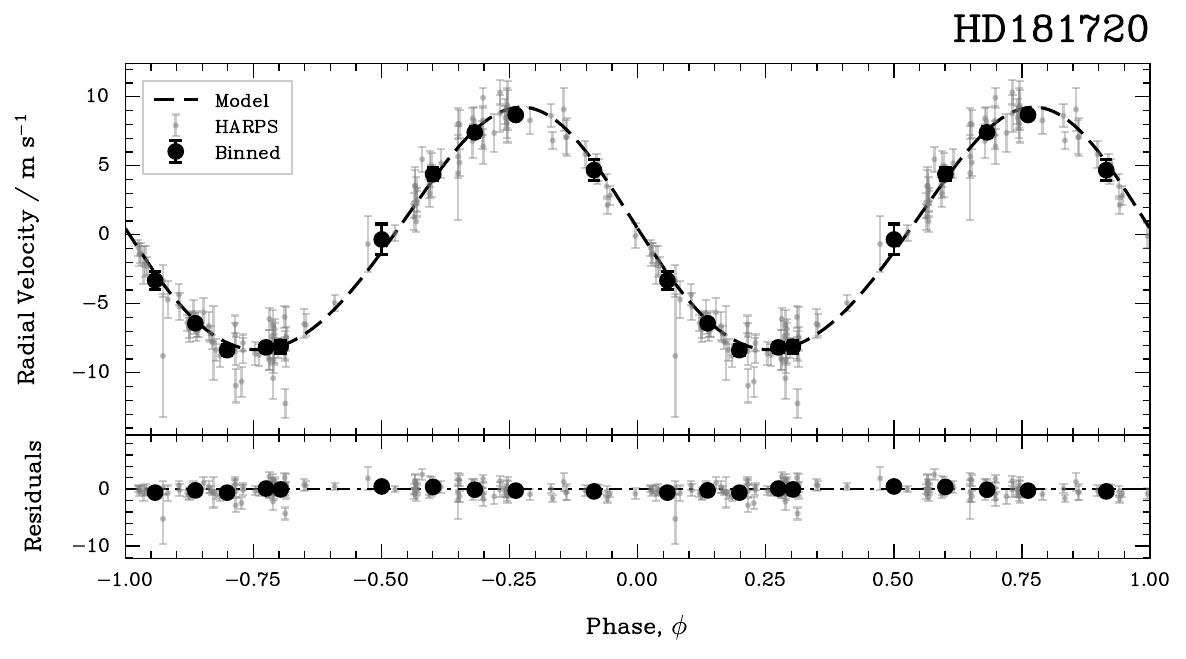}
    \includegraphics[width = 0.32\linewidth]{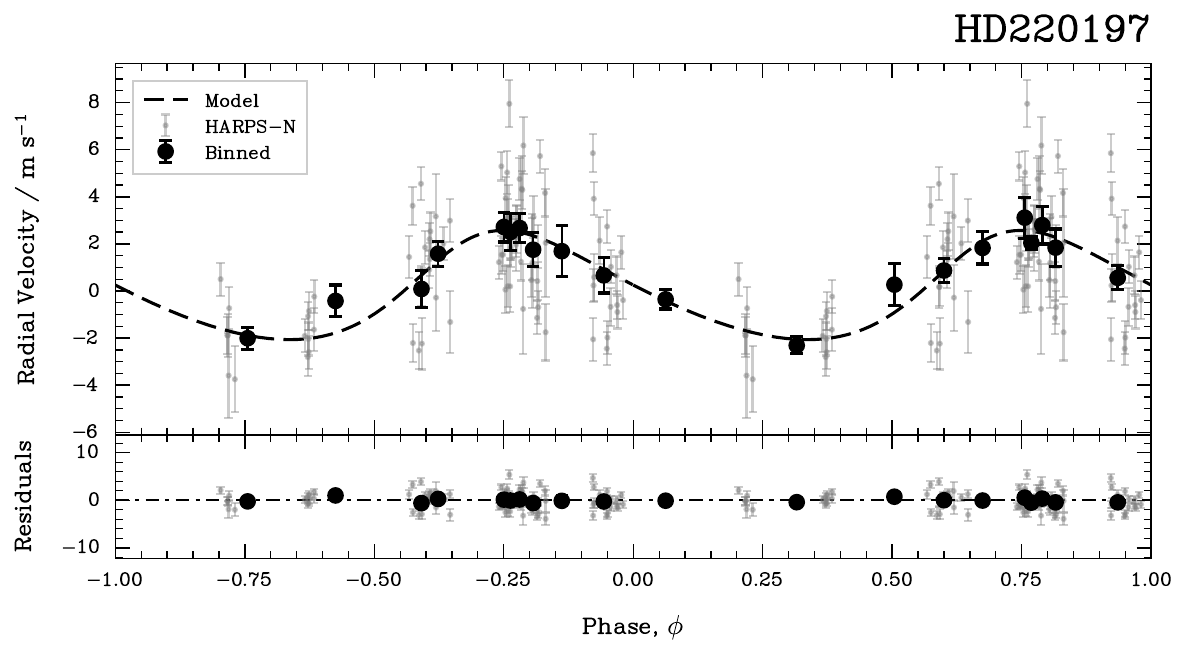}
    \includegraphics[width = 0.32\linewidth]{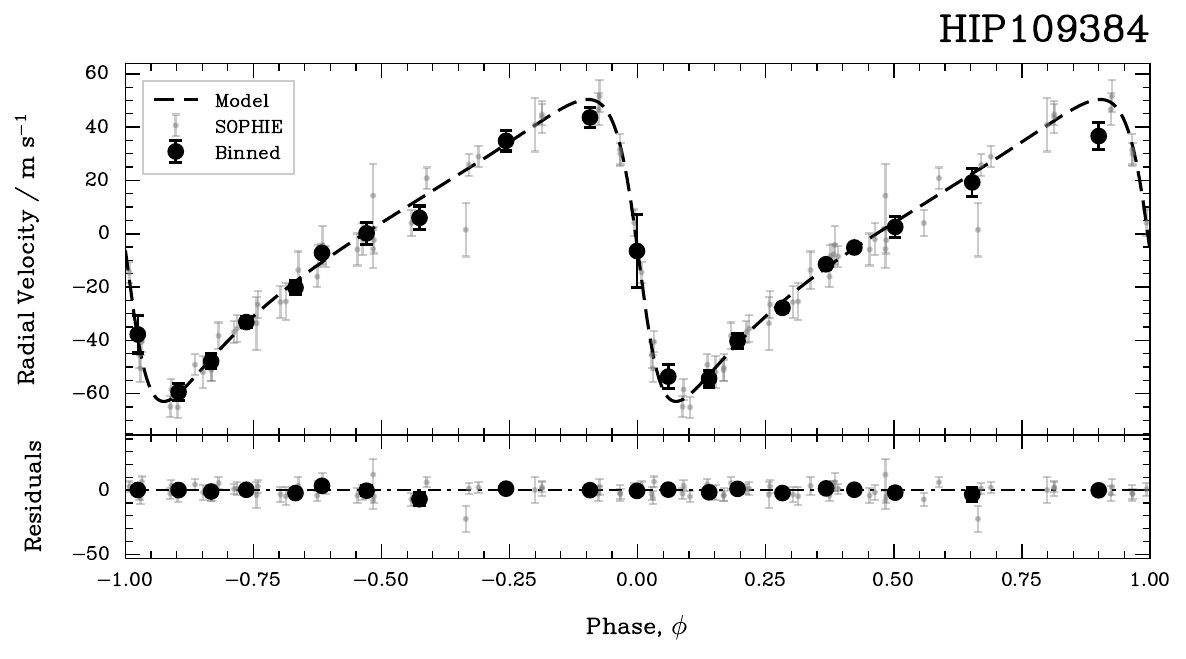}\\
    \includegraphics[width = 0.32\linewidth]{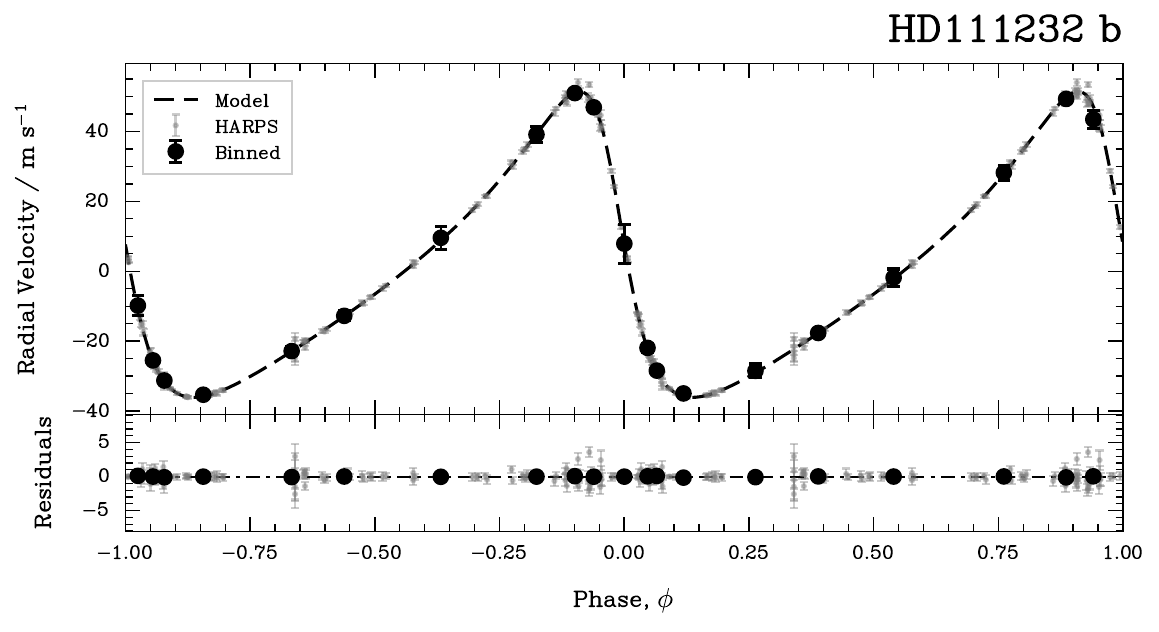}
    \includegraphics[width = 0.32\linewidth]{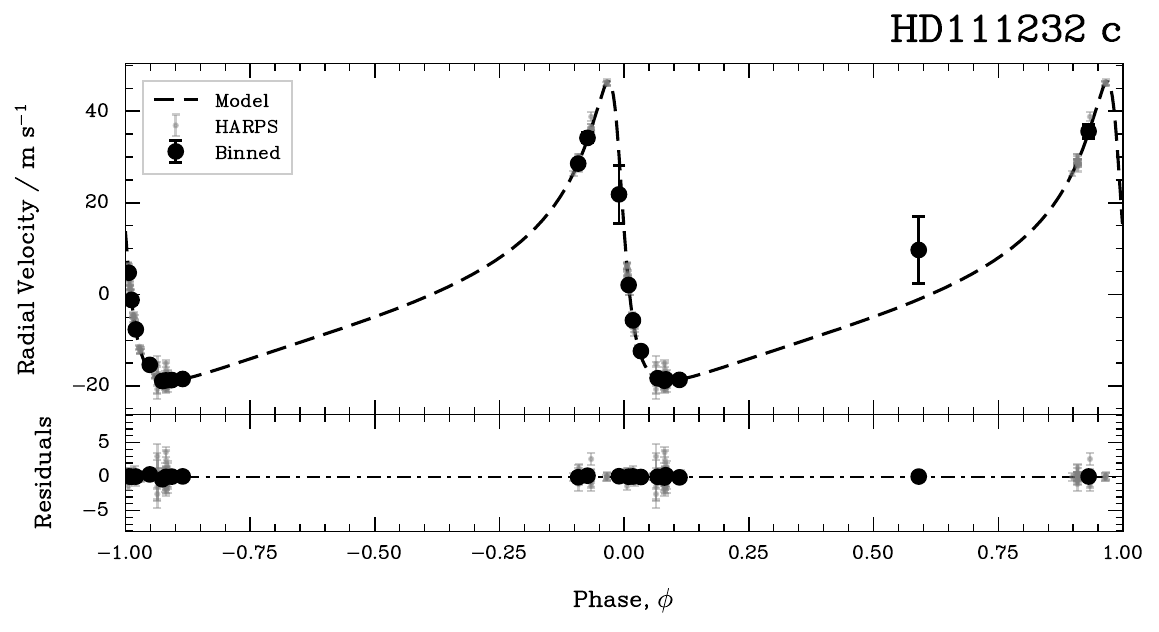}
    \includegraphics[width = 0.32\linewidth]{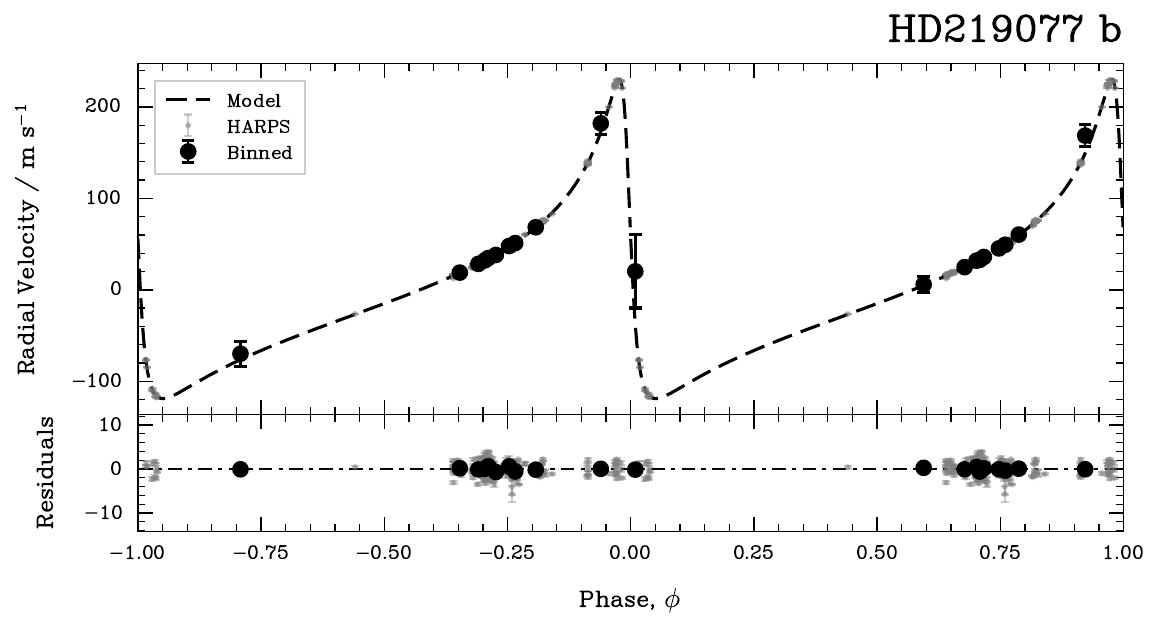}\\
    \includegraphics[width = 0.32\linewidth]{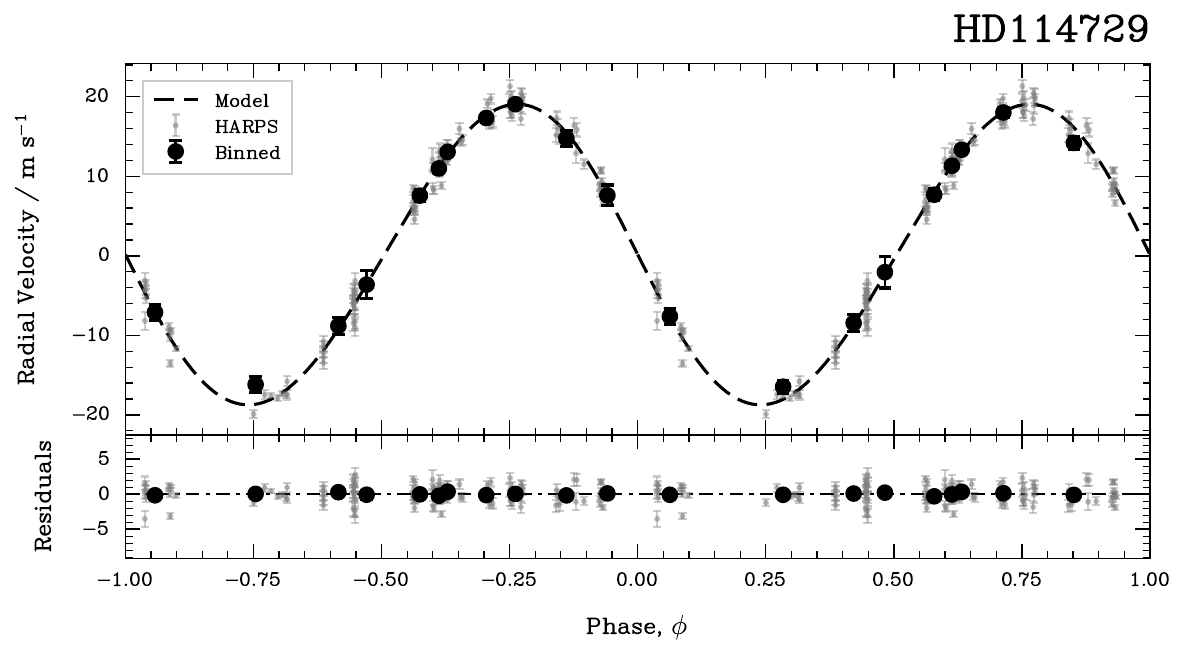}\\
    \includegraphics[width = 0.32\linewidth]{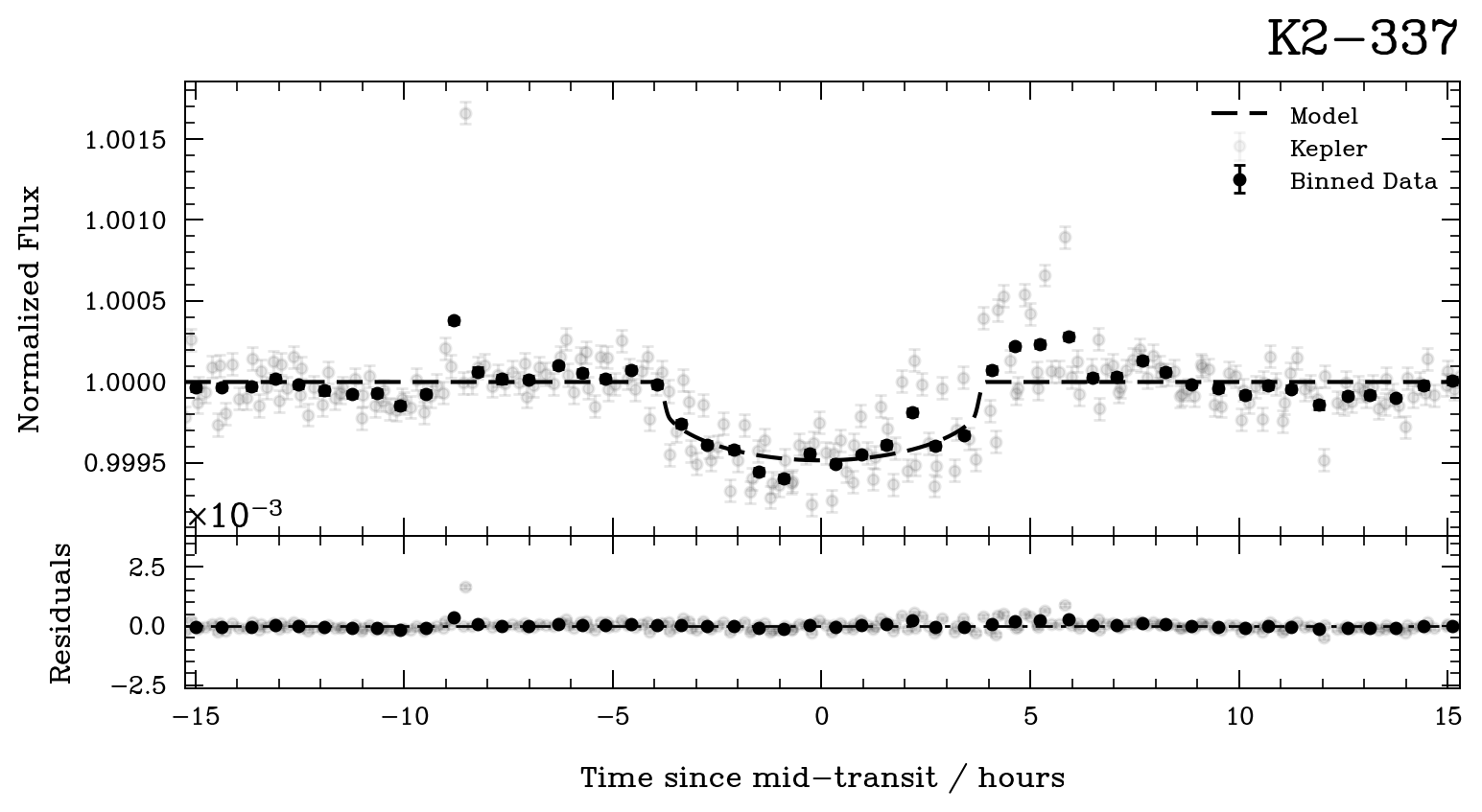}
    \includegraphics[width = 0.32\linewidth]{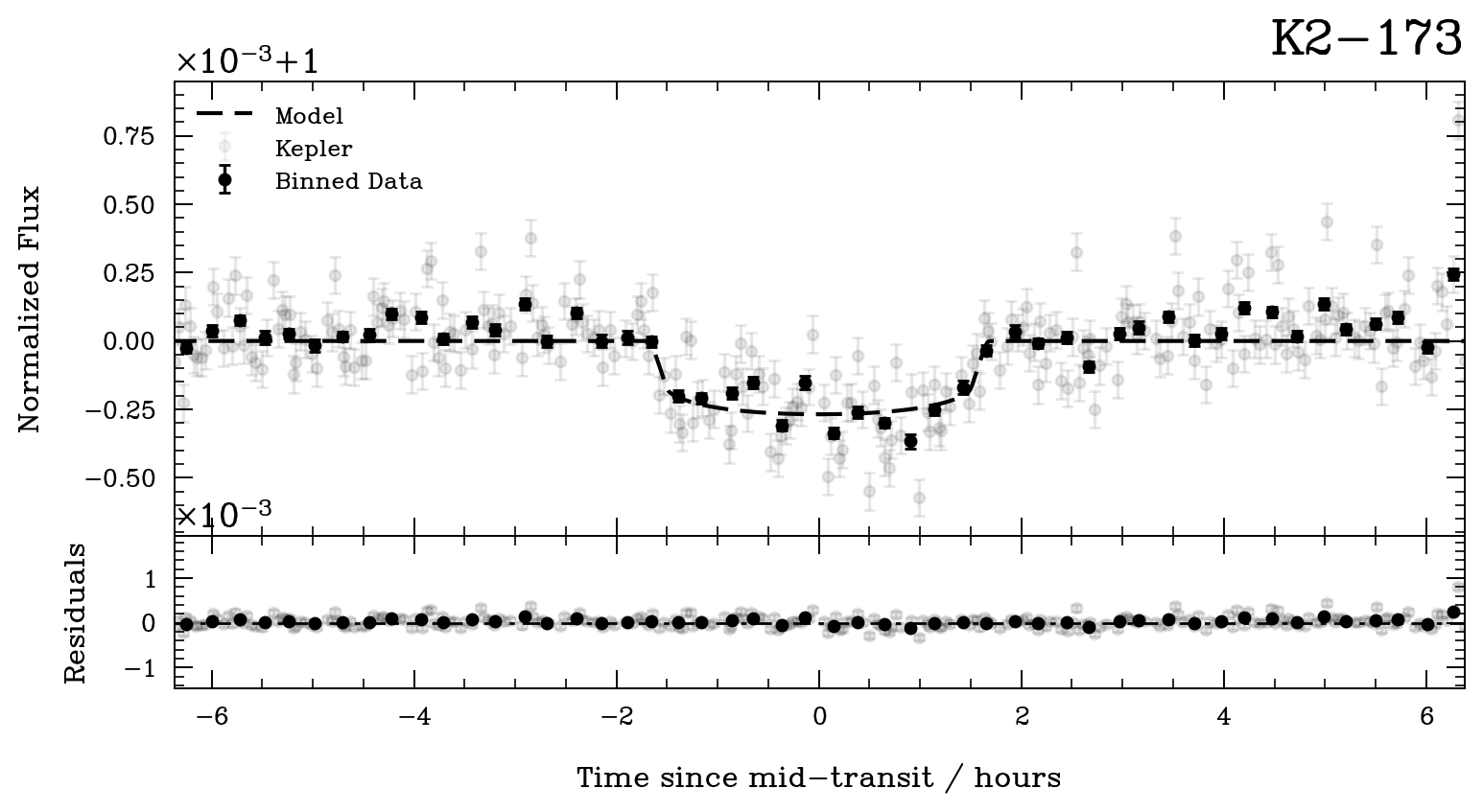}
    \includegraphics[width = 0.32\linewidth]{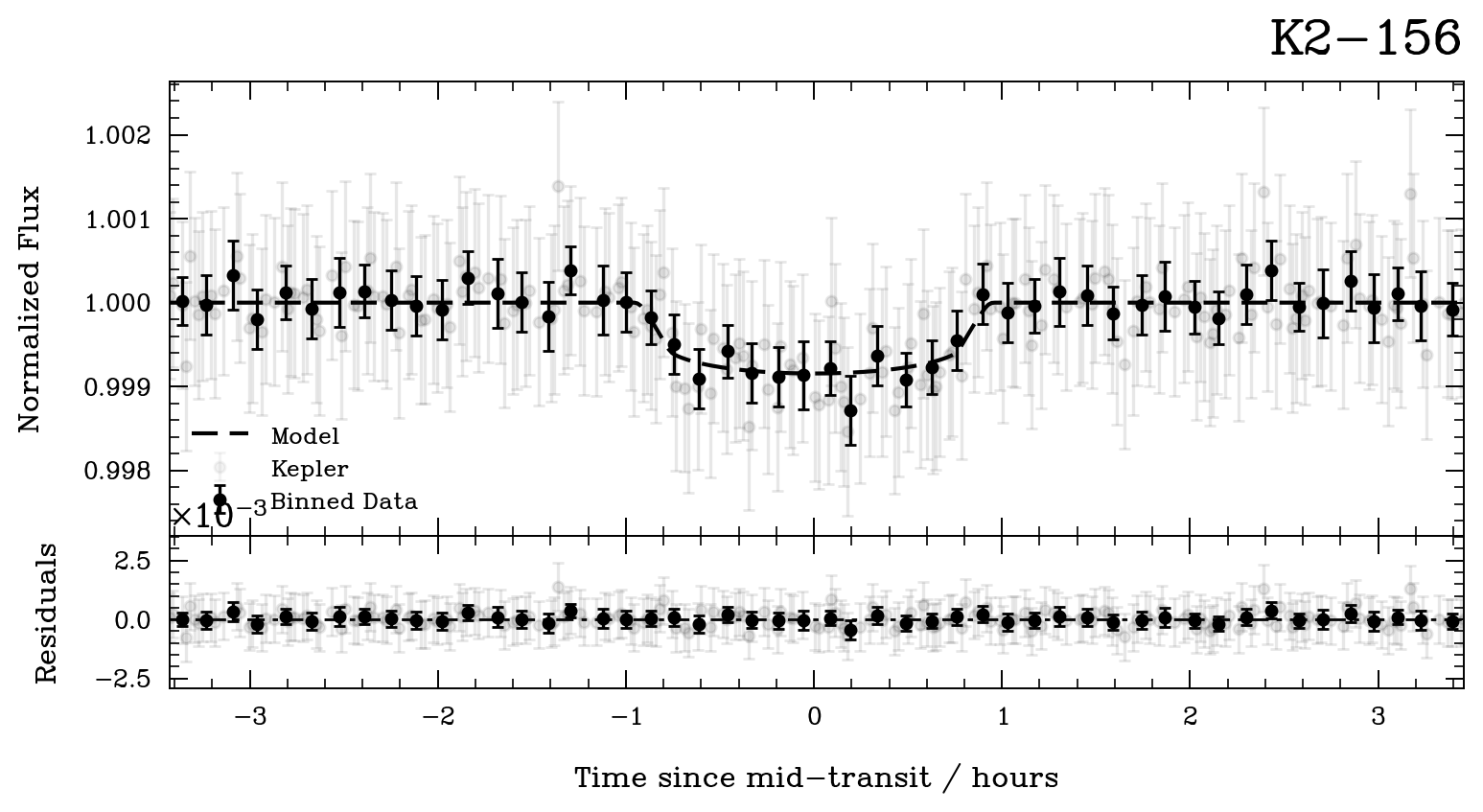}
    \includegraphics[width = 0.32\linewidth]{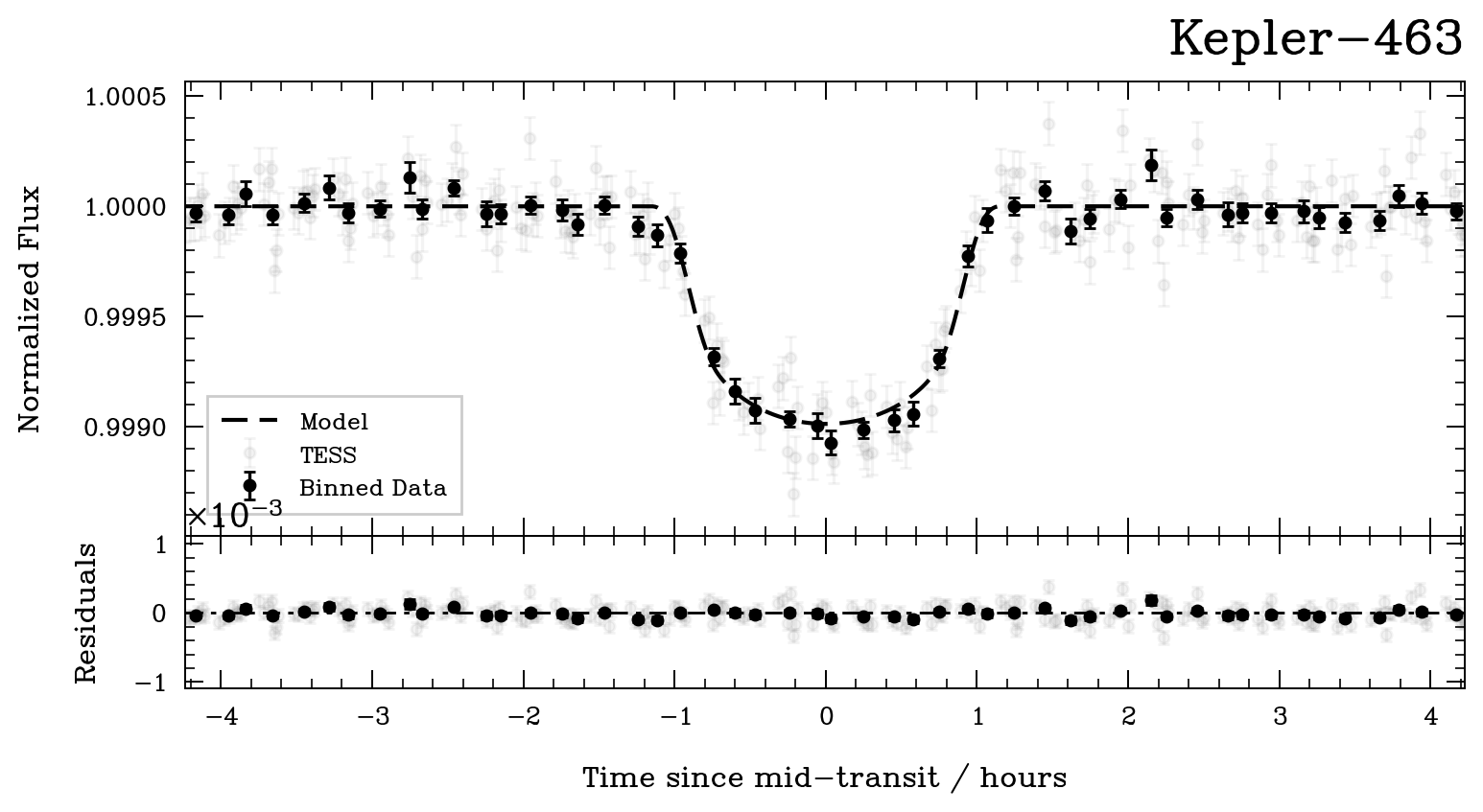}
    \includegraphics[width = 0.32\linewidth]{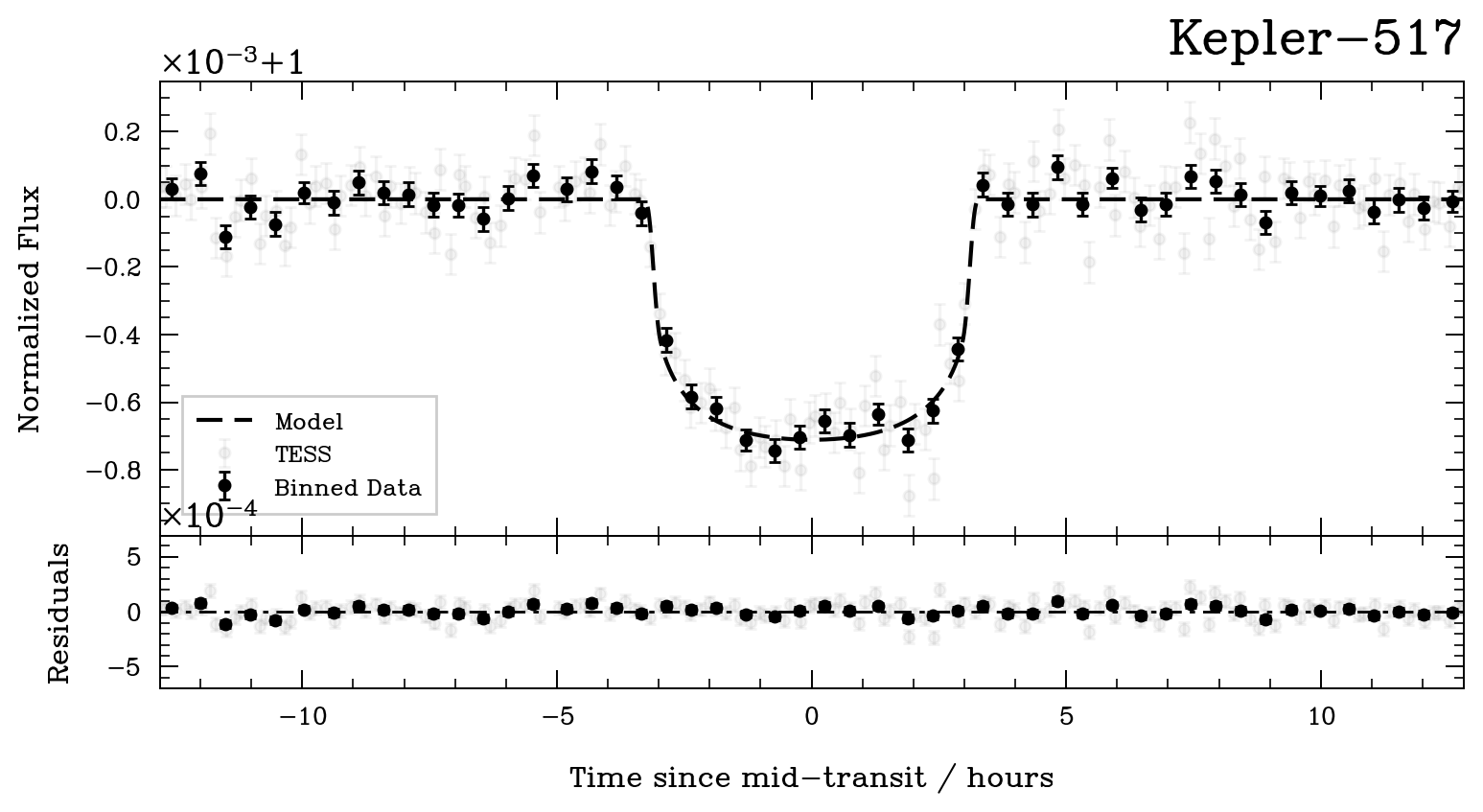}
    \includegraphics[width = 0.32\linewidth]{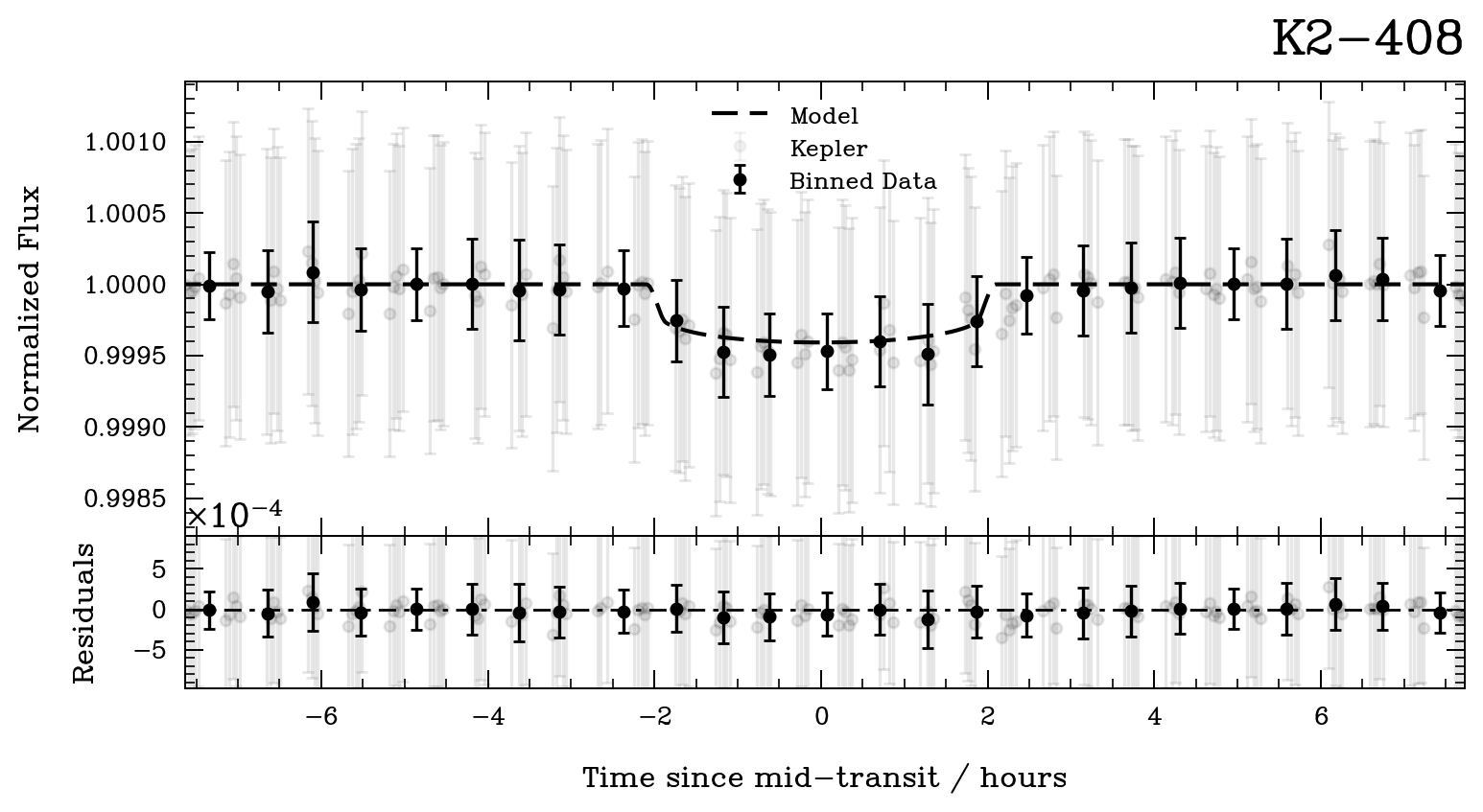} 
    \caption{Phase-folded radial velocity curves for HD6434 b, HD168746 b, HD175607 b, HD181720 b, HD220197 b, HIP109384 b, HD111232 b, HD111232 c, HD219077 b, HD114729 b, and light curves for K2-337 b, K2-173 b, K2-156 c, Kepler-463 b, Kepler-517 b, and K2-408 b, showing the best-fitting model over-plotted. The residuals of the fit are displayed at the bottom of each subplot. {Note: The planet designated here as K2-156 c ($P = 2.638\pm0.001$ days) is a newly identified transit signal, distinct from the validated ultra-short-period planet K2-156 b (P = 0.813 days) reported by \citet{2018AJ....155..136M}, and this signal was not reported in the other discovery or validation papers associated with this system \citep{2018AJ....156...78L, 2021PSJ.....2..152A}. As a transit-only detection, radial velocity follow-up is required to confirm the planetary nature of this candidate and measure its mass.}}
\end{figure*}

\begin{figure*}[!h]
    \centering
    \includegraphics[width = 0.32\linewidth]{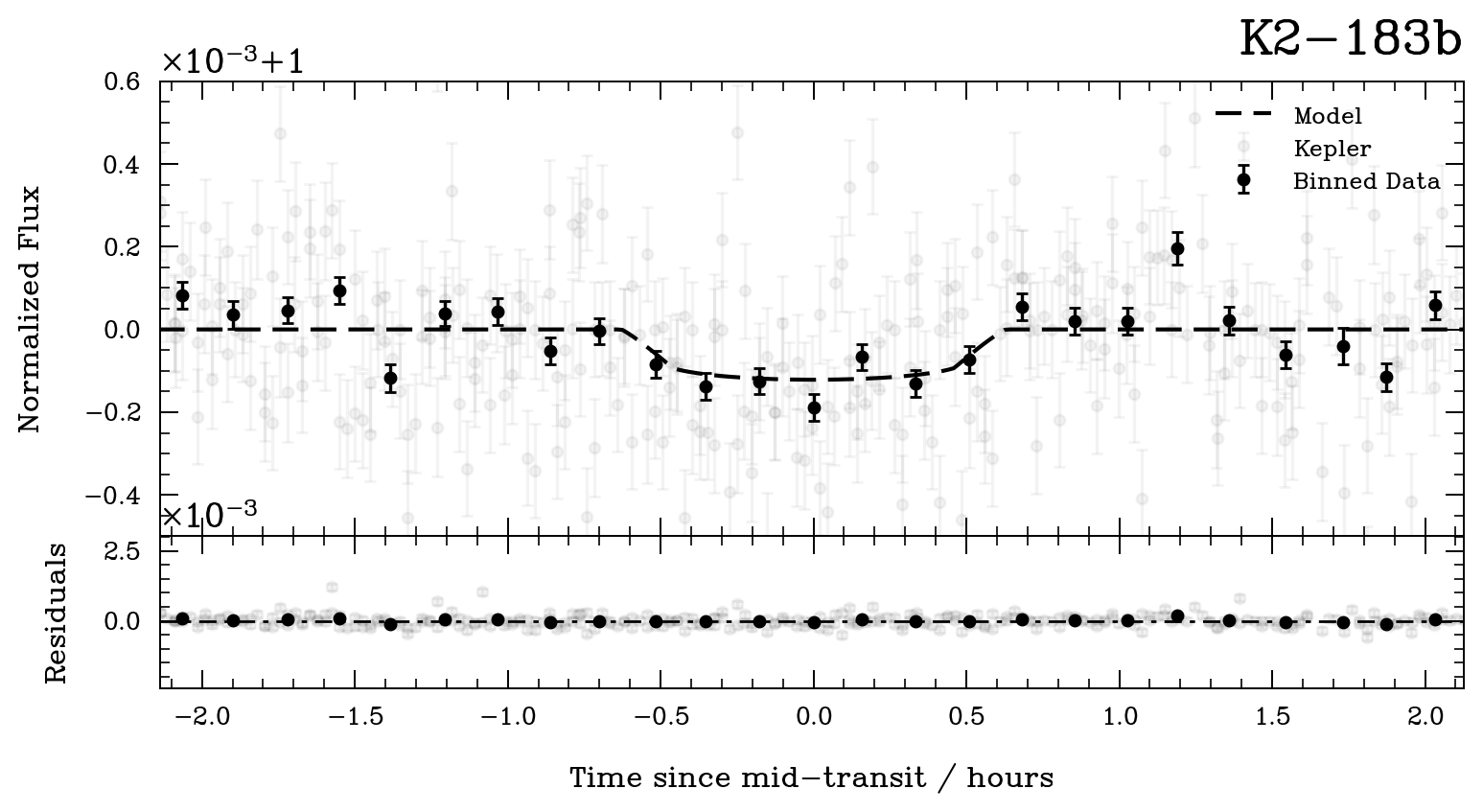}
    \includegraphics[width = 0.32\linewidth]{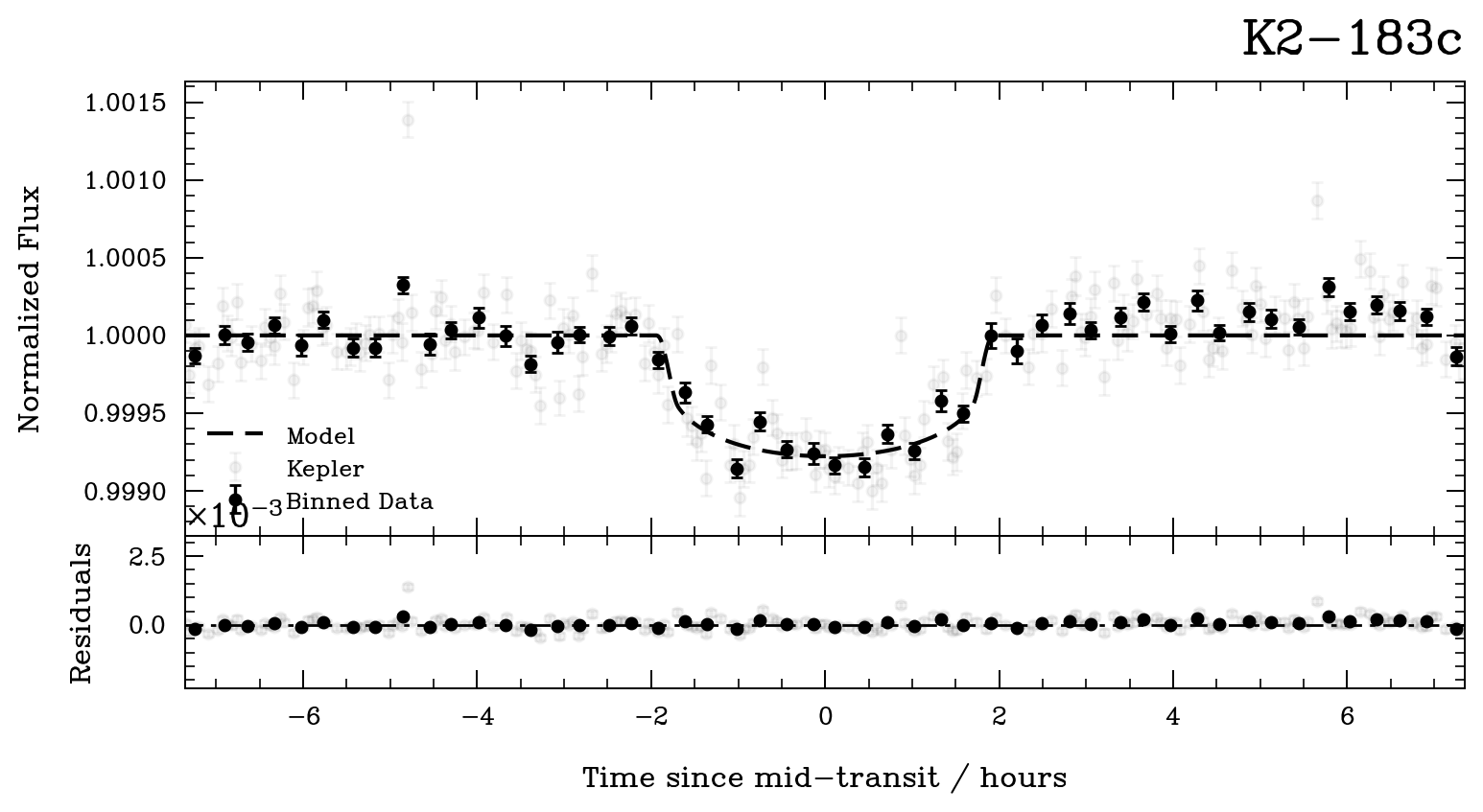}
    \includegraphics[width = 0.32\linewidth]{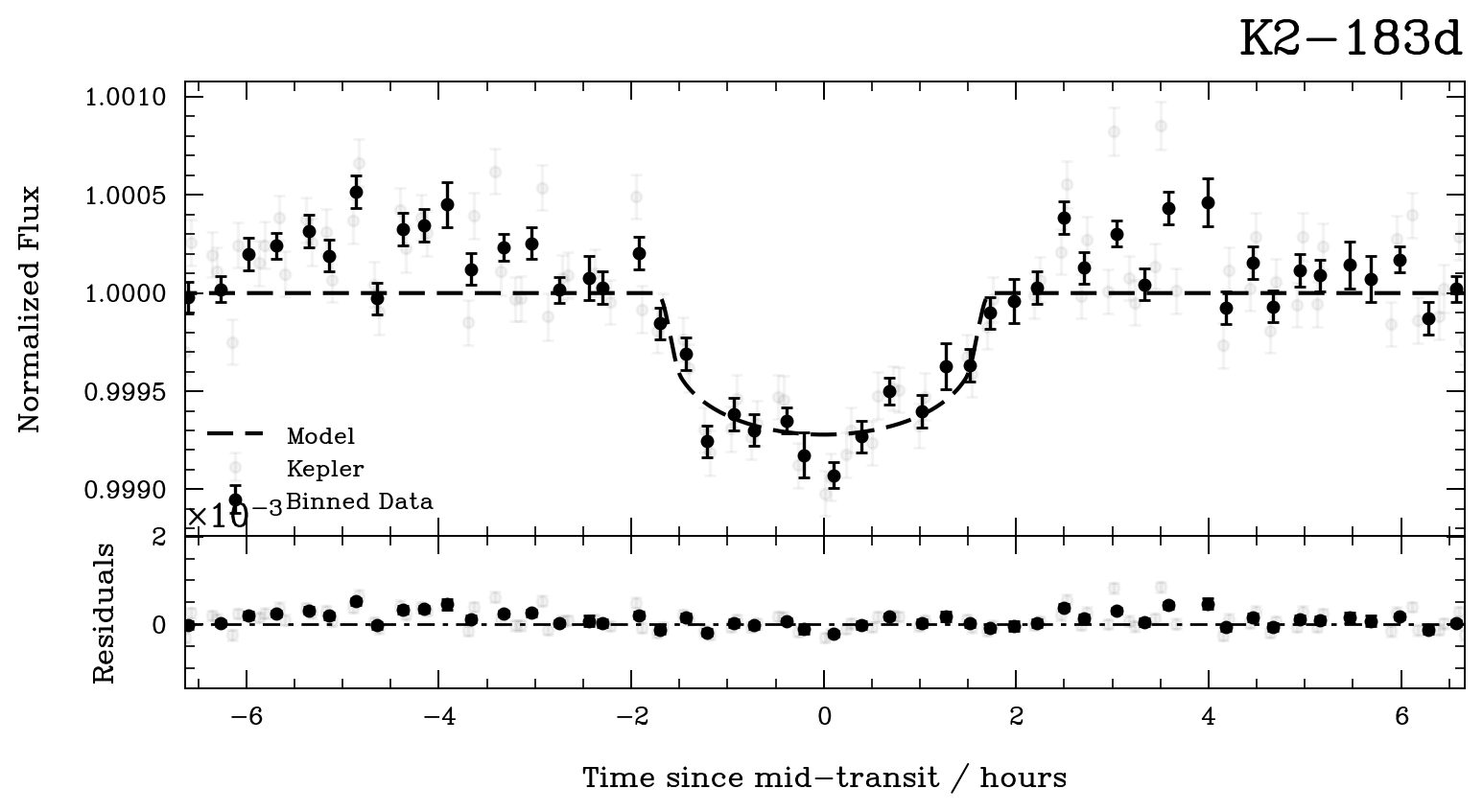}\\
    \includegraphics[width = 0.32\linewidth]{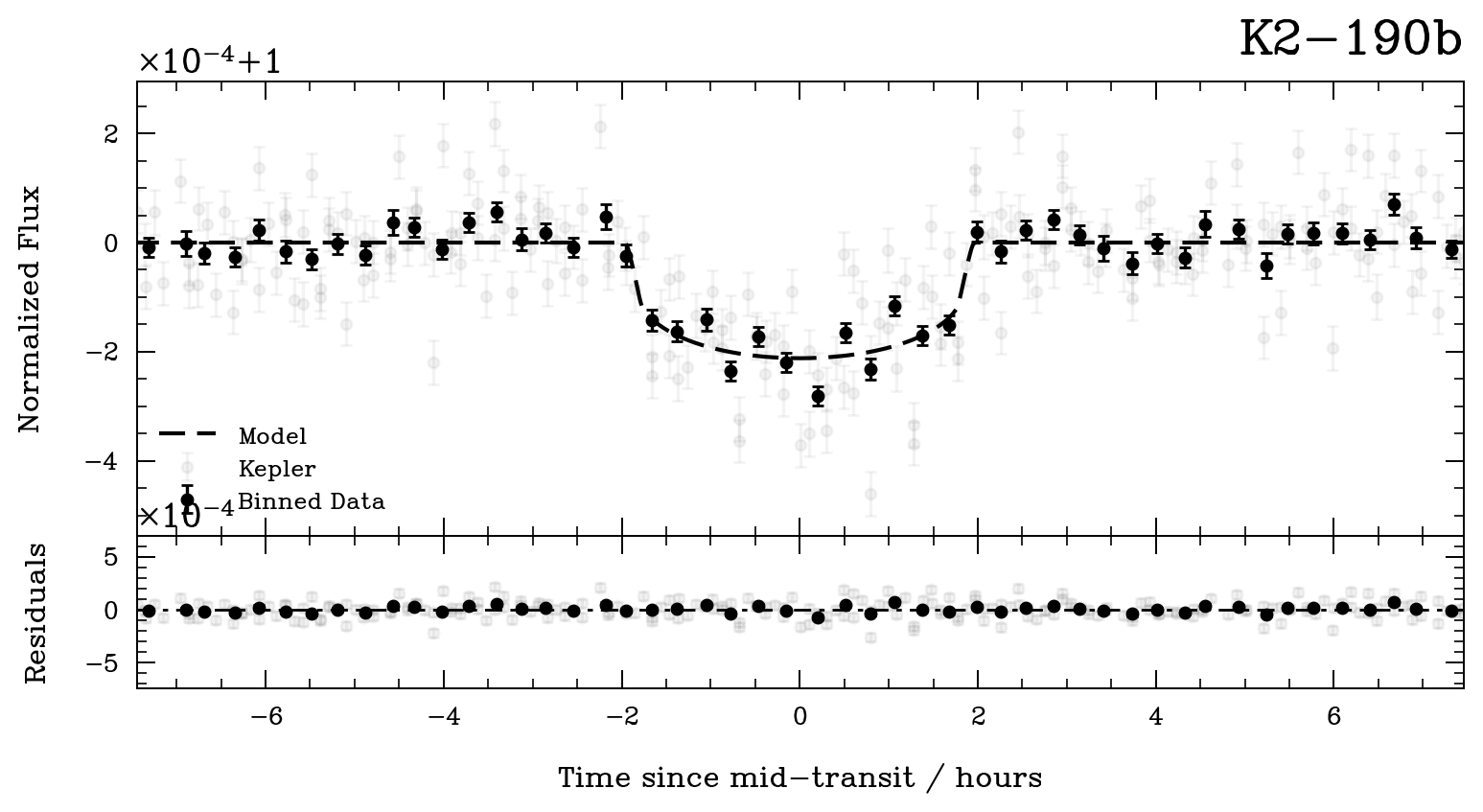}\\
    \includegraphics[width = 0.32\linewidth]{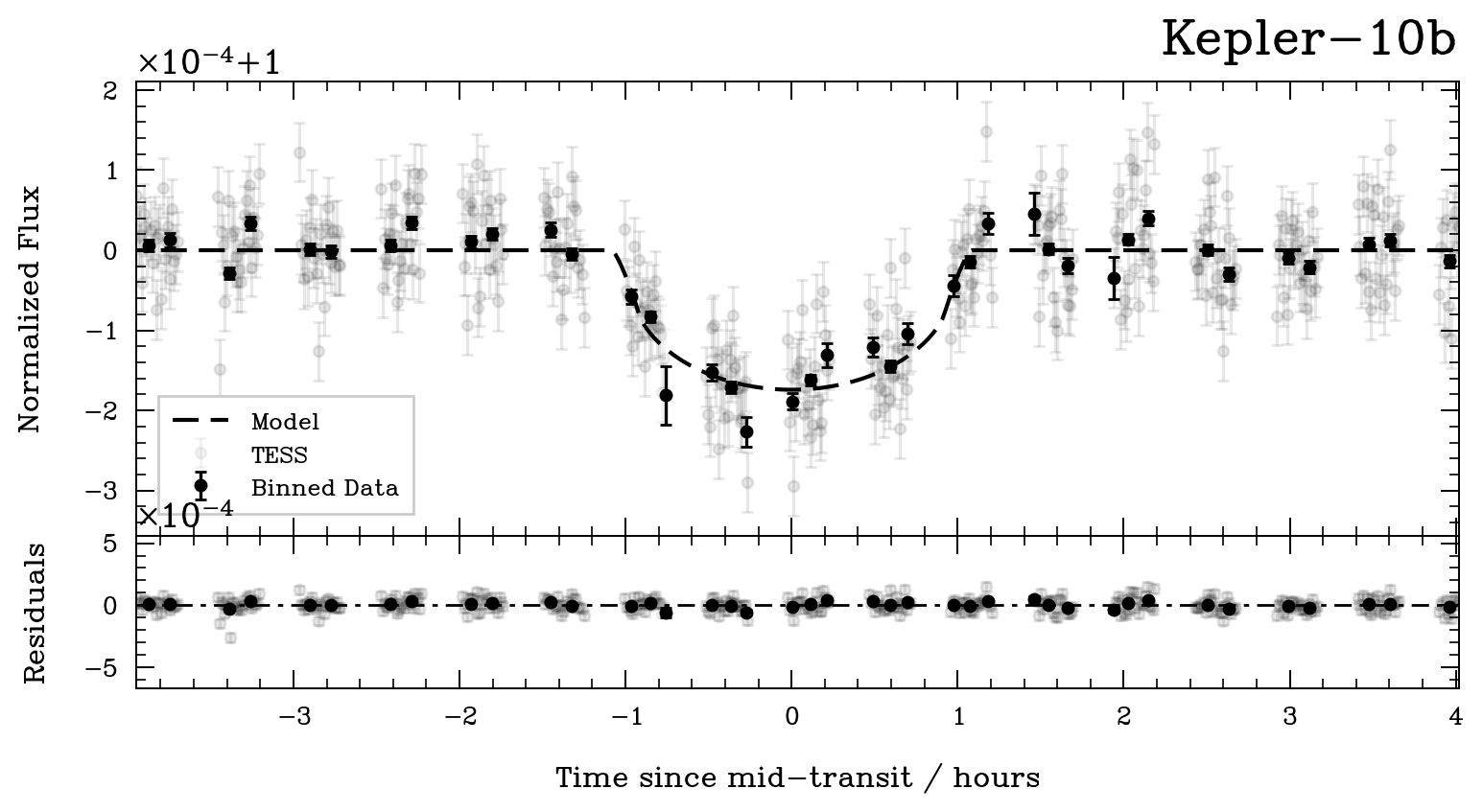}
    \includegraphics[width = 0.32\linewidth]{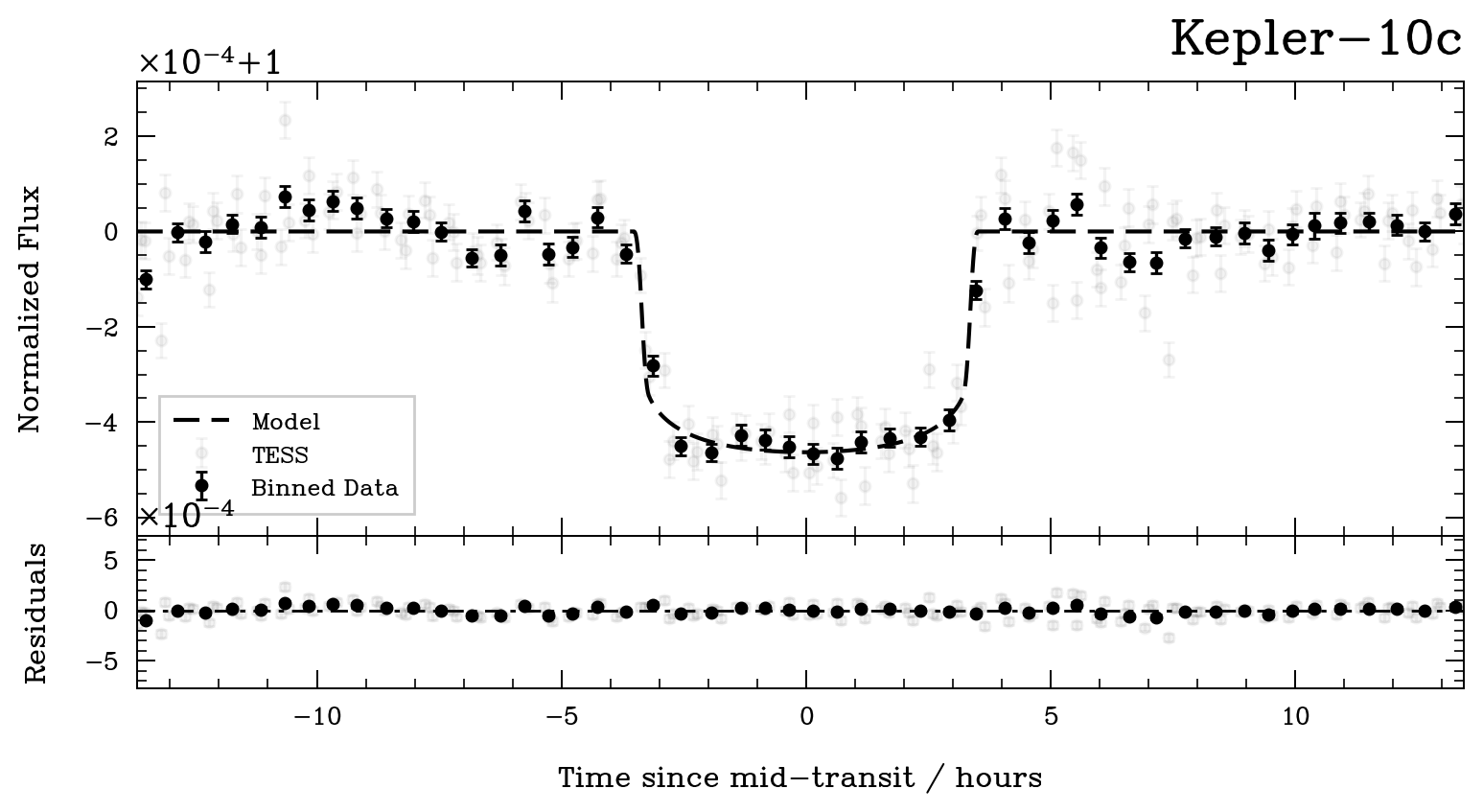}\\
    \includegraphics[width = 0.32\linewidth]{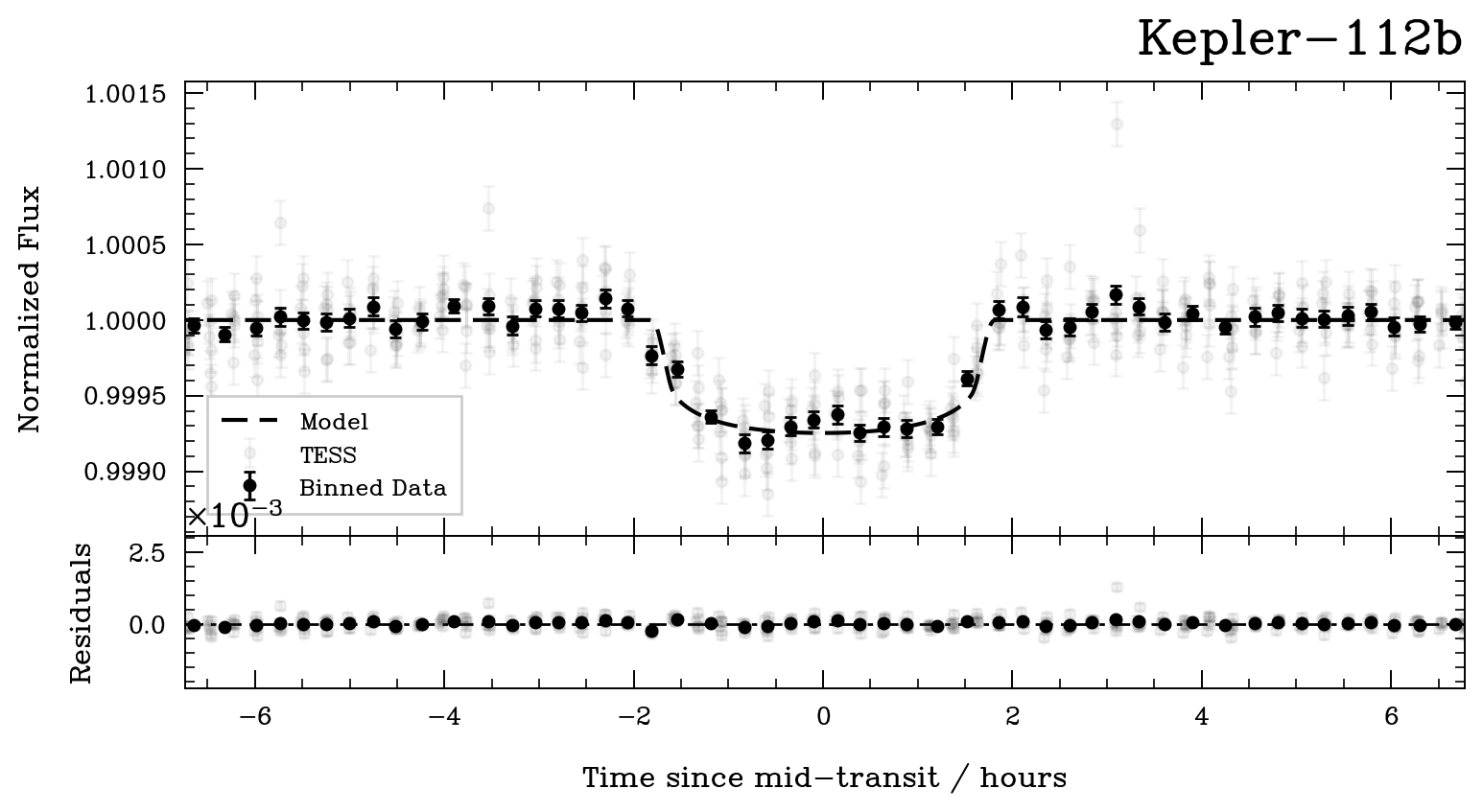}
    \includegraphics[width = 0.32\linewidth]{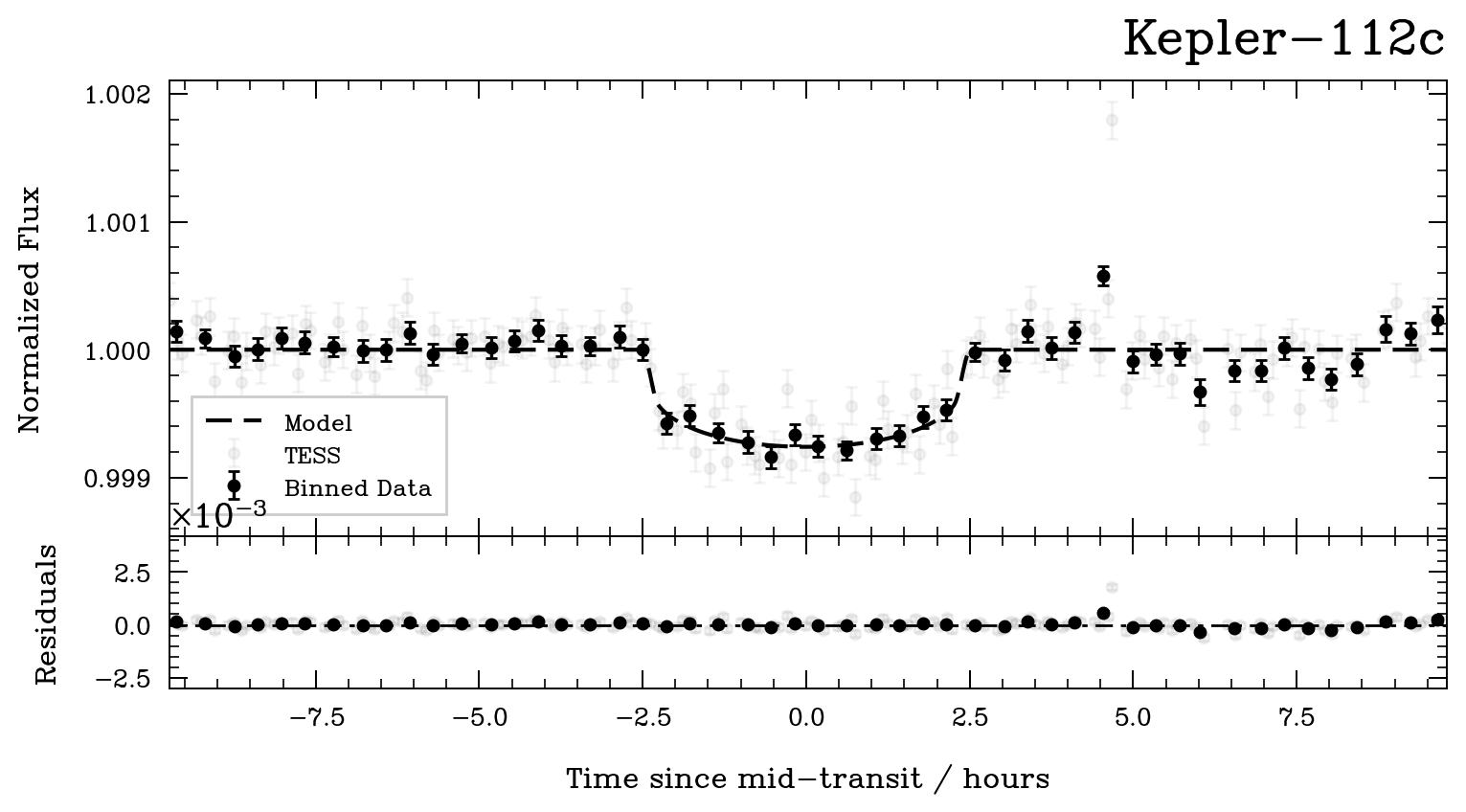}\\
    \includegraphics[width = 0.32\linewidth]{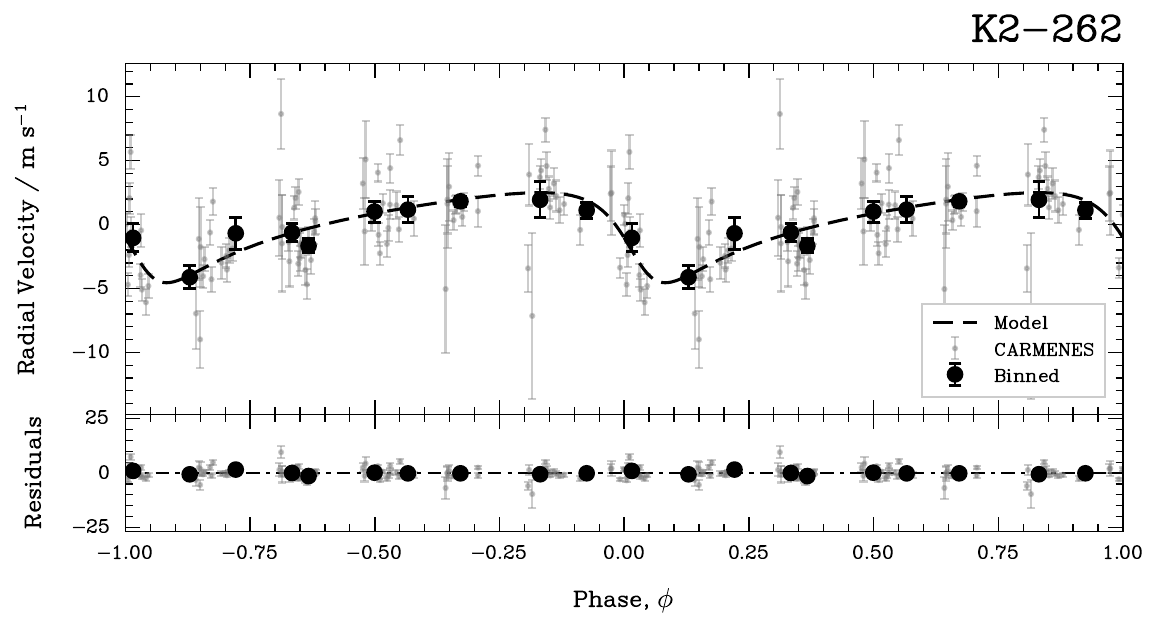}
    \includegraphics[width = 0.32\linewidth]{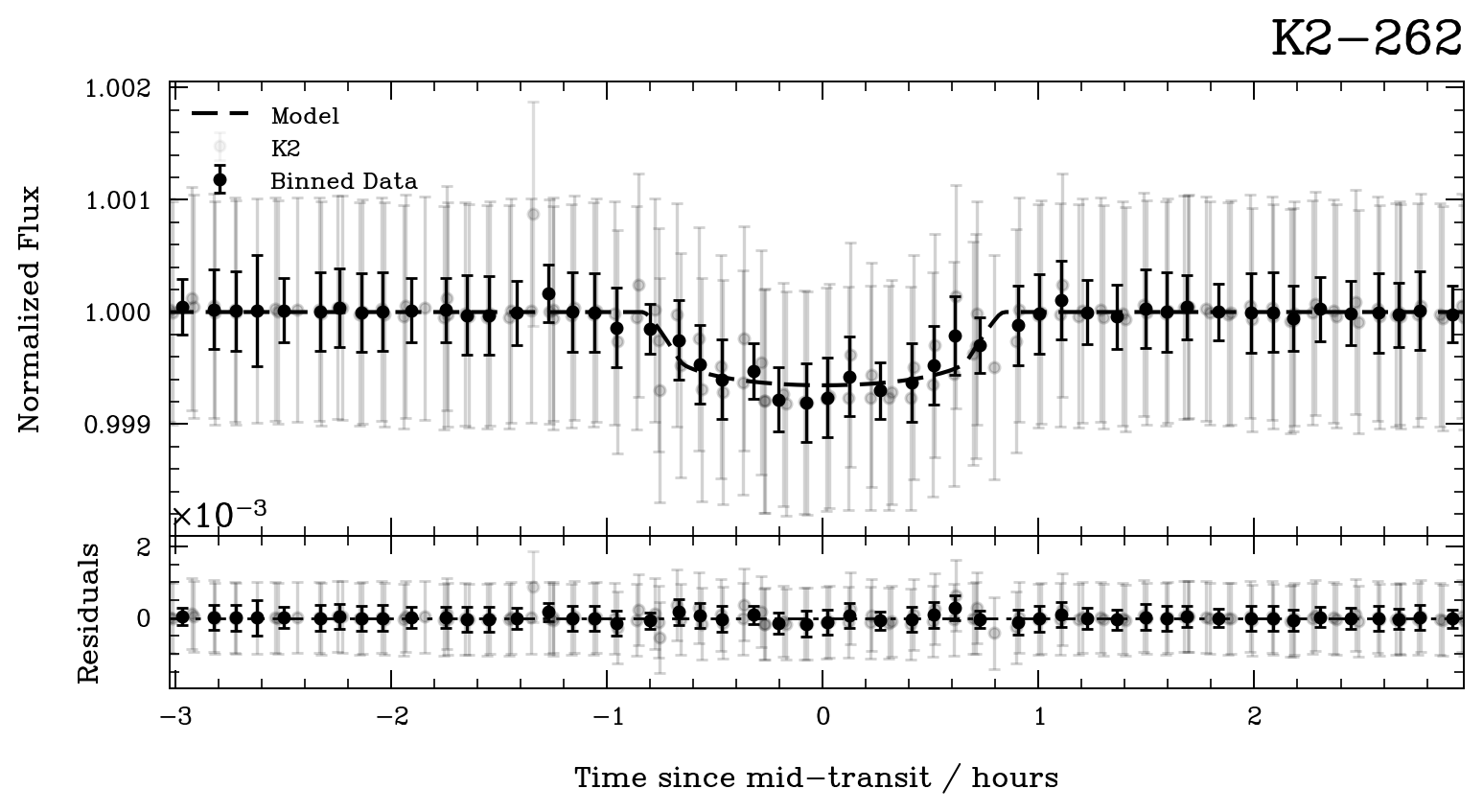}
    \caption{Phase-folded light curves for K2-183 b, K2-183 c, K2-183 d, K2-190 b, Kepler-10 b, Kepler-10 c, Kepler-112 b, Kepler-112 c, and K2-262 b, showing median-normalised photometry with the best-fitting transit model overplotted. The residuals of the fit are displayed at the bottom of each subplot.}
\end{figure*}

\begin{deluxetable*}{lccccccccccc}
\tablecaption{Orbital and physical parameters of thick disc exoplanets from radial velocity-only or transit-only analysis.}
\label{tab:single_method_planets}
\tabletypesize{\scriptsize}
\tablehead{
\colhead{\bf ID} & \colhead{$P$} & \colhead{$M_p$} & \colhead{$R_p$} & \colhead{$\rho_p$} & \colhead{$e$} & \colhead{$T_0$} & \colhead{$T_{\rm eq}$} & \colhead{$\Lambda$} & \colhead{$a/R_*$} & \colhead{$\gamma$} \\
\colhead{} & \colhead{(days)} & \colhead{(M$_{\rm Jup}$)} & \colhead{(R$_{\rm Jup}$)} & \colhead{(g cm$^{-3}$)} & \colhead{} & \colhead{(BJD)} & \colhead{(K)} & \colhead{} & \colhead{} & \colhead{(km s$^{-1}$)}
}
\startdata
HD111232 b & 997.33 $\pm$ 104.38 & 1.43 $\pm$ 0.67 & ... & ... & 0.47 $\pm$ 0.37 & 2354.34 $\pm$ 33.74 & ... & ... & ... & 104.63 $\pm$ 0.07; 104.59 $\pm$ 0.07 \\
HD111232 c & 34014.11 $\pm$ 1389.58 & 2.09 $\pm$ 1.73 & ... & ... & 0.72 $\pm$ 0.37 & 6505.74 $\pm$ 1853.00 & ... & ... & ... & 104.63 $\pm$ 0.07; 104.59 $\pm$ 0.07 \\
HD114729 b & 1141.17 $\pm$ 5.11 & 1.00 $\pm$ 0.11 & ... & ... & 0.04 $\pm$ 0.04 & 7264.77 $\pm$ 14.64 & ... & ... & ... & 64.990 $\pm$ 0.001 \\
HD136352 b & 11.42 $\pm$ 0.12 & 0.15 $\pm$ 0.12 & ... & ... & 0.08 $\pm$ 0.06 & 3507.53 $\pm$ 4.78 & ... & ... & ... & -68.84 $\pm$ 0.19 \\
HD136352 c & 31.55 $\pm$ 0.41 & 0.06 $\pm$ 0.07 & ... & ... & 0.43 $\pm$ 0.18 & 4195.91 $\pm$ 3.67 & ... & ... & ... & -68.85 $\pm$ 0.22 \\
HD136352 d & 107.53 $\pm$ 0.42 & 0.03 $\pm$ 0.04 & ... & ... & 0.94 $\pm$ 0.07 & 4205.91 $\pm$ 3.87 & ... & ... & ... & -68.66 $\pm$ 0.12 \\
HD150433 b & 1003.02 $\pm$ 18.08 & 0.16 $\pm$ 0.02 & ... & ... & 0.08 $\pm$ 0.08 & 5111.75 $\pm$ 35.64 & ... & ... & ... & -40.11 $\pm$ $3.5\times10^{-4}$ \\
HD168746 b & 6.41 $\pm$ 0.84 & 0.05 $\pm$ 0.09 & ... & ... & 0.59 $\pm$ 0.42 & 8285.38 $\pm$ 636.74 & ... & ... & ... & -25.54 $\pm$ 0.01 \\
HD175607 b & 29.01 $\pm$ 0.90 & 0.01 $\pm$ 0.01 & ... & ... & 0.47 $\pm$ 0.36 & 5553.32 $\pm$ 28.90 & ... & ... & ... & -91.890 $\pm$ 0.001 \\
HD181720 b$^\dagger$ & 970.32 $\pm$ 12.01 & 0.41 $\pm$ 0.03 & ... & ... & 0.06 $\pm$ 0.05 & 7303.96 $\pm$ 498.46 & ... & ... & ... & -45.335 $\pm$ 0.003 \\
HD219077 b & 5468.41 $\pm$ 38.96 & 9.68 $\pm$ 0.22 & ... & ... & 0.775 $\pm$ 0.004 & 5997.26 $\pm$ 2.72 & ... & ... & ... & -30.804 $\pm$ 0.005; -31.053 $\pm$ 0.005 \\
HD220197 b$^\dagger$ & 1748.26 $\pm$ 34.06 & 0.12 $\pm$ 0.09 & ... & ... & 0.18 $\pm$ 0.12 & 6619.96 $\pm$ 16.68 & ... & ... & ... & -40.234 $\pm$ 0.001 \\
HD4308 b & 15.62 $\pm$ 0.49 & 0.01 $\pm$ 0.02 & ... & ... & 0.15 $\pm$ 0.32 & 3411.21 $\pm$ 636.49 & ... & ... & ... & 95.255 $\pm$ 0.002 \\
HD4308 c & 850.15 $\pm$ 25.26 & 0.15 $\pm$ 0.07 & ... & ... & 0.47 $\pm$ 0.34 & 3664.39 $\pm$ 199.74 & ... & ... & ... & 95.255 $\pm$ 0.002 \\
HD4308 d & 1504.04 $\pm$ 55.59 & 0.14 $\pm$ 0.08 & ... & ... & 0.37 $\pm$ 0.30 & 4428.15 $\pm$ 223.90 & ... & ... & ... & 95.255 $\pm$ 0.002 \\
HD6434 b$^\dagger$ & 22.01 $\pm$ 0.03 & 0.48 $\pm$ 0.05 & ... & ... & 0.16 $\pm$ 0.05 & 8400.21 $\pm$ 627.38 & ... & ... & ... & 23.10 $\pm$ 0.01 \\
HIP109384 b$^\dagger$ & 499.21 $\pm$ 1.80 & 1.55 $\pm$ 0.08 & ... & ... & 0.55 $\pm$ 0.03 & 6331.61 $\pm$ 2.46 & ... & ... & ... & -63.671 $\pm$ 0.002 \\
K2-156 c$^\ddagger$ & 2.64 $\pm$ $6.7\times10^{-4}$ & ... & 0.16 $\pm$ 0.02 & ... & ... & 2776.72 $\pm$ 0.01 & 1045 $\pm$ 78 & ... & 10.78 $\pm$ 1.46 & ... \\
K2-173 b & 5.87 $\pm$ $4.2\times10^{-4}$ & ... & 0.15 $\pm$ 0.01 & ... & ... & 2240.008 $\pm$ 0.003 & 1075 $\pm$ 59 & ... & 13.54 $\pm$ 1.29 & ... \\
K2-183 b & 0.47 $\pm$ $1.9\times10^{-4}$ & ... & 0.09 $\pm$ 0.02 & ... & ... & 2340.00 $\pm$ 0.01 & 2199 $\pm$ 203 & ... & 3.11 $\pm$ 0.55 & ... \\
K2-183 c & 10.793 $\pm$ 0.001 & ... & 0.21 $\pm$ 0.01 & ... & ... & 2314.779 $\pm$ 0.004 & 837 $\pm$ 59 & ... & 21.48 $\pm$ 2.77 & ... \\
K2-183 d & 22.627 $\pm$ 0.003 & ... & 0.21 $\pm$ 0.02 & ... & ... & 2311.16 $\pm$ 0.01 & 571 $\pm$ 44 & ... & 46.16 $\pm$ 6.40 & ... \\
K2-190 b & 10.101 $\pm$ 0.001 & ... & 0.12 $\pm$ 0.09 & ... & ... & 2391.76 $\pm$ 0.01 & 904 $\pm$ 57 & ... & 19.62 $\pm$ 2.29 & ... \\
K2-337 b & 16.29 $\pm$ $4.3\times10^{-4}$ & ... & 0.25 $\pm$ 0.01 & ... & ... & 2312.934 $\pm$ 0.003 & 1042 $\pm$ 40 & ... & 15.96 $\pm$ 1.06 & ... \\
K2-408 b & 20.98 $\pm$ 0.01 & ... & 0.14 $\pm$ 0.03 & ... & ... & 3158.52 $\pm$ 0.03 & 620 $\pm$ 84 & ... & 36.79 $\pm$ 9.19 & ... \\
Kepler-10 b & 0.84 $\pm$ $1.6\times10^{-4}$ & ... & 0.12 $\pm$ 0.01 & ... & ... & 170.11 $\pm$ 0.01 & 2284 $\pm$ 130 & ... & 3.12 $\pm$ 0.33 & ... \\
Kepler-10 c & 45.297 $\pm$ 0.002 & ... & 0.21 $\pm$ 0.01 & ... & ... & 183.969 $\pm$ 0.004 & 576 $\pm$ 30 & ... & 49.12 $\pm$ 4.73 & ... \\
Kepler-112 b & 8.41 $\pm$ $2.6\times10^{-4}$ & ... & 0.21 $\pm$ 0.05 & ... & ... & 134.010 $\pm$ 0.002 & 938 $\pm$ 143 & ... & 17.67 $\pm$ 4.10 & ... \\
Kepler-112 c & 28.575 $\pm$ 0.002 & ... & 0.20 $\pm$ 0.05 & ... & ... & 143.831 $\pm$ 0.005 & 598 $\pm$ 36 & ... & 43.39 $\pm$ 4.36 & ... \\
Kepler-463 b & 8.98 $\pm$ $1.6\times10^{-4}$ & ... & 0.30 $\pm$ 0.03 & ... & ... & 136.649 $\pm$ 0.001 & 968 $\pm$ 136 & ... & 17.06 $\pm$ 5.41 & ... \\
{Kepler-517 b} & 60.926 $\pm$ 0.002 & ... & 0.24 $\pm$ 0.03 & ... & ... & 173.371 $\pm$ 0.002 & 479 $\pm$ 26 & ... & 70.96 $\pm$ 6.81 & ... \\
TOI-126 b & 3.04 $\pm$ $2.2\times10^{-4}$ & ... & 1.24 $\pm$ 0.07 & ... & ... & 1326.73 $\pm$ $9.6\times10^{-4}$ & 1435 $\pm$ 65 & ... & 8.69 $\pm$ 0.67 & ... \\
\enddata
\tablecomments{Columns indicate: $P$ = Orbital period in days, $M_p$ = Companion's mass in Jupiter masses (for radial velocity-only systems this represents the minimum mass, $M_p \sin i$), $R_p$ = Companion's radius in Jupiter radii, $\rho_p$ = Companion's bulk density from mass and radius in g cm$^{-3}$, $e$ = Eccentricity, $T_0$ = Transit midpoint or time of periastron passage in Barycentric Julian Date (BJD$-$2450000), $T_{\rm eq}$ = Equilibrium temperature in Kelvin, $\Lambda$ = Jeans escape parameter that indicates atmospheric retention capability, defined by $\Lambda\equiv(G M_p m_p)(k_B T_{\rm eq} R_p)$, where $G$ is the gravitational constant, $M_p$ the companion's mass, $m_p$ the mass of the escaping particle (typically the proton mass when assessing atmospheric hydrogen escape), $k_B$ Boltzmann's constant, $T_{\rm eq}$ the companion's equilibrium temperature, and $R_p$ the companion's radius, $a/R_\star$ = Scaled semi-major axis, $\gamma$ = Systemic radial velocity for the corresponding instrument(s) in km s$^{-1}$ (multiple values separated by semicolons indicate measurements from different instruments). Values marked with '...' indicate parameters that were not measured or are not applicable for that particular detection method. $^\dagger$See Appendix \ref{app:gaussian_process} for a discussion on the use of GP terms in the radial velocity analysis for these systems. {$\ddagger$See Figure \ref{fig:XXX}}}
\end{deluxetable*}

\begin{deluxetable*}{lccccccccccc}
\tablecaption{Orbital and physical parameters of thick disc exoplanets from joint radial velocity and transit analysis.}
\label{tab:rv_transit_planets}
\tabletypesize{\scriptsize}
\tablehead{
\colhead{\bf ID} & \colhead{$P$} & \colhead{$M_p$} & \colhead{$R_p$} & \colhead{$\rho_p$} & \colhead{$e$} & \colhead{$T_0$} & \colhead{$T_{\rm eq}$} & \colhead{$\Lambda$} & \colhead{$a/R_*$} & \colhead{$\gamma$} \\
\colhead{} & \colhead{(days)} & \colhead{(M$_{\rm Jup}$)} & \colhead{(R$_{\rm Jup}$)} & \colhead{(g cm$^{-3}$)} & \colhead{} & \colhead{(BJD)} & \colhead{(K)} & \colhead{} & \colhead{} & \colhead{(km s$^{-1}$)}
}
\startdata
K2-262 b & 6.00 $\pm$ $3.1\times10^{-4}$ & 0.021 $\pm$ 0.004 & 0.16 $\pm$ 0.02 & 6.51 $\pm$ 2.72 & 0.48 $\pm$ 0.18 & 3352.348 $\pm$ 0.005 & 792 $\pm$ 77 & 35.38 & 17.66 $\pm$ 3.18 & $...$ \\
TOI-1926 b & 1.338 $\pm$ 0.001 & 0.06 $\pm$ 0.01 & 0.23 $\pm$ 0.03 & 5.63 $\pm$ 2.95 & 0.21 $\pm$ 0.18 & 2175.37 $\pm$ 0.01 & 1694 $\pm$ 249 & 31.96 & 4.60 $\pm$ 1.34 & 111.203 $\pm$ 0.002 \\
TOI-1927 b & 4.11 $\pm$ $2.0\times10^{-4}$ & 0.43 $\pm$ 0.11 & 1.38 $\pm$ 0.14 & 0.19 $\pm$ 0.09 & ... & 3263.24 $\pm$ $8.9\times10^{-4}$ & 1111 $\pm$ 35 & 58.71 & 11.45 $\pm$ 0.66 & 68.71 $\pm$ 0.02; 68.73 $\pm$ 0.03 \\
TOI-2643 b & 9.17 $\pm$ 0.01 & 0.17 $\pm$ 0.16 & 1.07 $\pm$ 0.07 & 0.17 $\pm$ 0.17 & 0.29 $\pm$ 0.32 & 2287.09 $\pm$ 0.01 & 1117 $\pm$ 219 & 31.71 & 9.71 $\pm$ 4.59 & 10.37 $\pm$ 0.04; 10.36 $\pm$ 0.04 \\
\enddata
\tablecomments{Columns indicate: $P$ = Orbital period in days, $M_p$ = Companion's mass in Jupiter masses, $R_p$ = Companion's radius in Jupiter radii, $\rho_p$ = Companion's bulk density from mass and radius in g cm$^{-3}$, $e$ = Orbital eccentricity, $T_0$ = Transit midpoint or time of periastron passage in Barycentric Julian Date (BJD$-$2450000), $T_{\rm eq}$ = Equilibrium temperature in Kelvin, $\Lambda$ = Jeans escape parameter that indicates atmospheric retention capability, defined by $\Lambda\equiv(G M_p m_p)(k_B T_{\rm eq} R_p)$, where $G$ is the gravitational constant, $M_p$ the companion's mass, $m_p$ the mass of the escaping particle (typically the proton mass when assessing atmospheric hydrogen escape), $k_B$ Boltzmann's constant, $T_{\rm eq}$ the companion's equilibrium temperature, and $R_p$ the companion's radius, $a/R_\star$ = Scaled semi-major axis, $\gamma$ = Systemic radial velocity for the corresponding instrument(s) in km s$^{-1}$ (multiple values separated by semicolons indicate measurements from different instruments). Values marked with '...' indicate parameters that were not measured or are not applicable for that particular detection method.}
\end{deluxetable*}

\begin{deluxetable*}{lccccc}
\tablecaption{Priors for HD6434, HD168746, HD150433, HD181720, HD136352, HD175607, HD220197, and HIP109384 systems. Radial velocity priors are listed by instrument. In all cases, the parameters $\sqrt{e}\cos\omega$ and $\sqrt{e}\sin\omega$ are sampled uniformly over the interval $[-1, +1]$. $^\alpha$: HARPS; $^\gamma$: SOPHIE; $^\gamma$: HARPS-N; $^\delta$: ESPRESSO and HARPS.}
\tablehead{\colhead{ID} & \colhead{$T_0$} & \colhead{$P$} & \colhead{$K$} & \colhead{RV}}
\startdata
HD4308 b$^\alpha$ & $\mathcal{U}[14,16]$ & $\mathcal{U}[2390,4390]$ & $\mathcal{U}[0,0.05]$ & $\mathcal{U}[94.24, 96.27]$ \\
HD4308 c$^\alpha$ & $\mathcal{U}[800,900]$ & $\mathcal{U}[3190,5190]$ & $\mathcal{U}[0,0.05]$ & $\mathcal{U}[94.24, 96.27]$ \\
HD4308 d$^\alpha$ & $\mathcal{U}[1400,1600]$ & $\mathcal{U}[3390,5390]$ & $\mathcal{U}[0,0.05]$ & $\mathcal{U}[94.24, 96.27]$ \\
HD6434 b$^\alpha$ & $\mathcal{U}[7325,9325]$ & $\mathcal{U}[18,26]$ & $\mathcal{U}[0,5]$ & $\mathcal{U}[22.06, 24.13]$ \\
HD114729 b$^\alpha$ & $\mathcal{U}[5868,7868]$ & $\mathcal{U}[915,1373]$ & $\mathcal{U}[0,5]$ & $\mathcal{U}[63.96, 66.01]$ \\
HD168746 b$^\alpha$ & $\mathcal{U}[7198,9198]$ & $\mathcal{U}[5,8]$ & $\mathcal{U}[0,50]$ & $\mathcal{U}[-26.57, -24.51]$ \\
HD150433 b$^\alpha$ & $\mathcal{U}[4821,6821]$ & $\mathcal{U}[900,1200]$ & $\mathcal{U}[0,3.68]$ & $\mathcal{U}[-41.12, -39.11]$ \\
HD181720 b$^\alpha$ & $\mathcal{U}[6512,8512]$ & $\mathcal{U}[807,1210]$ & $\mathcal{U}[0,3.69]$ & $\mathcal{U}[-46.35, -44.32]$ \\
HD136352 b$^\alpha$ & $\mathcal{U}[3500,3550]$ & $\mathcal{U}[11,12]$ & $\mathcal{U}[0,10]$ & $\mathcal{U}[-69.72, -67.70]$ \\
HD136352 c$^\alpha$ & $\mathcal{U}[4100,4200]$ & $\mathcal{U}[29,32]$ & $\mathcal{U}[0,10]$ & $\mathcal{U}[-69.72, -67.70]$ \\
HD136352 d$^\alpha$ & $\mathcal{U}[4200,4300]$ & $\mathcal{U}[100,110]$ & $\mathcal{U}[0,10]$ & $\mathcal{U}[-69.72, -67.70]$ \\
HD175607 b$^\alpha$ & $\mathcal{U}[5500,5600]$ & $\mathcal{U}[27,31]$ & $\mathcal{U}[0,10]$ & $\mathcal{U}[-92.90, -90.88]$ \\
HIP109384 b$^\beta$ & $\mathcal{U}[6300,6400]$ & $\mathcal{U}[490,510]$ & $\mathcal{U}[0,10]$ & $\mathcal{U}[-64.74, -62.63]$ \\
HD220197 b$^\gamma$ & $\mathcal{U}[6600,6650]$ & $\mathcal{U}[1700,1800]$ & $\mathcal{U}[0,10]$ & $\mathcal{U}[-41.24, -39.22]$ \\
HD219077 b$^\gamma$ & $\mathcal{U}[4000,6000]$ & $\mathcal{U}[5300,5500]$ & $\mathcal{U}[0,0.50]$ & $\mathcal{U}[-31.93, -29.57]$; $\mathcal{U}[-31.91, -29.88]$ \\
HD111232 b$^\delta$ & $\mathcal{U}[900,1200]$ & $\mathcal{U}[2300,2400]$ & $\mathcal{U}[0,0.05]$ & $\mathcal{U}[103.68, 105.75]$; $\mathcal{U}[103.38, 105.78]$ \\
HD111232 c$^\delta$ & $\mathcal{U}[32000,36000]$ & $\mathcal{U}[2000,8000]$ & $\mathcal{U}[0,0.05]$ & $\mathcal{U}[103.68, 105.75]$; $\mathcal{U}[103.38, 105.78]$ \\
\enddata
\end{deluxetable*}

\begin{deluxetable*}{lcccc}
\tablecaption{Priors for Kepler-10, Kepler-112, Kepler-1898, Kepler-463, Kepler-517, TOI-1258, TOI-2847, TOI-329, TOI-6707, K2-156, K2-173, K2-183, K2-190, K2-337, K2-408, and TOI-126 systems. In all cases, where applicable, the impact parameter $b$ is sampled within the interval $[0, 1]$, and the limb-darkening coefficients $q_1$ and $q_2$ are sampled within $[0, 1]$.}
\tablehead{\colhead{ID} & \colhead{$T_0$} & \colhead{$P$} & \colhead{$a$} & \colhead{$R_p$}}
\startdata
K2-156 c & $\mathcal{U}[2776.60, 2776.80]$ & $\mathcal{U}[2.63, 2.65]$ & $\mathcal{U}[0.01, 5]$ & $\mathcal{U}[0,0.40]$ \\
K2-173 b & $\mathcal{U}[2239.90, 2240.10]$ & $\mathcal{U}[5.86, 5.88]$ & $\mathcal{U}[0.01, 5]$ & $\mathcal{U}[0,0.40]$ \\
K2-183 b & $\mathcal{U}[2340,2340.20]$ & $\mathcal{U}[0.45, 0.47]$ & $\mathcal{U}[0.01, 5]$ & $\mathcal{U}[0,0.40]$ \\
K2-183 c & $\mathcal{U}[2314.68, 2314.88]$ & $\mathcal{U}[10.78, 10.80]$ & $\mathcal{U}[0.01, 5]$ & $\mathcal{U}[0,0.40]$ \\
K2-183 d & $\mathcal{U}[2311.05, 2311.25]$ & $\mathcal{U}[22.61, 22.63]$ & $\mathcal{U}[0.01, 5]$ & $\mathcal{U}[0,0.40]$ \\
K2-190 b & $\mathcal{U}[2391.66, 2391.86]$ & $\mathcal{U}[10.09, 10.11]$ & $\mathcal{U}[0.01, 5]$ & $\mathcal{U}[0,0.10]$ \\
K2-337 b & $\mathcal{U}[2312.82, 2313.02]$ & $\mathcal{U}[16.29, 16.31]$ & $\mathcal{U}[0.01, 5]$ & $\mathcal{U}[0,0.40]$ \\
K2-408 b & $\mathcal{U}[3158.43, 3158.63]$ & $\mathcal{U}[20.97, 20.99]$ & $\mathcal{U}[0.01, 5]$ & $\mathcal{U}[0,0.40]$ \\
Kepler-10    b & $\mathcal{U}[169.97, 170.17]$ & $\mathcal{U}[0.83, 0.85]$ & $\mathcal{U}[0.01, 5]$ & $\mathcal{U}[0,0.40]$ \\
Kepler-10 c & $\mathcal{U}[183.89, 184.09]$ & $\mathcal{U}[45.28, 45.30]$ & $\mathcal{U}[0.01, 5]$ & $\mathcal{U}[0,0.40]$ \\
Kepler-112 b & $\mathcal{U}[133.93, 134.13]$ & $\mathcal{U}[8.40, 8.42]$ & $\mathcal{U}[0.01, 5]$ & $\mathcal{U}[0,0.40]$ \\
Kepler-112 c & $\mathcal{U}[143.77, 143.97]$ & $\mathcal{U}[28.56, 28.58]$ & $\mathcal{U}[0.01, 5]$ & $\mathcal{U}[0,0.40]$ \\
Kepler-463 b & $\mathcal{U}[136.56, 136.76]$ & $\mathcal{U}[8.97, 8.99]$ & $\mathcal{U}[0.01, 5]$ & $\mathcal{U}[0,0.40]$ \\
{Kepler-517 b} & $\mathcal{U}[173.24,173.44]$ & $\mathcal{U}[60.92, 60.94]$ & $\mathcal{U}[0.01, 5]$ & $\mathcal{U}[0,0.40]$ \\
TOI-126 b & $\mathcal{U}[1326.60, 1326.80]$ & $\mathcal{U}[3.01, 3.04]$ & $\mathcal{U}[0.01, 5]$ & $\mathcal{U}[0,0.40]$ \\
\enddata
\end{deluxetable*}

\begin{deluxetable*}{lccccccc}
\tablecaption{Priors for HAT-P-26, TOI-1247, TOI-1926, TOI-1927, TOI-2643, and K2-262 systems. In all cases, where applicable, the impact parameter $b$ is sampled within $[0, 1]$, the limb-darkening coefficients $q_1$ and $q_2$ are sampled within $[0, 1]$, and the coupled eccentricity-argument of periastron parameters, $\sqrt{e}\cos\omega$ and $\sqrt{e}\sin\omega$, are sampled within $[-1, +1]$. Radial velocity priors are provided for each instrument as follows: $^\alpha$ HARPS; $^\beta$ CORALIE and HARPS; $^\gamma$ HARPS and FEROS; $^\delta$ CARMENES, HIRES, and PFS.}
\tablehead{\colhead{ID} & \colhead{$T_0$} & \colhead{$P$} & \colhead{$a$} & \colhead{$R_p$} & \colhead{$K$} & \colhead{RV}}
\startdata
TOI-1926 b$^\alpha$ & $\mathcal{U}[2175.27, 2175.47]$ & $\mathcal{U}[1.33, 1.35]$ & $\mathcal{U}[0.01, 5]$ & $\mathcal{U}[0,0.10]$ & $\mathcal{U}[0,0.05]$ & $\mathcal{U}[110.69, 111.72]$ \\
TOI-1927 b$^\beta$ & $\mathcal{U}[3263.14, 3263.34]$ & $\mathcal{U}[4.10, 4.12]$ & $\mathcal{U}[0.01, 5]$ & $\mathcal{U}[0,0.50]$ & $\mathcal{U}[0,0.90]$ & $\mathcal{U}[68.11, 69.30]$; $\mathcal{U}[68.16, 69.28]$ \\
TOI-2643 b$^\gamma$ & $\mathcal{U}[2286.99, 2287.19]$ & $\mathcal{U}[9.16, 9.18]$ & $\mathcal{U}[0.01, 5]$ & $\mathcal{U}[0,0.10]$ & $\mathcal{U}[0,0.05]$ & $\mathcal{U}[9.84, 10.91]$; $\mathcal{U}[9.82, 10.91]$ \\
K2-262 b$^\delta$ & $\mathcal{U}[3352.24, 3352.44]$ & $\mathcal{U}[5.99, 6.01]$ & $\mathcal{U}[0.01, 5]$ & $\mathcal{U}[0,0.10]$ & $\mathcal{U}[0,0.05]$ & $\mathcal{U}[-0.50, 0.52]$; $\mathcal{U}[-0.51, 0.51]$; $\mathcal{U}[-0.51, 0.51]$ \\
\enddata
\end{deluxetable*}

\clearpage\section{On Gaussian Process Regression Phase Diagrams and Orbital Parameters}\label{app:gaussian_process}

\begin{figure*}[!h]
    \centering
    \includegraphics[width = 0.32\linewidth]{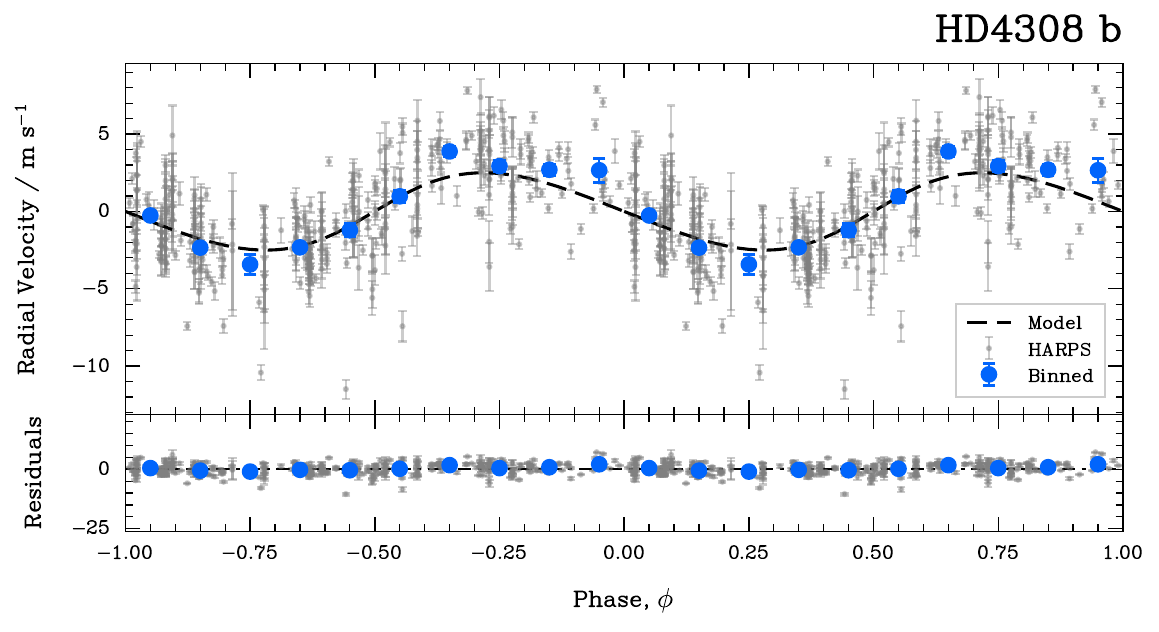}
    \includegraphics[width = 0.32\linewidth]{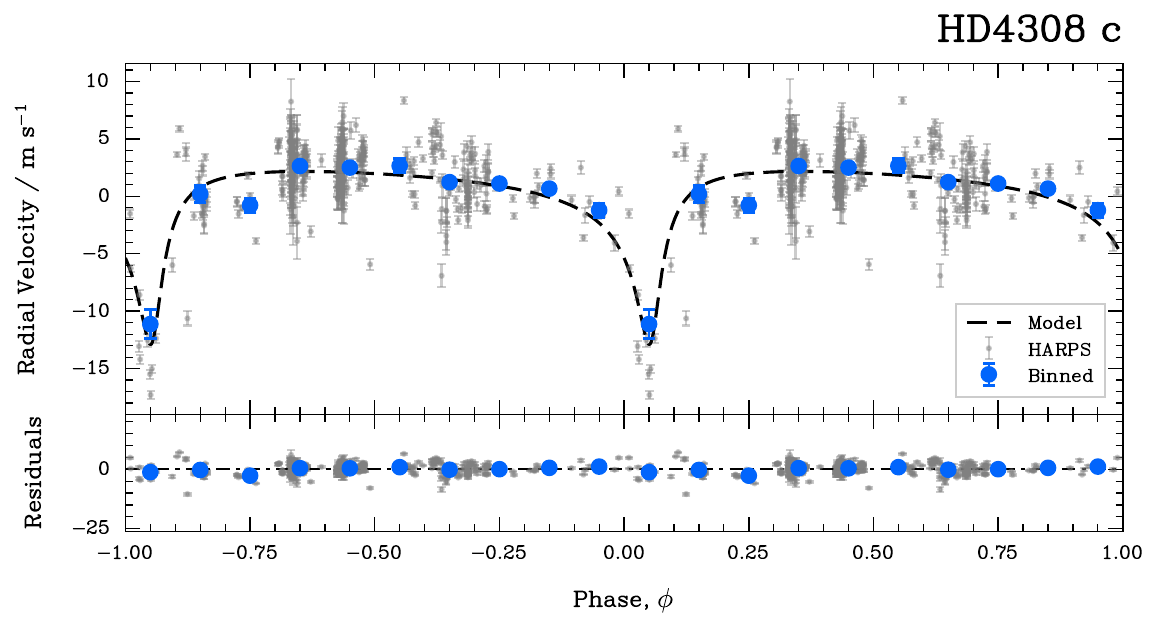}
    \includegraphics[width = 0.32\linewidth]{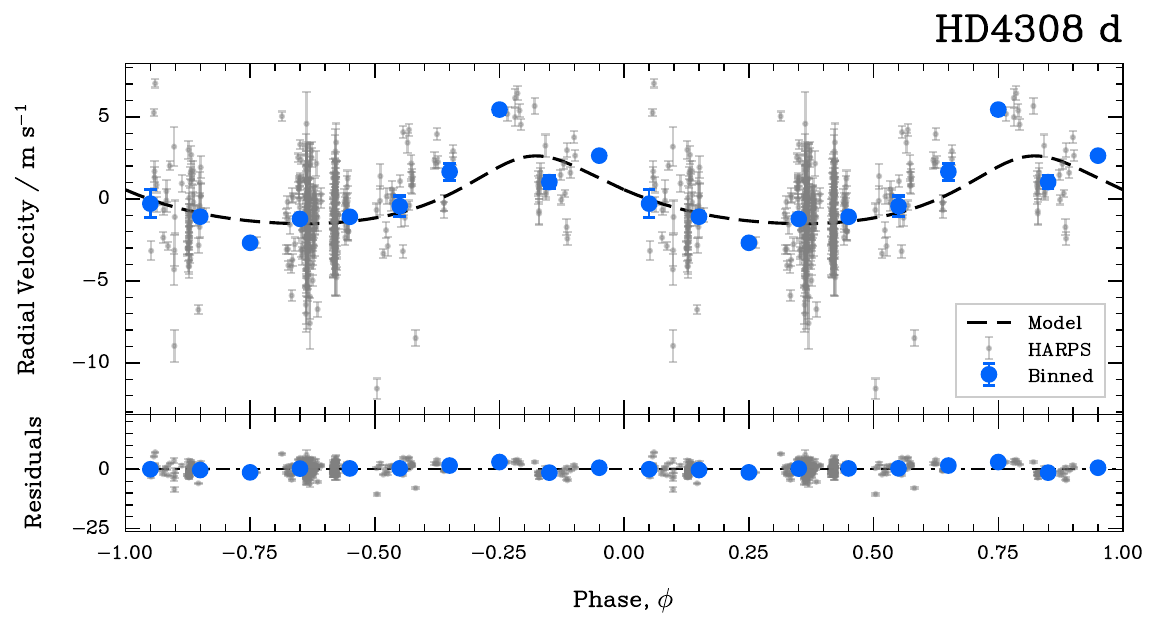}\\
    \includegraphics[width = 0.32\linewidth]{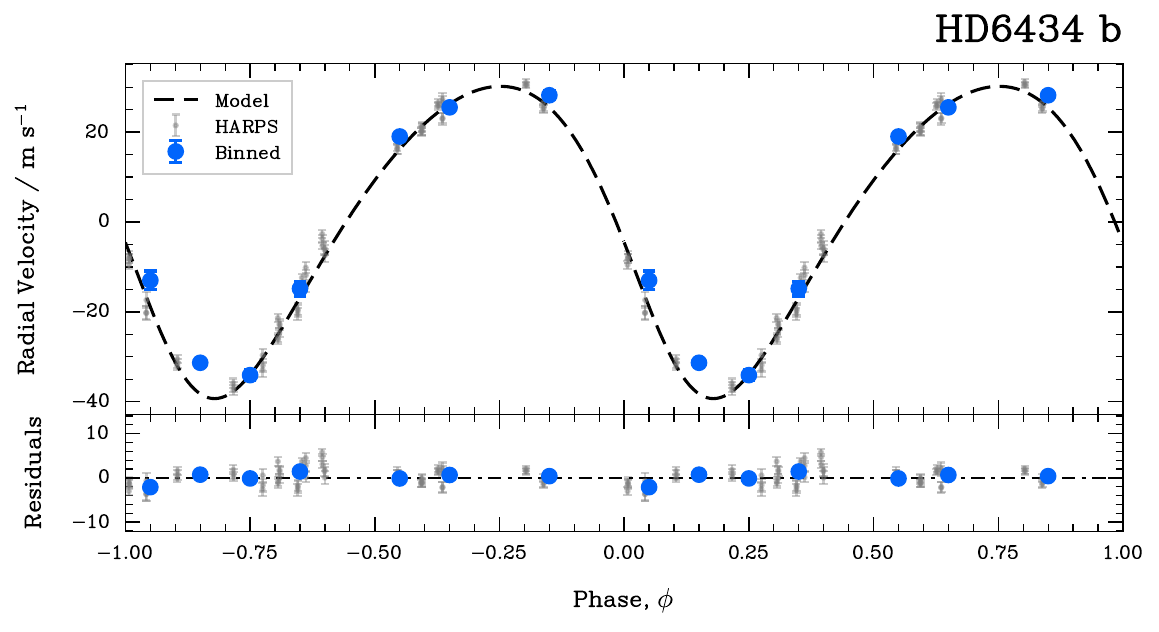}
    \includegraphics[width = 0.32\linewidth]{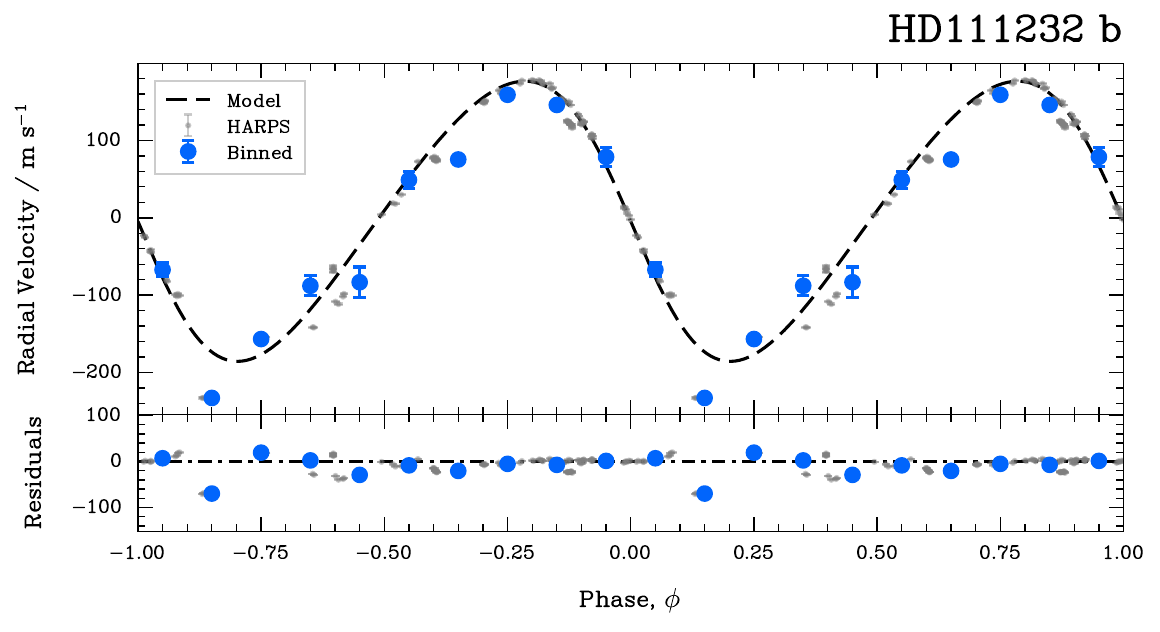}
    \includegraphics[width = 0.32\linewidth]{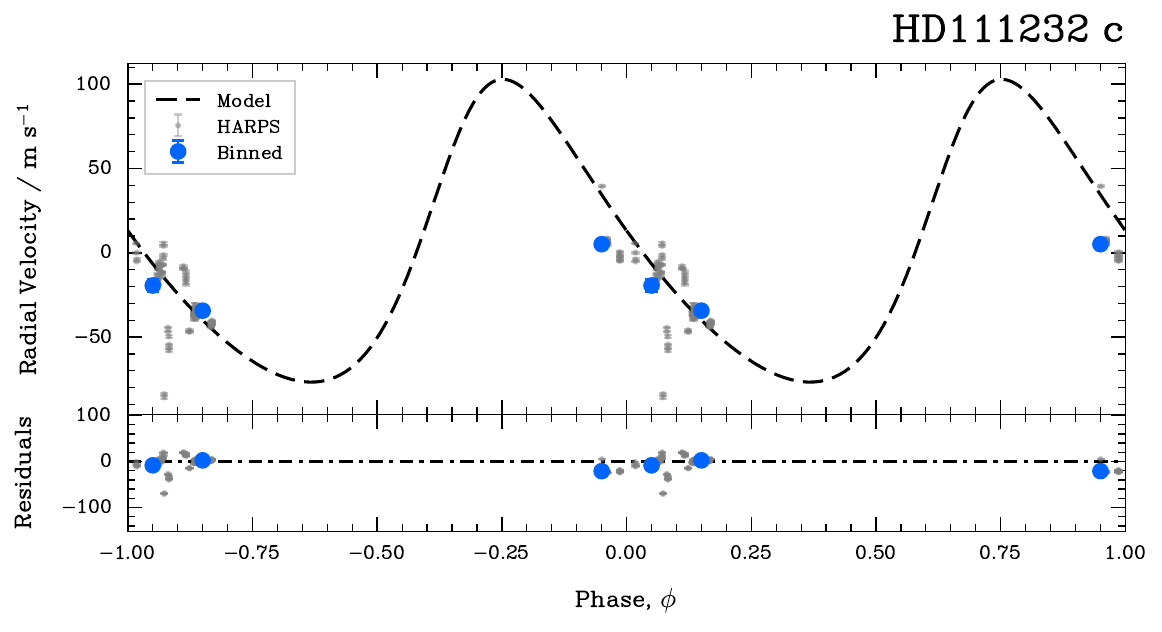}\\
    \includegraphics[width = 0.32\linewidth]{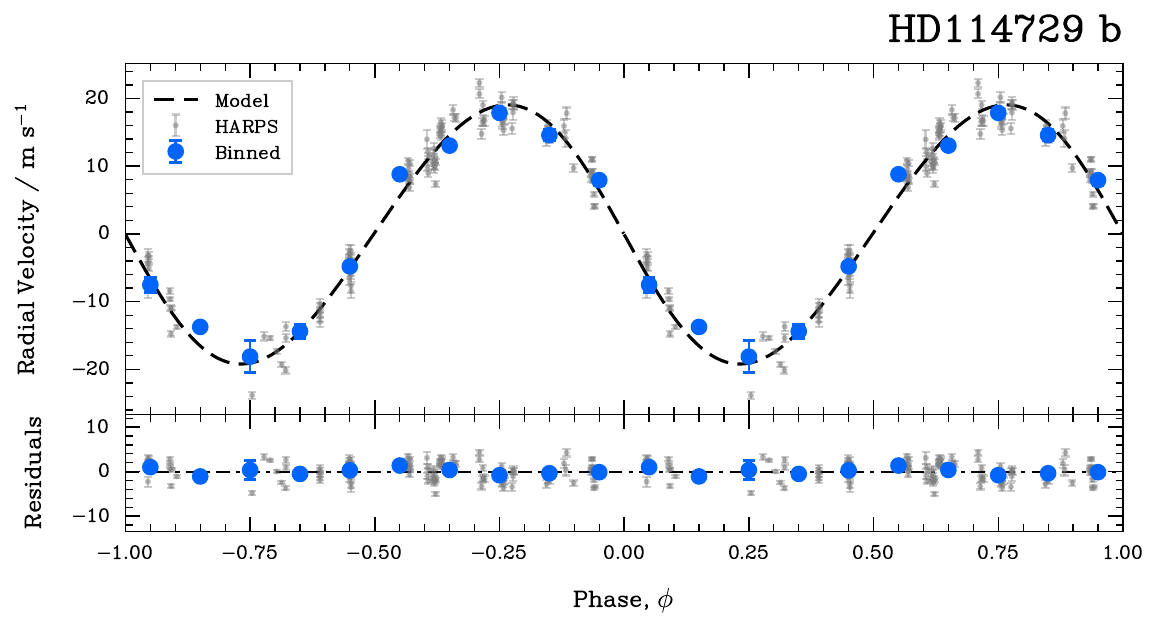}
    \includegraphics[width = 0.32\linewidth]{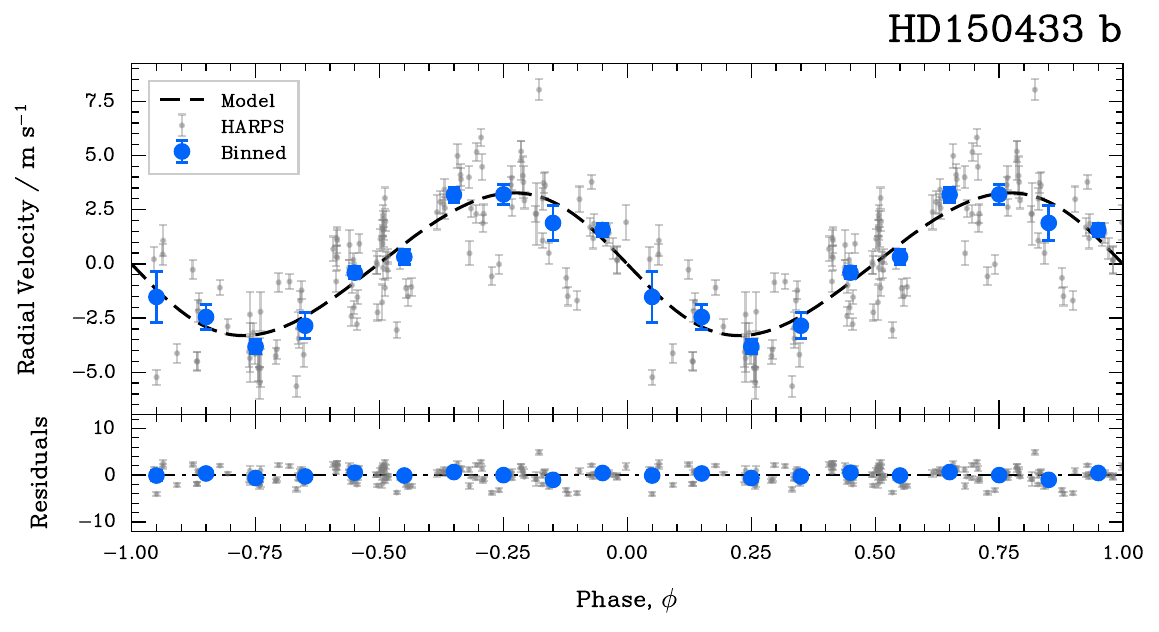}
    \includegraphics[width = 0.32\linewidth]{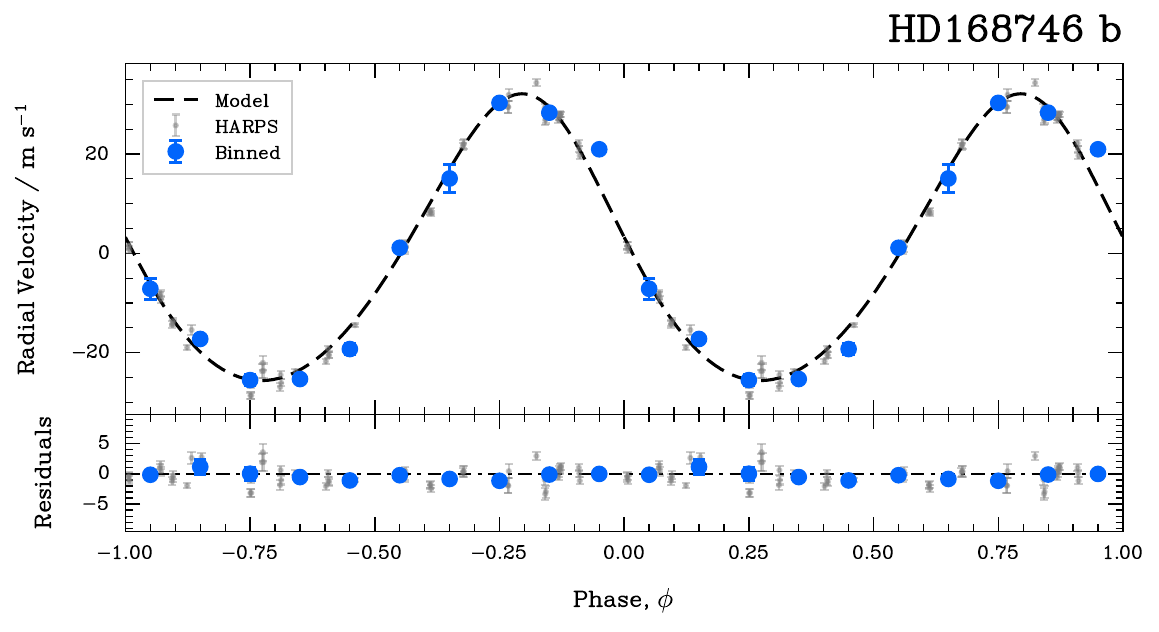}\\
    \includegraphics[width = 0.32\linewidth]{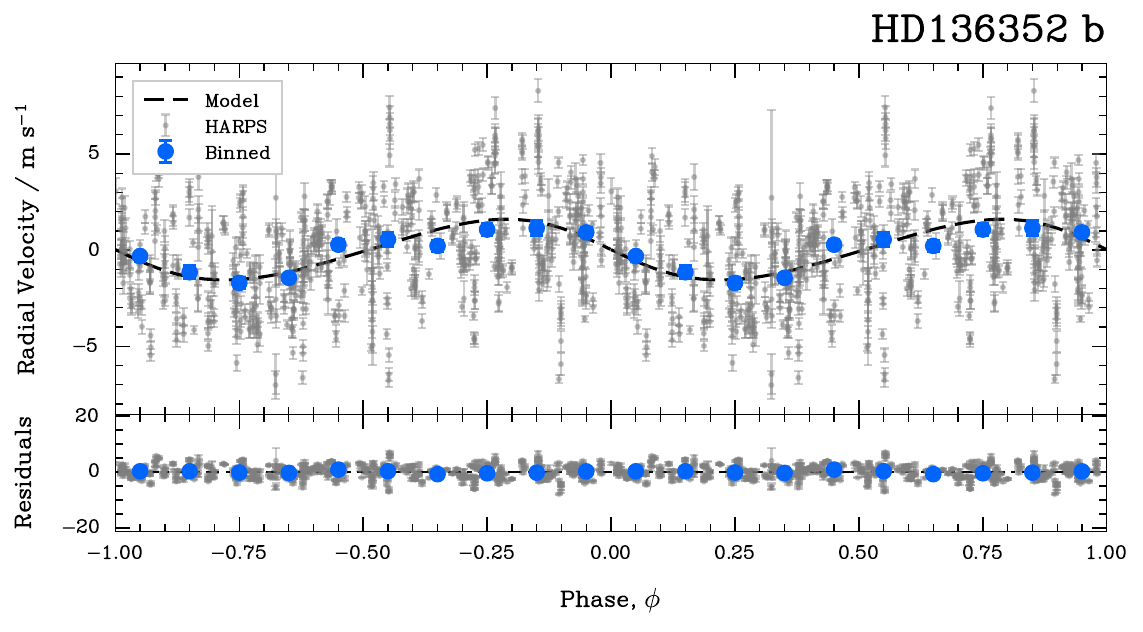}
    \includegraphics[width = 0.32\linewidth]{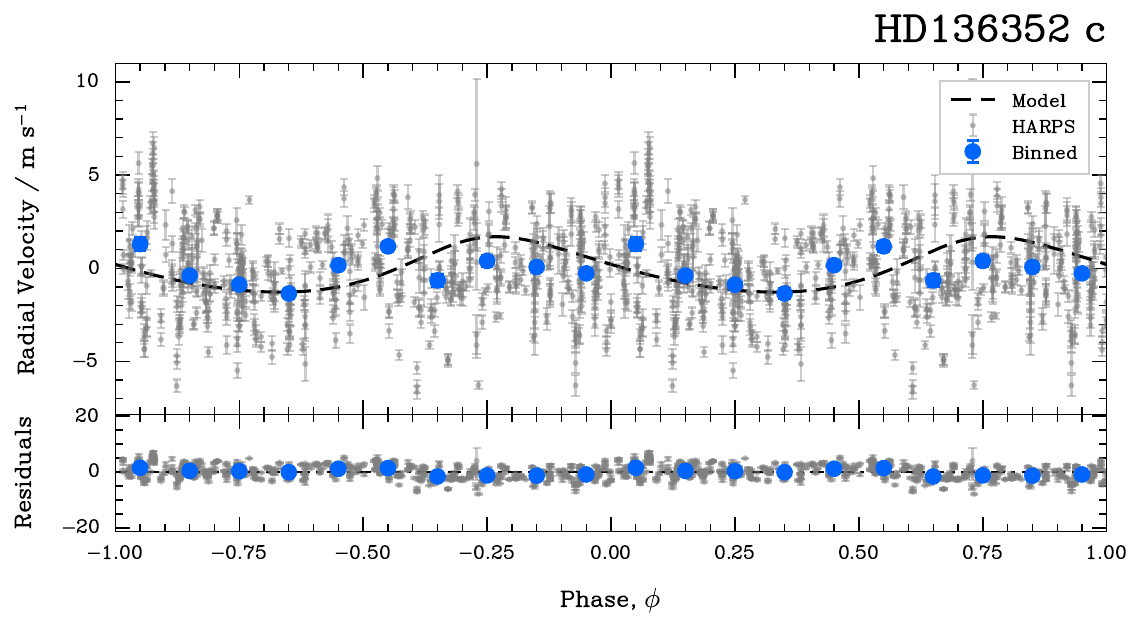}
    \includegraphics[width = 0.32\linewidth]{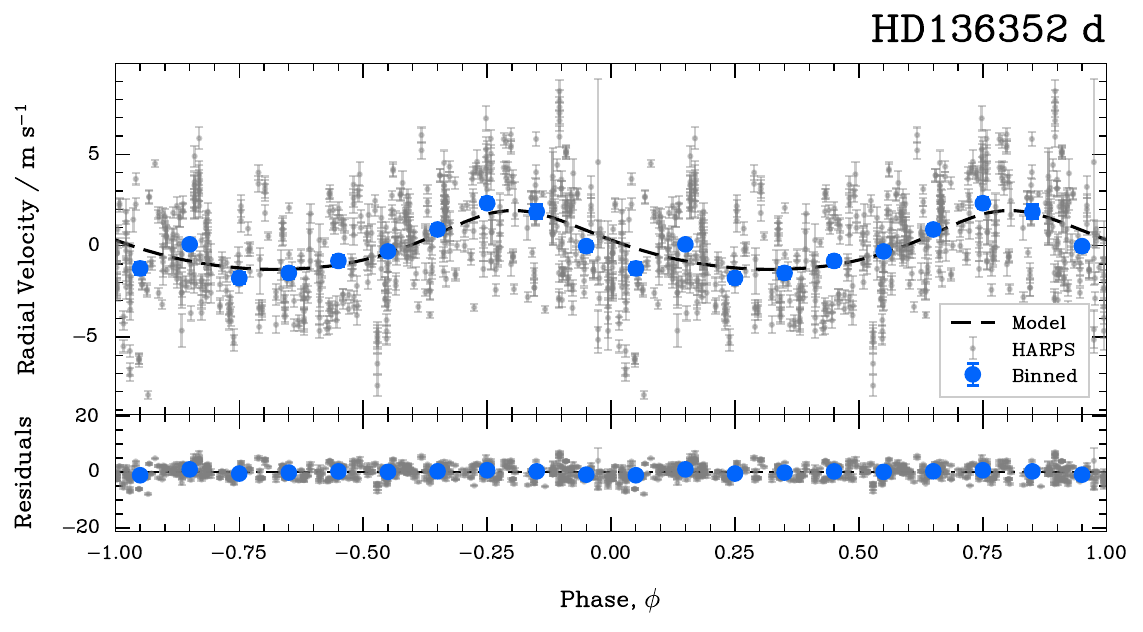}\\
    \includegraphics[width = 0.32\linewidth]{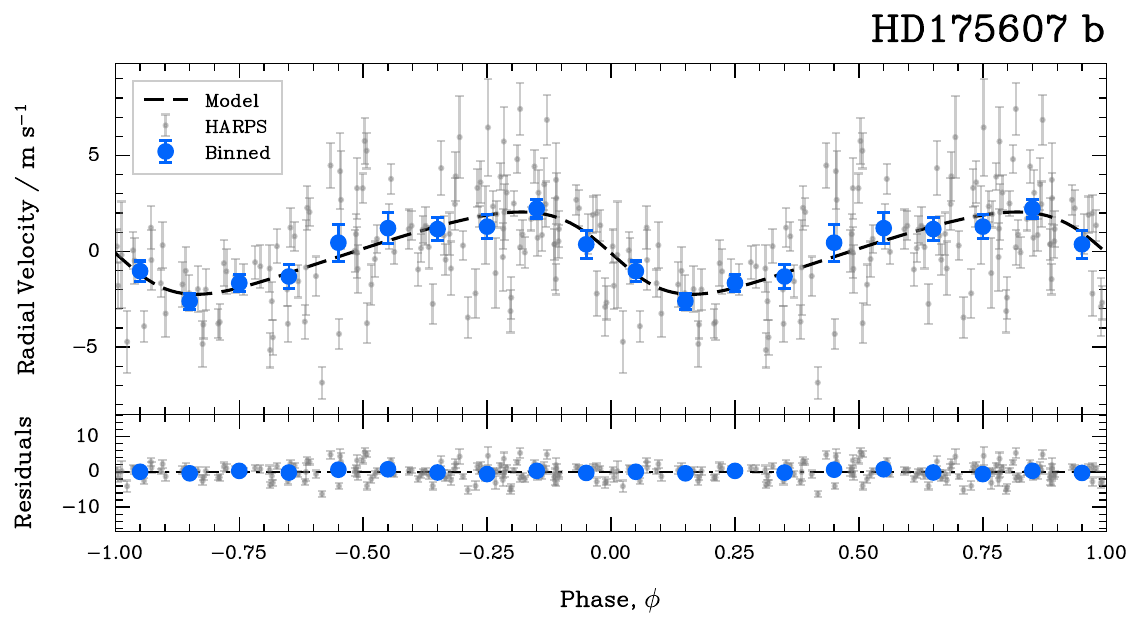}
    \includegraphics[width = 0.32\linewidth]{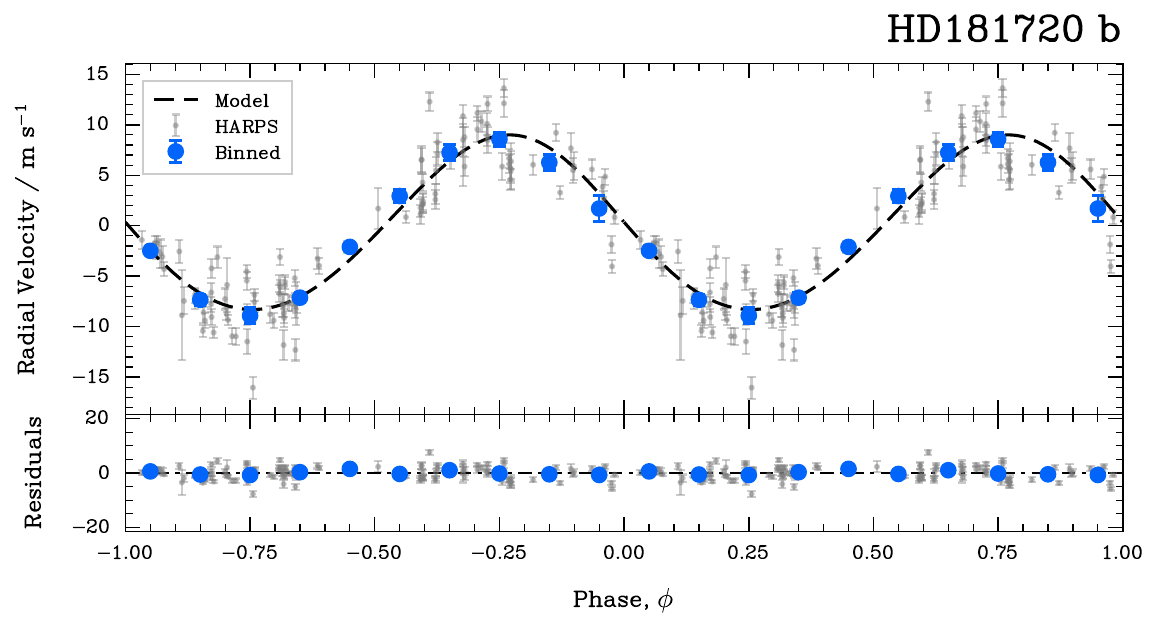}
    \includegraphics[width = 0.32\linewidth]{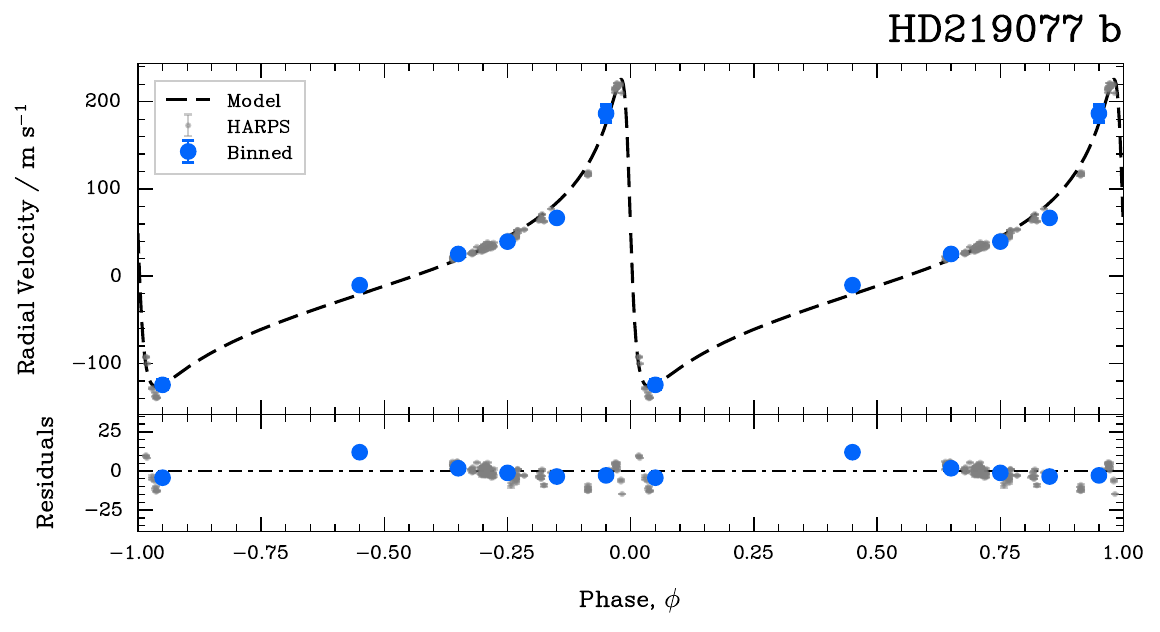}\\
    \includegraphics[width = 0.32\linewidth]{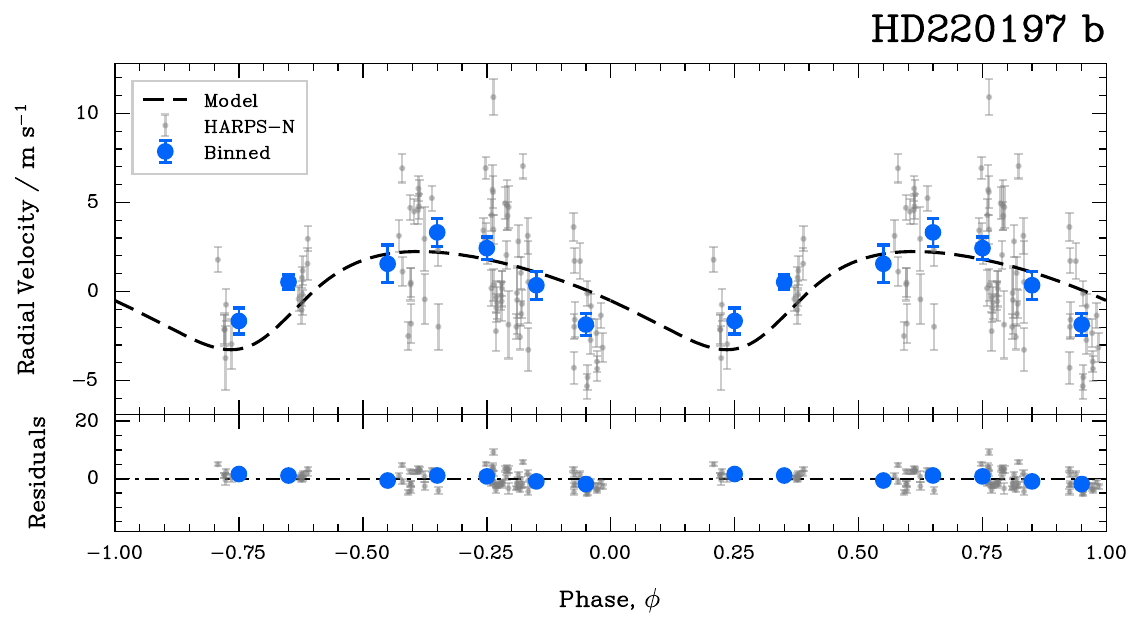}
    \includegraphics[width = 0.32\linewidth]{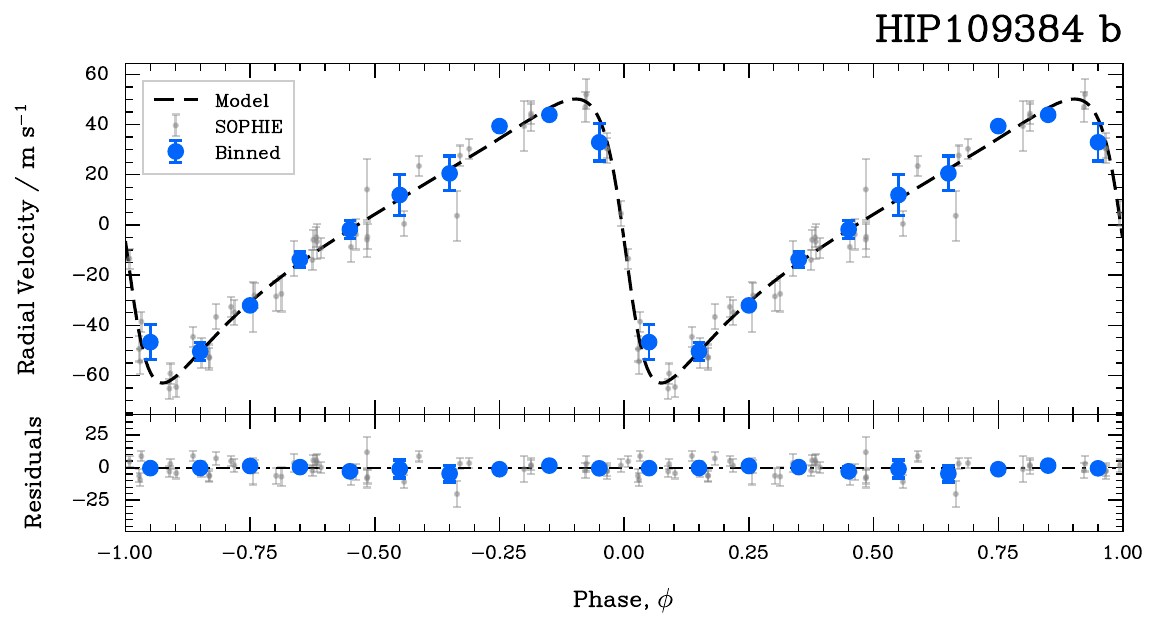}
    \caption{Keplerian-only radial velocity curves for systems HD4308, HD6434, HD111232, HD114729, HD150433, HD168746, HD136352, HD175607, HD181720, HD219077, HD220197, and HIP109384 showing the best-fitting model over-plotted. The residuals of the fit are displayed at the bottom of each subplot.}
    \label{fig:d1}
\end{figure*}

\begin{figure*}[h]
    \centering
    \includegraphics[width = 0.49\linewidth]{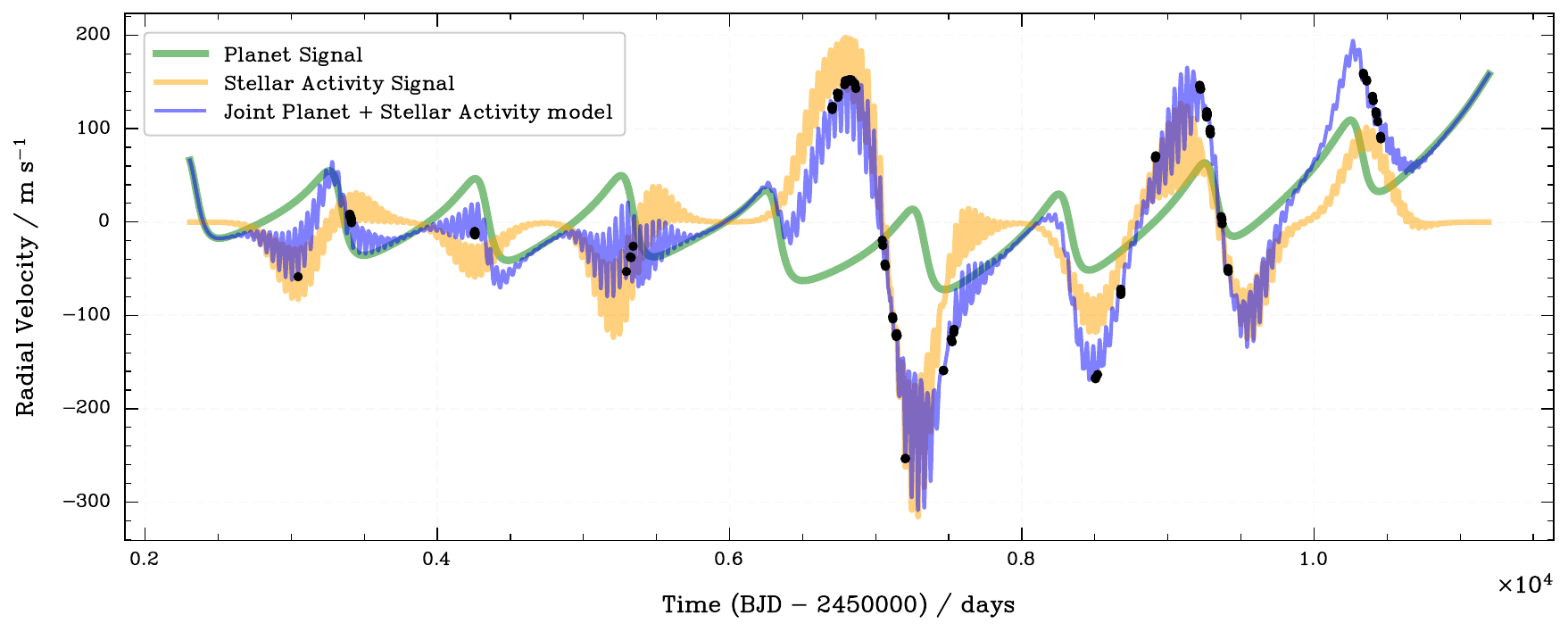}
    \includegraphics[width = 0.49\linewidth]{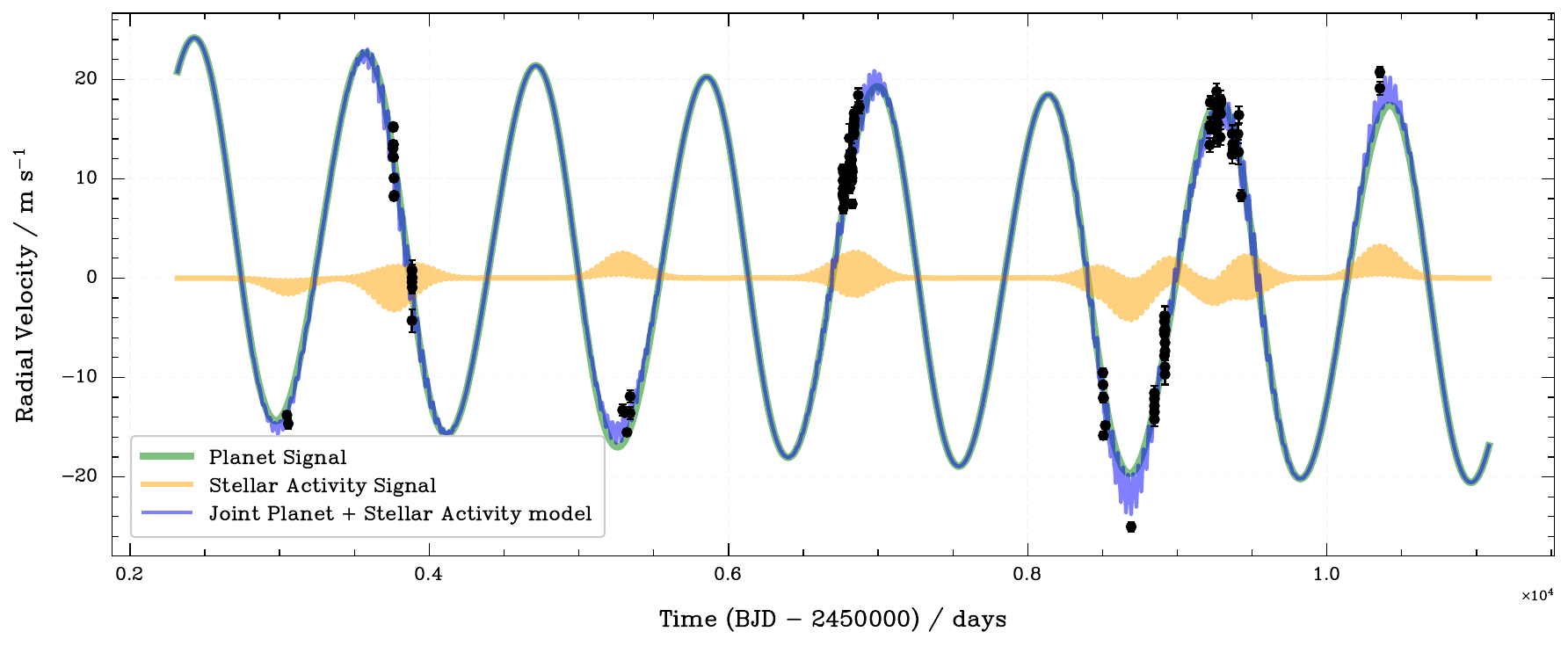}\\
    \includegraphics[width = 0.49\linewidth]{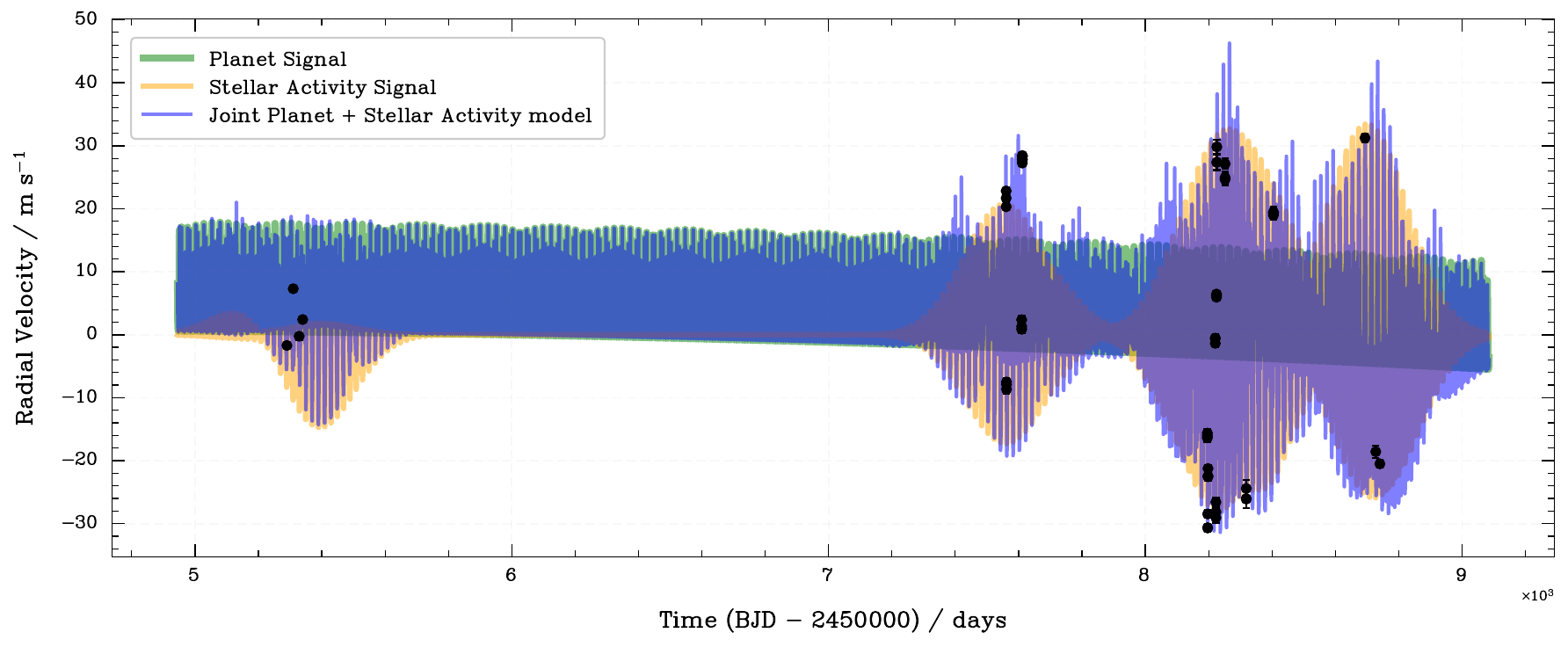}
    \includegraphics[width = 0.49\linewidth]{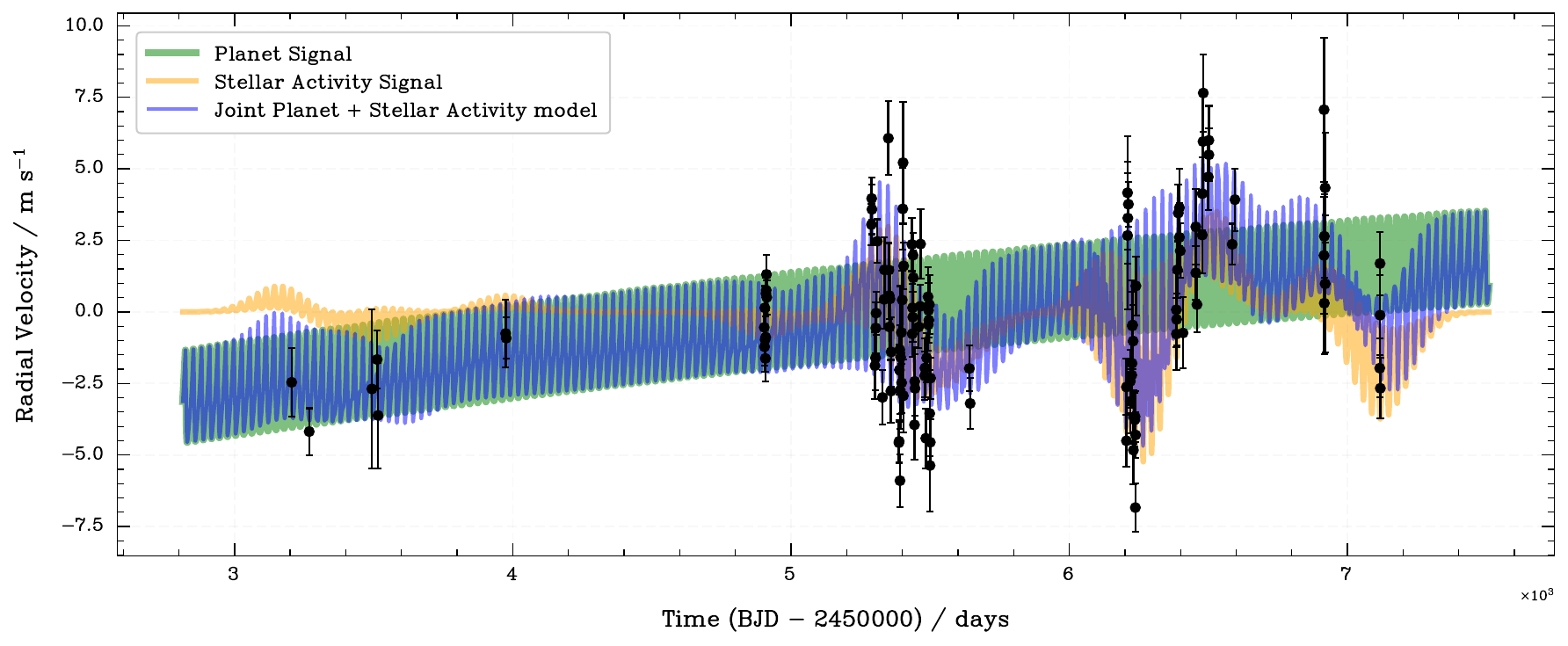}
    \includegraphics[width = 0.49\linewidth]{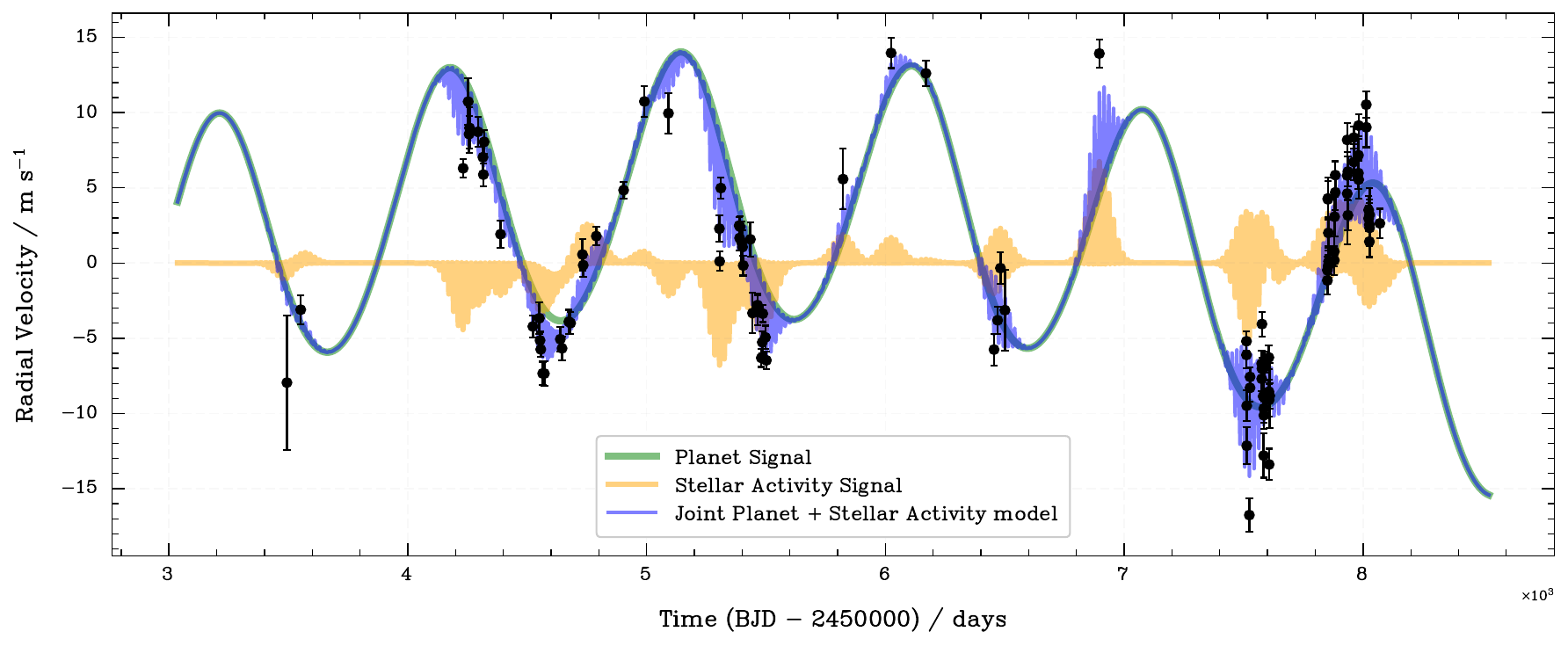}
    \includegraphics[width = 0.49\linewidth]{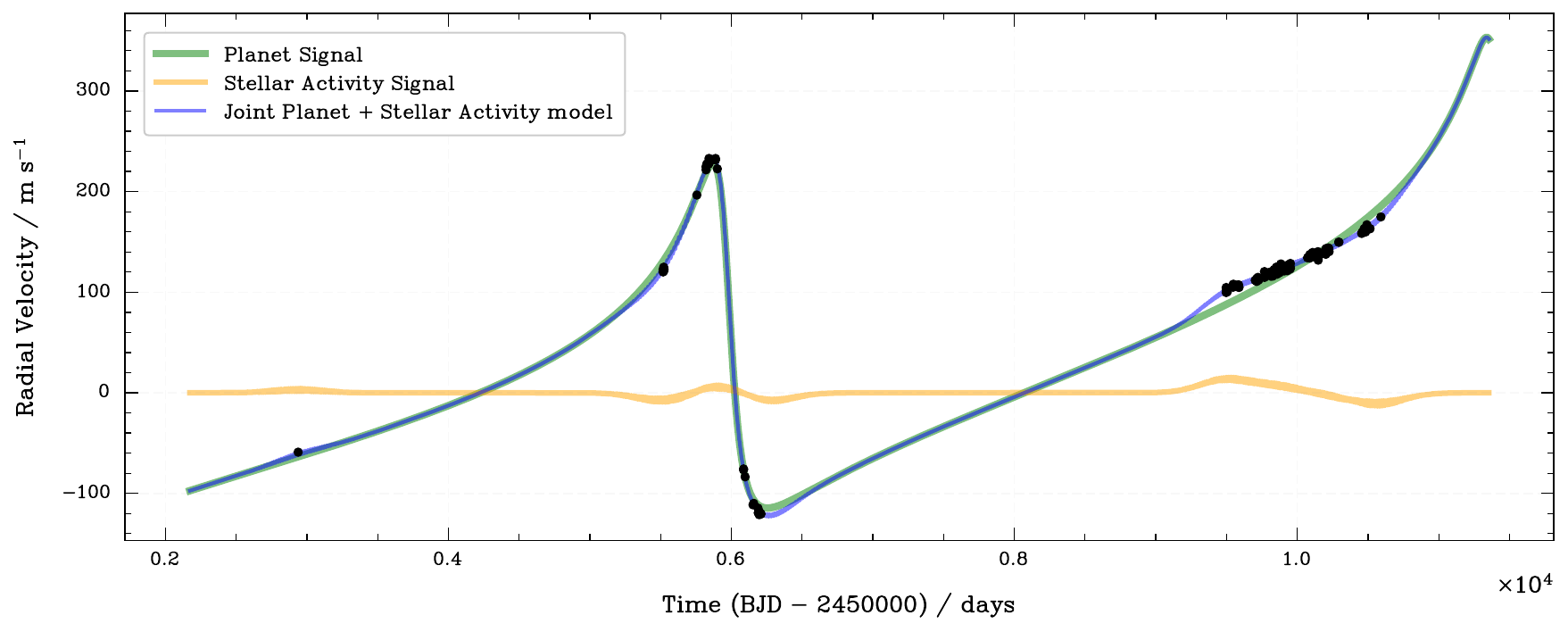}
    \includegraphics[width = 0.49\linewidth]{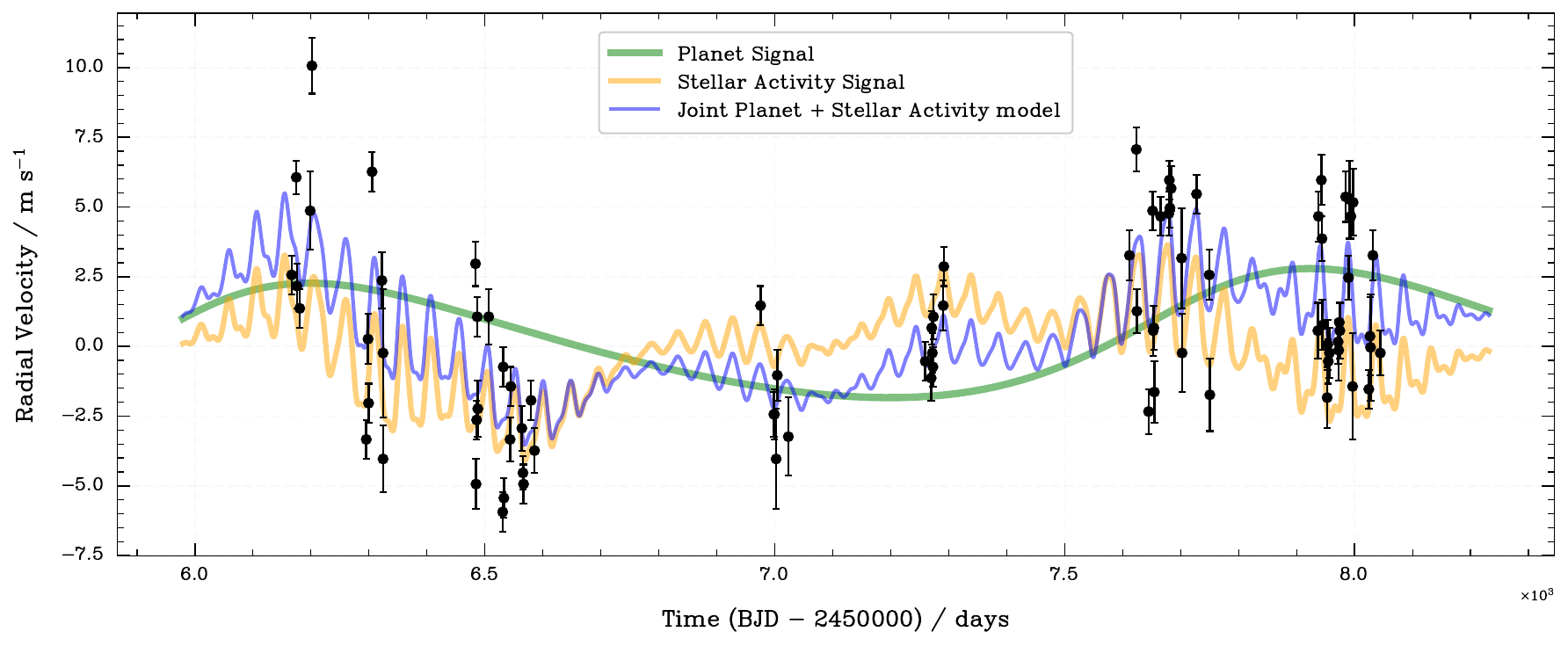}
    \includegraphics[width = 0.49\linewidth]{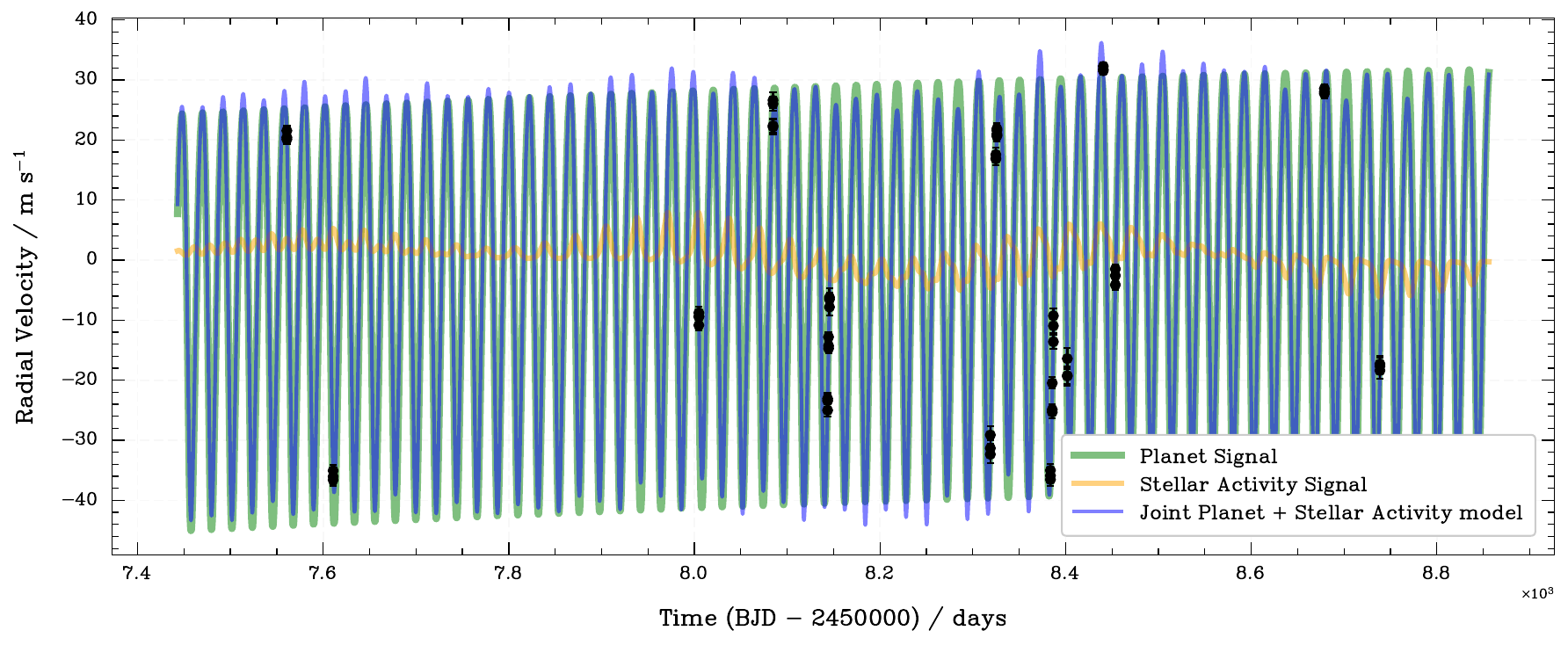}
    \includegraphics[width = 0.49\linewidth]{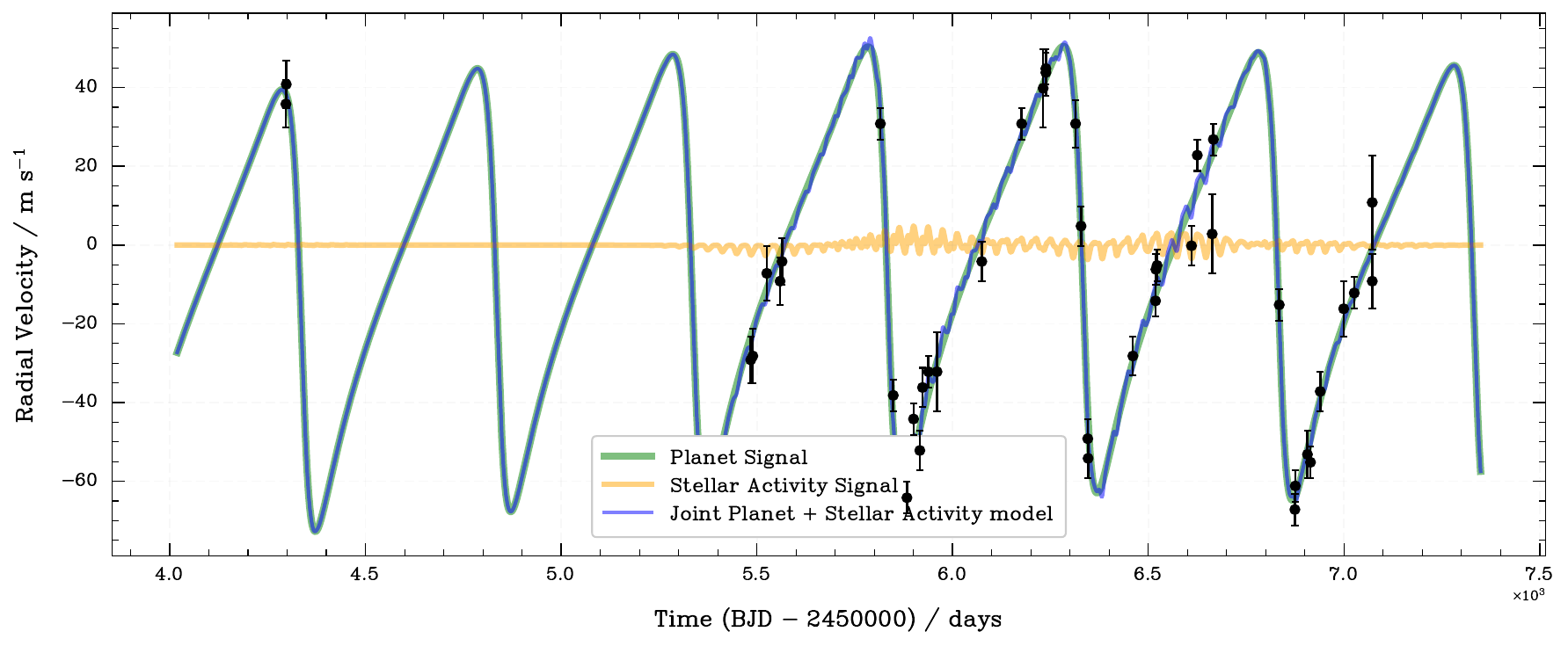}
    \includegraphics[width = 0.49\linewidth]{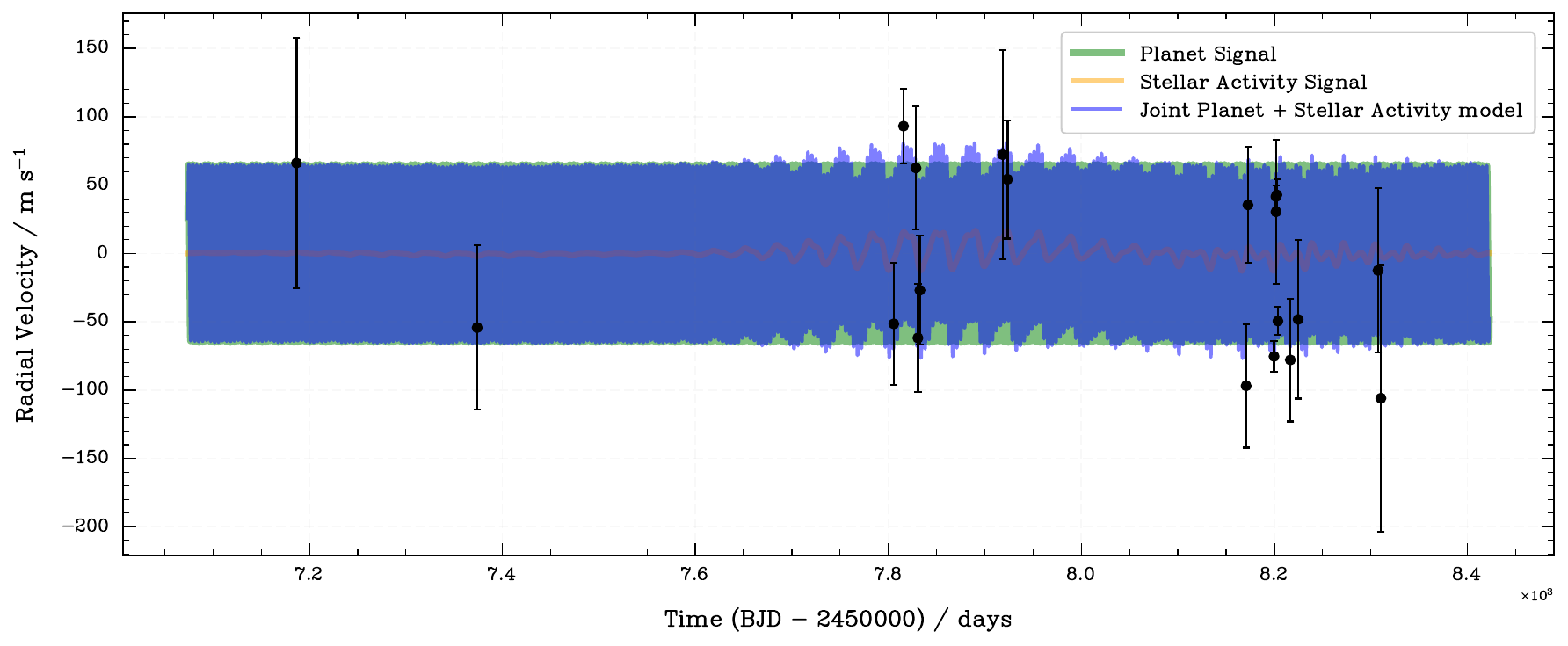}
    \includegraphics[width = 0.49\linewidth]{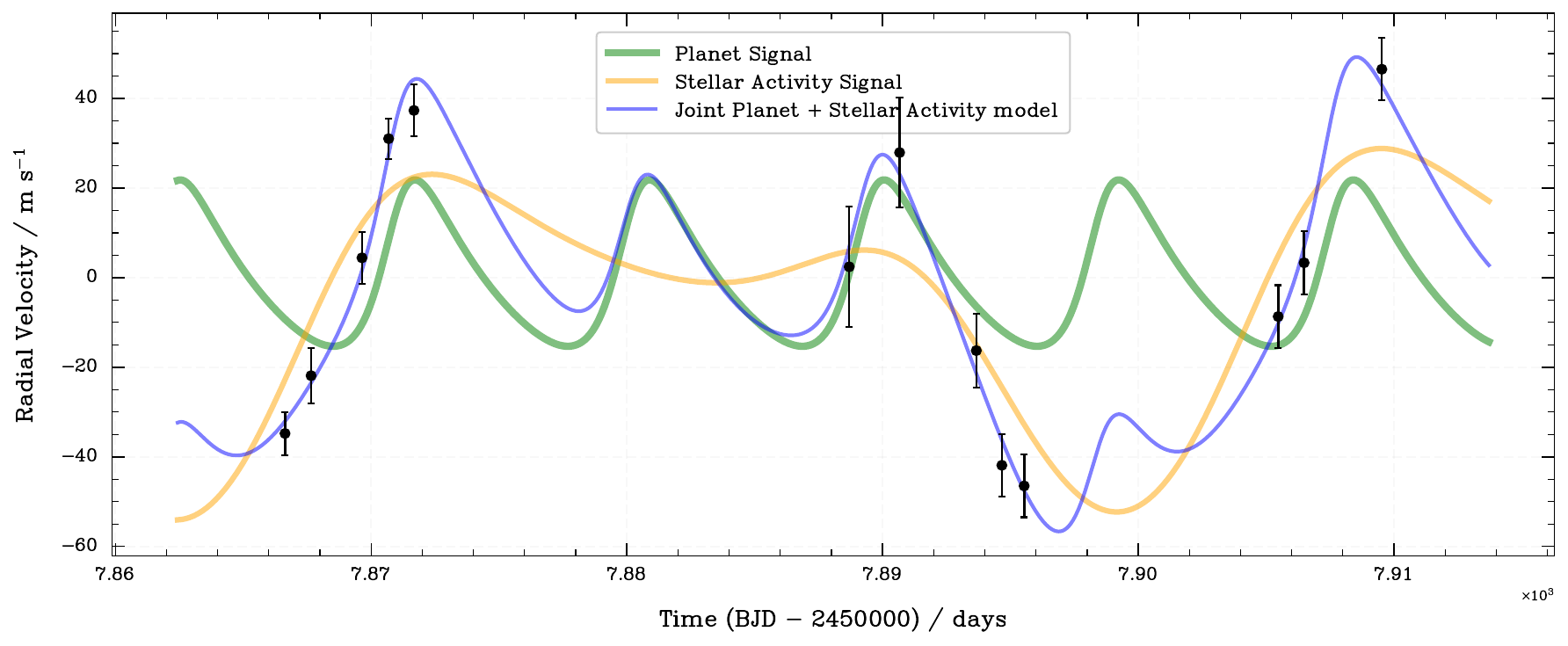}
    \caption{Best-fit joint Keplerian and stellar activity models for HD111232, HD114729, HD168746, HD175607, HD181720, HD219077, HD220197, HD6434, HIP109384, TOI-1927, and TOI-2643. The planet's signal is represented in green, stellar activity modulations are shown in yellow, and the joint model is depicted in blue.}
    \label{fig:gps1}
\end{figure*}

In this Section, we compare the results obtained using a Keplerian-only model with those obtained using a Keplerian model combined with GP regression for the systems analysed in this work following Section \ref{sec:rv_transit_analysis}. Figure \ref{fig:d1} shows the phase-folded radial velocity diagrams for the systems analysed using only a Keplerian framework, whilst Figure \ref{fig:gps1} shows the best-fit models decomposed into the planetary signal (green curves in all panels), stellar activity signal (in yellow), and their combined contribution (in blue). Table \ref{tab:d1} presents the resulting GP hyper-parameters for the analysed thick disc hosts $A_0$, $A_1$, $\lambda_e$, $\lambda_p$, and $P_{\rm GP}$ (see Equations \ref{eq:1}-\ref{eq:3}), and Table \ref{tab:d2} presents the orbital and physical parameters of the thick disc exoplanets derived from the radial velocity-only analysis without GP regression.

\begin{itemize}

    \item {\it HD4308 (b, c, d)}: For all three planets in the {HD4308} system, the inclusion of a GP component is favoured, with $\Delta {\rm BIC} \equiv {\rm BIC}_{\rm GP} - {\rm BIC}_{\rm Keplerian~only} \simeq -20$ and reduced $\chi-$squared values increasing from $\chi^2_{{\rm red}} \sim 0.74$ to $\chi^2_{{\rm red}} \sim 1.0$. The inferred masses are not consistent within $1\sigma$ but overlap at approximately $2\sigma$, and the GP-based solutions should therefore be adopted.

    \item {\it HD136352 (b, c, d)}: The GP model is overwhelmingly favoured, with $\Delta {\rm BIC} \gg -100$, and reduced $\chi-$squared values decreasing from $\chi^2_{{\rm red}} \sim 1.31$ to $\chi^2_{{\rm red}} \sim 0.85-0.95$; the inferred masses are inconsistent at both the $1\sigma$ and $2\sigma$ levels, indicating that the Keplerian-only solutions are strongly biased, and only the GP-based results should be considered physically meaningful.

    \item {\it HIP109384 b}: The GP model is disfavoured, with $\Delta {\rm BIC} \simeq +21$ and reduced $\chi-$squared increasing from $\chi^2_{{\rm red}} \sim 1.07$ to $\chi^2_{{\rm red}} \sim 1.31$; the inferred masses are consistent within $1\sigma$, and the Keplerian-only model should therefore be preferred.

    \item {\it HD168746 b}: {Although the GP model yields a formally lower BIC ($\Delta{\rm BIC} \simeq -91$) and a better reduced  $\chi^2$ ($\chi^2_{\rm red} \sim 0.97$ vs $\sim 1.47$), we would suggest the adoption of a non-GP, Keplerian-only solution. The inferred GP period ($P_{\rm GP} = 19.9^{+6.1}_{-7.1}$ days) is inconsistent with stellar rotation: given $v\sin i < 1$ km s$^{-1}$ and  $R_\star = 1.08~R_\odot$, the stellar rotation period must exceed $\sim$50 days as expect for old ones. Instead, $P_{\rm GP} \approx 3~P_{\rm orb}$, indicating the GP is absorbing power at a harmonic of the orbital frequency rather than modelling genuine stellar activity. As a consequence, the GP solution returns a planet mass consistent with zero ($M_p\sin i = 0.046^{+0.140}_{-0.035}$~M$_{\rm Jup}$), while the non-GP solution ($M_p\sin i = 0.25 \pm 0.02$~M$_{\rm Jup}$) is in excellent agreement with all published values \citep[$0.23-0.27$~M$_{\rm Jup}$;][]{2002A&A...388..632P, 2006ApJ...646..505B, 2021ApJS..255....8R, 2017AJ....153..136S}. With only 52 RV measurements, the GP is prone to overfitting, and the phase-folded RV curve shows a clean,  well-constrained Keplerian with residuals of $\lesssim 2$ m s$^{-1}$, with no evidence of stellar activity contamination.}

    \item {\it HD150433 b}: The GP model is mildly favoured, with $\Delta {\rm BIC} \simeq -5$, the reduced $\chi-$squared remaining close to unity in both cases, and inferred masses consistent within $1\sigma$, suggesting that the GP-based result should be adopted.

    \item {\it HD175607 b}: The GP model is favoured, with $\Delta {\rm BIC} \simeq -16$ and reduced $\chi-$squared increasing slightly from $\chi^2_{{\rm red}} \sim 1.01$ to $\chi^2_{{\rm red}} \sim 1.16$; the inferred masses are consistent within $1\sigma$, and the GP-based solution should therefore be preferred.

    \item {\it HD219077 b}: The GP model is strongly favoured, with $\Delta {\rm BIC} \simeq -232$ and reduced $\chi-$squared decreasing from $\chi^2_{{\rm red}} \sim 1.39$ to $\chi^2_{{\rm red}} \sim 1.05$; the inferred masses are fully consistent within $1\sigma$, and the GP-based parameters should be adopted.

    \item {\it HD111232 (b, c)}: For both planets in the {HD111232} system, the GP model is decisively favoured, with $\Delta {\rm BIC} \gtrsim -1.4 \times 10^{4}$ and reduced $\chi-$squared dropping from $\chi^2_{{\rm red}} \sim 161$ to $\chi^2_{{\rm red}} \sim 1.34$; {the inferred masses from the GP and Keplerian-only models are inconsistent at the $1-\sigma$ level, demonstrating that stellar activity is essential for this system. The GP-corrected semi-amplitude ($K = 43.8^{+4.9}_{-11.8}$ m s$^{-1}$) is substantially lower than literature values derived without activity modelling (e.g., \citealt{2022ApJS..262...21F}), yielding revised minimum masses of $m\sin{i} = 1.43\pm0.67~M_{\rm Jup}$ and $m\sin{i} = 2.09^{+2.28}_{-1.18}~M_{\rm Jup}$ for planets $b$ and $c$, respectively. We also note that the inferred GP period of $P_{\rm GP} = 43.3_{-2.6}^{+3.2}$ days is consistent with the expected rotation period for old and subsequently thick disc stars (e.g., \citealt{2018A&A...619A..73L, 2020MNRAS.495L..61L}).}

    \item {\it HD6434 b}: The GP model is disfavoured, with $\Delta {\rm BIC} \simeq +8$, the reduced $\chi-$squared values being comparable, and the inferred masses consistent within $1\sigma$, indicating that the Keplerian-only solution should be adopted.

    \item {\it HD181720 b}: {The Keplerian-only model is favoured, with $\Delta {\rm BIC} \simeq +21$ and reduced $\chi^2$ increasing from $\chi^2_{\rm red} \sim 0.99$ to $\sim 1.22$ when including a GP; the inferred masses are consistent within $1\sigma$, and the no-GP solution is therefore preferred. Our re-analysis yields, however, near-circular orbit ($e = 0.046^{+0.055}_{-0.033}$), lower than the value reported by \cite{2010A&A...512A..47S} ($e = 0.26 \pm 0.06$), in which we attribute to the significantly expanded dataset: \cite{2010A&A...512A..47S} fitted 28 radial-velocity measurements (see their Table 2), whereas our reanalysis uses 110 measurements spanning a longer baseline, providing greatly improved orbital phase coverage. Notably, \cite{2010A&A...512A..47S} themselves noted residual structure in the data and cautioned that their orbital parameters may be subject to revision. The phase-folded RV curve is consistent with a near-circular orbit, showing a symmetric Keplerian profile incompatible with a moderately eccentric solution.}

    \item {\it HD114729 b}: The inclusion of a GP component is only weakly favoured, with $\Delta {\rm BIC} \simeq -4.5$ and reduced $\chi-$squared increasing from $\chi^2_{{\rm red}} \sim 0.95$ to $\chi^2_{{\rm red}} \sim 1.12$. The inferred planetary masses are fully consistent within $1\sigma$, and the orbital parameters remain statistically stable between the two models, indicating that correlated noise does not significantly bias the Keplerian solution.

    \item {\it HD220197 b}: The inclusion of a GP component is disfavoured, with $\Delta {\rm BIC} \simeq +14$ and reduced $\chi-$squared increasing from $\chi^2_{{\rm red}} \sim 1.03$ to $\chi^2_{{\rm red}} \sim 1.14$ when the GP is included. The inferred planetary masses are fully consistent within $1\sigma$, and the orbital parameters remain statistically compatible between the two models, indicating that correlated noise does not significantly influence the shaping of the radial velocity signal. The Keplerian-only solution should therefore be preferred for this system.

\end{itemize}

These results collectively demonstrate that radial velocity variability in old, thick disc planet hosts is governed by complex connections between planetary signals and residual correlated noise, such that no single noise-modelling prescription is universally applicable across the sample. The statistical preference for GP components in several systems, in contrast to their rejection in others, indicates that the relevance of correlated noise must be evaluated on an individual system basis rather than assumed a priori. Importantly, the frequent impact of GP modelling on the inferred planetary masses, including cases where the masses differ beyond the $2\sigma$ level, shows that stellar activity, granulation, and long-timescale photospheric or instrumental correlations remain physically significant even at advanced stellar ages, and neglecting such correlated noises in old stars can therefore lead to systematically biased mass estimates, particularly for low-amplitude and multi-planet systems, with direct implications for population-level inferences on planet formation, composition, and long-term dynamical stability in the early Galactic disc. 

\begin{deluxetable*}{lccccc}
\tablecaption{Gaussian Process Regression parameters for thick disc exoplanet hosts. For every system in which a GP model was employed, the hyper-parameters were initialised under uniform priors defined as: $A_0\sim\mathcal{U}(0, 0.6)$, $A_1\sim\mathcal{U}(0, 0.6)$, $\lambda_e\sim\mathcal{U}(1, 200)$, $\lambda_p\sim\mathcal{U}(0.2, 0.7)$, and $P_{\rm GP}\sim\mathcal{U}(1, 50)$.}
\label{tab:gp_parameters}
\tabletypesize{\scriptsize}
\tablehead{
\colhead{\bf ID} & \colhead{$A_0$} & \colhead{$A_1$} & \colhead{$\lambda_e$} & \colhead{$\lambda_p$} & \colhead{$P_{\rm GP}$} \\
\colhead{} & \colhead{} & \colhead{} & \colhead{(days)} & \colhead{(days)} & \colhead{}
}
\startdata
HD111232 & 0.0634 $\pm$ 0.0094 & 0.0272 $\pm$ 0.0248 & 162.1 $\pm$ 22.5 & 0.7 $\pm$ 0.0 & 43.292 $\pm$ 2.933 \\
HD114729 & 0.0022 $\pm$ 0.0008 & 0.0056 $\pm$ 0.0046 & 134.4 $\pm$ 59.7 & 0.5 $\pm$ 0.1 & 34.464 $\pm$ 4.596 \\
HD136352b & 0.3988 $\pm$ 0.0455 & 0.3538 $\pm$ 0.0550 & 136.6 $\pm$ 20.2 & 0.5 $\pm$ 0.0 & 34.740 $\pm$ 3.008 \\
HD136352c & 0.0811 $\pm$ 0.0741 & 0.4468 $\pm$ 0.0725 & 73.0 $\pm$ 14.6 & 0.7 $\pm$ 0.0 & 39.481 $\pm$ 4.661 \\
HD136352d & 0.2875 $\pm$ 0.0422 & 0.2851 $\pm$ 0.0736 & 153.2 $\pm$ 8.6 & 0.4 $\pm$ 0.1 & 19.712 $\pm$ 5.748 \\
HD150433 & 0.0016 $\pm$ 0.0003 & 0.0023 $\pm$ 0.0019 & 30.1 $\pm$ 10.3 & 0.6 $\pm$ 0.1 & 21.566 $\pm$ 1.694 \\
HD168746 & 0.0134 $\pm$ 0.0121 & 0.0439 $\pm$ 0.0335 & 129.0 $\pm$ 49.3 & 0.4 $\pm$ 0.2 & 19.903 $\pm$ 6.591 \\
HD175607 & 0.0023 $\pm$ 0.0005 & 0.0046 $\pm$ 0.0033 & 105.7 $\pm$ 42.0 & 0.6 $\pm$ 0.1 & 30.454 $\pm$ 1.473 \\
HD181720 & 0.0018 $\pm$ 0.0007 & 0.0032 $\pm$ 0.0024 & 45.9 $\pm$ 39.0 & 0.4 $\pm$ 0.2 & 19.784 $\pm$ 4.763 \\
HD219077 & 0.0059 $\pm$ 0.0008 & 0.0017 $\pm$ 0.0016 & 186.2 $\pm$ 15.9 & 0.7 $\pm$ 0.0 & 16.918 $\pm$ 0.508 \\
HD220197 & 0.0024 $\pm$ 0.0006 & 0.0056 $\pm$ 0.0053 & 99.0 $\pm$ 57.2 & 0.6 $\pm$ 0.1 & 47.792 $\pm$ 3.475 \\
HD4308 & 0.0044 $\pm$ 0.0009 & 0.0082 $\pm$ 0.0062 & 56.4 $\pm$ 15.9 & 0.5 $\pm$ 0.1 & 31.238 $\pm$ 0.486 \\
HD6434 & 0.0037 $\pm$ 0.0026 & 0.0113 $\pm$ 0.0129 & 105.7 $\pm$ 65.1 & 0.5 $\pm$ 0.1 & 33.341 $\pm$ 10.583 \\
HIP109384 & 0.0032 $\pm$ 0.0024 & 0.0153 $\pm$ 0.0157 & 106.3 $\pm$ 62.1 & 0.5 $\pm$ 0.1 & 36.959 $\pm$ 12.097 \\
TOI-1927 & 0.0207 $\pm$ 0.0206 & 0.0775 $\pm$ 0.0867 & 105.3 $\pm$ 65.3 & 0.5 $\pm$ 0.2 & 33.515 $\pm$ 14.118 \\
TOI-2643 & 0.0432 $\pm$ 0.0468 & 0.1393 $\pm$ 0.1434 & 113.3 $\pm$ 62.4 & 0.6 $\pm$ 0.1 & 36.927 $\pm$ 6.604 \\
\enddata
\tablecomments{The hyper-parameters $A_0$ and $A_1$ represent the amplitude of the activity-induced variations and contributions from the derivative of the GP, respectively, capturing correlations in radial velocities slopes, $P_{\rm GP}$ corresponds to the characteristic recurrence timescale of active regions, typically associated to the stellar rotation period, while $\lambda_p$ and $\lambda_e$ control the smoothness of the periodic component and the evolutionary timescale of active regions, respectively (see Section \ref{sec:rv_transit_analysis}).}
\label{tab:d1}
\end{deluxetable*}

\begin{deluxetable*}{lccccc}
\tablecaption{Orbital and physical parameters of thick disc exoplanets from radial velocity-only analysis without considering Gaussian Process Regression.}
\label{tab:single_method_planets_nogp}
\tabletypesize{\scriptsize}
\tablehead{
\colhead{\bf ID} & \colhead{$P$} & \colhead{$M_p$} & \colhead{$e$} & \colhead{$T_0$} & \colhead{$\gamma$} \\
\colhead{} & \colhead{(days)} & \colhead{(M$_{\rm Jup}$)} & \colhead{} & \colhead{(BJD)} & \colhead{(km s$^{-1}$)}
}
\startdata
HD111232 b & 1165.98 $\pm$ 79.62 & 8.26 $\pm$ 1.94 & 0.13 $\pm$ 0.19 & 2385.32 $\pm$ 15.15 & 104.43 $\pm$ 0.14; 104.39 $\pm$ 0.16 \\
HD111232 c & 33926.82 $\pm$ 1338.18 & 11.46 $\pm$ 14.14 & 0.23 $\pm$ 0.26 & 4723.12 $\pm$ 1950.47 & 104.43 $\pm$ 0.14; 104.39 $\pm$ 0.16 \\
HD114729 b & 1142.03 $\pm$ 114.31 & 1.04 $\pm$ 0.33 & 0.05 $\pm$ 0.38 & 7259.29 $\pm$ 92.09 & 64.989 $\pm$ 0.003 \\
HD136352 b & 11.58 $\pm$ 0.08 & 0.02 $\pm$ 0.01 & 0.11 $\pm$ 0.10 & 3513.48 $\pm$ 1.73 & -68.71 $\pm$ $9.3\times10^{-4}$ \\
HD136352 c & 32.08 $\pm$ 3.83 & 0.01 $\pm$ 0.01 & 0.17 $\pm$ 0.34 & 4175.51 $\pm$ 15.03 & -68.71 $\pm$ $9.3\times10^{-4}$ \\
HD136352 d & 107.64 $\pm$ 0.20 & 0.034 $\pm$ 0.003 & 0.20 $\pm$ 0.19 & 4283.77 $\pm$ 8.35 & -68.71 $\pm$ $9.3\times10^{-4}$ \\
HD150433 b & 1008.85 $\pm$ 11.95 & 0.15 $\pm$ 0.01 & 0.07 $\pm$ 0.06 & 6123.25 $\pm$ 25.08 & -40.11 $\pm$ $2.2\times10^{-4}$ \\
HD168746 b & 6.40 $\pm$ 0.57 & 0.25 $\pm$ 0.02 & 0.12 $\pm$ 0.06 & 8470.64 $\pm$ 445.10 & -25.542 $\pm$ 0.004 \\
HD175607 b & 29.06 $\pm$ 0.06 & 0.025 $\pm$ 0.005 & 0.25 $\pm$ 0.19 & 5556.95 $\pm$ 29.15 & -91.89 $\pm$ $3.1\times10^{-4}$ \\
HD181720 b & 972.85 $\pm$ 9.67 & 0.41 $\pm$ 0.02 & 0.05 $\pm$ 0.04 & 7275.18 $\pm$ 29.87 & -45.33 $\pm$ $4.2\times10^{-4}$ \\
HD219077 b & 5481.96 $\pm$ 56.29 & 9.15 $\pm$ 4.29 & 0.80 $\pm$ 0.05 & 5996.28 $\pm$ 600.92 & -30.79 $\pm$ 0.07; -31.04 $\pm$ 0.07 \\
HD220197 b & 1742.14 $\pm$ 33.49 & 0.14 $\pm$ 0.05 & 0.27 $\pm$ 0.10 & 6614.04 $\pm$ 14.83 & -40.23 $\pm$ $5.1\times10^{-4}$ \\
HD4308 b & 15.62 $\pm$ 0.21 & 0.03 $\pm$ 0.01 & 0.10 $\pm$ 0.09 & 3396.65 $\pm$ 547.07 & 95.26 $\pm$ $4.3\times10^{-4}$ \\
HD4308 c & 833.30 $\pm$ 2.89 & 0.22 $\pm$ 0.03 & 0.75 $\pm$ 0.05 & 4678.59 $\pm$ 15.38 & 95.26 $\pm$ $4.3\times10^{-4}$ \\
HD4308 d & 1475.93 $\pm$ 38.17 & 0.10 $\pm$ 0.03 & 0.27 $\pm$ 0.22 & 4228.80 $\pm$ 163.51 & 95.26 $\pm$ $4.3\times10^{-4}$ \\
HD6434 b & 22.01 $\pm$ 0.01 & 0.48 $\pm$ 0.04 & 0.17 $\pm$ 0.01 & 8356.87 $\pm$ 396.10 & 23.096 $\pm$ 0.002 \\
HIP109384 b & 499.02 $\pm$ 1.52 & 1.55 $\pm$ 0.06 & 0.55 $\pm$ 0.02 & 6331.53 $\pm$ 1.86 & -63.671 $\pm$ 0.001 \\
\enddata
\tablecomments{Columns indicate: $P$ = Orbital period in days, $M_p$ = Companion's mass in Jupiter masses, $e$ = Eccentricity, $T_0$ = Transit midpoint or time of periastron passage in Barycentric Julian Date (BJD$-$2450000), $\gamma$ = Systemic radial velocity for the corresponding instrument(s) in km s$^{-1}$ (multiple values separated by semicolons indicate measurements from different instruments).}
\label{tab:d2}
\end{deluxetable*}

\end{document}